\RequirePackage{fix-cm}
\RequirePackage{amsmath}
\documentclass{svjour3}                     
\smartqed  
\usepackage{amssymb}
\usepackage{graphicx}
\usepackage{threeparttable}
\usepackage{tabularx}
\usepackage{multirow}
\usepackage{xspace}
\usepackage{natbib}
\usepackage{svg}
\usepackage{flushend}
\usepackage{framed}
\usepackage{mdframed}
\usepackage{paralist}
\usepackage{marvosym}
\usepackage{array}
\usepackage{wasysym}
\usepackage{newtxtext,newtxmath}
\usepackage{comment}    
\usepackage{rotating}  
\usepackage{afterpage}
\usepackage{blindtext}
\usepackage{adjustbox}
\usepackage{ifthen}
\usepackage[normalem]{ulem} 
\usepackage{algorithmic}
\usepackage{textcomp}
\usepackage{url}
\usepackage{subfig}
\usepackage{todonotes}
\usepackage{array}
\usepackage{enumitem}
\newcolumntype{C}[1]{>{\centering\arraybackslash}p{#1}}
\usepackage{tikz}
\usepackage{pdflscape} 
\newcolumntype{P}[1]{>{\centering\arraybackslash}p{#1}}
\newcolumntype{M}[1]{>{\centering\arraybackslash}m{#1}}

\definecolor{mircolor}{rgb}{0.4,0.6,0.2}

{
 }

\tikzstyle{mybox} = [draw=black, very thick, rectangle, rounded corners, inner ysep=5pt, inner xsep=5pt]

\newcommand{\seclabel}[1]{\label{sec:#1}}
\newcommand{\secref}[1]{\autoref{sec:#1}}
\newcommand{\figlabel}[1]{\label{fig:#1}}
\newcommand{\figref}[1]{\autoref{fig:#1}}
\newcommand{\tablabel}[1]{\label{tab:#1}}
\newcommand{\tabref}[1]{\autoref{tab:#1}}

\newcommand{\lstref}[1]{\autoref{lst:#1}}

\newcommand{\ie}{\emph{i.e.},\xspace}
\newcommand{\eg}{\emph{e.g.},\xspace}
\newcommand{\etal}{\emph{et al.}\xspace}
\newcommand{\etc}{\emph{etc.}\xspace}

\newcommand{\rqI}{What is the class commenting trend of developers over the Pharo versions, and in particular, do developers change comments of old classes?}
\newcommand{\rqII}{What types of information are present in Pharo class comments?}
\newcommand{\rqIII}{To what extent do developer commenting practices adhere to the class comment template over Pharo versions?}

\newcommand{\repFolder}[1]{Folder ``\href{https://github.com/poojaruhal/CommentAnalysisInPharo/tree/master/#1}{RP/#1}'' in the Replication package}
\newcommand{\repFile}[1]{File ``\href{https://github.com/poojaruhal/CommentAnalysisInPharo/tree/master/#1}{RP/#1}'' in the Replication package}

\usepackage[T1]{fontenc}
\usepackage{textcomp}
\usepackage{listings}
\definecolor{source}{gray}{0.9}
\usepackage{booktabs} 
\NewDocumentCommand{\rot}{O{30} O{1em} m}{\makebox[#2][l]{\rotatebox{#1}{#3}}}%
\usepackage{hyperref}
\hypersetup{
  colorlinks   = true, 
  urlcolor     = blue, 
  linkcolor    = blue, 
  citecolor   = blue 
}

\journalname{EMSE}

\begin{document}

\title{What do class comments tell us? An investigation of comment evolution and practices in Pharo Smalltalk}

\titlerunning{What do class comments tell us in Pharo Smalltalk?} 
\author{Pooja Rani         \and
        Sebastiano Panichella \and
        Manuel Leuenberger \and
        Mohammad Ghafari \and
        Oscar Nierstrasz
        }

\institute{Pooja Rani, Manuel Leuenberger, Oscar Nierstrasz  \at
               Software Composition Group, University of Bern, 3012 Bern, Switzerland\\
             \href{http://scg.unibe.ch/staff/}{http://scg.unibe.ch/staff} \\
           \and
          Sebastiano Panichella  \at
              Zurich University of Applied Science, Switzerland \\
                 \email{panc@zhaw.ch}  
            \and
            Mohammad Ghafari  \at
            School of Computer Science, University of Auckland, New Zealand\\
                \email{m.ghafari@auckland.ac.nz}  
}

\date{Received: date / Accepted: date}

\maketitle

\begin{abstract}
Previous studies have characterized code comments in various programming languages, showing
how high quality of code comments is crucial to support program comprehension activities, and to improve the effectiveness of maintenance tasks.
However, very few studies have focused on understanding developer practices to write comments. 
None of them has compared such developer practices to the standard comment guidelines to study the extent to which developers follow the guidelines.
This paper reports the first empirical study investigating commenting practices in Pharo Smalltalk.
First, we analyze class comment evolution over seven Pharo versions.
Then, we quantitatively and qualitatively investigate the information types embedded in class comments.
Finally, we study the adherence of developer commenting practices to the official \emph{class comment template} over Pharo versions.
   
The results of this study show that there is a rapid increase in class comments in the initial three Pharo versions, while in subsequent versions developers added comments to both new and old classes, thus maintaining a similar code to comment ratio.
We furthermore found three times as many information types in class comments as those suggested by the template.
However, the information types suggested by the template tend to be present more often than other types of information.
Additionally, we find that a substantial proportion of comments follow the writing style of the template in writing these information types, but they are written and formatted in a non-uniform way.
This suggests the need to standardize the commenting guidelines for formatting the text, and to provide headers for the different information types to ensure a consistent style and to identify the information easily.
Given the importance of high-quality code comments, we draw numerous implications for developers and researchers to improve the support for comment quality assessment tools.

\end{abstract}


\section{Introduction}
\label{intro}

Software understanding is an integral and required activity across multiple tasks in the software development life-cycle, and is  critical to any software maintenance task~\citep{Sieg15a,Haid10a}.
To understand a software system, developers usually refer to both the software documentation and the code itself~\citep{Bavo13b}, with code comments representing one of the most-used forms of documentation artifact for code comprehension~\citep{Souz05a}.
A study by Maalej \etal~\citep{Maal14a} shows that developers trust source code and code comments more than other forms of documentation for sharing program knowledge, and they consult comments when they try to answer their questions.

Given the relevance of code comments for program comprehension and maintenance activities~\citep{Wood81a,Tenn85a,Tenn88a,Hart93a,Souz06a,Lidw10a,Corn09a}, researchers have analyzed comments to detect low-quality comments~\citep{Stei13b,Liu15b}, identify existing inconsistency between comments and their related code elements~\citep{Rato17a,Wen19a,Styl09a,Petr15a,Zhou17a}, and they have examined the co-evolution of comments and code~\citep{Jian06a,Flur07b,Flur09a,Ibra12a}.
However, very few studies have focused on analyzing the information embedded in the source code comments~\citep{Padi09a,Haou11a,Stei13b,Pasc17a,Zhan18a}, and none of them specifically analyzed \emph{class comments}, or
to what extent these \emph{class commenting practices} adhere to the coding style guidelines.

Class comments in object-oriented programming play an important role in obtaining a high-level overview of classes~\citep{Clin15a} and are helpful for understanding complex programs~\citep{Nurv03a}.
However, different programming languages provide different notations and guidelines for writing comments in their code~\citep{Faro15a}, and embed different kinds of information into the comments~\citep{Ying05a,Padi09a,Pasc17a,Zhan18a}.
For instance in Java, a statically-typed language, a class comment provides an overview of high-level design of a class \eg the purpose of the class, what the class does, and other classes it interacts with~\citep{Nurv03a}.
On the other hand, in Pharo Smalltalk, a dynamically-typed live language and environment, a class comment contains high-level design information as well as low-level implementation details, \eg the application programming interfaces (APIs) the class provides, the instance variables it has,  and its key implementation features.
To write these class comments in an informative and consistent manner, different programming languages provide various coding style guidelines, such as the Oracle style guideline, PEP257.
However, to what extent Pharo class commenting practices vary from other systems and to what extent developers follow its style guidelines in their comments is not known.

In this paper, we conjecture that code commenting practices (\eg comment content and style) in different programming languages tend to evolve over time, as a result of the natural program language development and ecosystem evolution.
Thus, the \emph{goal} of our work is to investigate this conjecture, observing the way developers adapt to commenting practices over time, focusing on  Pharo, a modern Smalltalk environment.
First, we discuss the key characteristics that make Pharo ideal for our investigation of class commenting practices in object-oriented programming languages:
\begin{itemize}
    \item Class comments are a primary source of documentation in Pharo.
    \item  As a descendant of Smalltalk-80, Pharo has a long history of class comments being separated from the source code~\citep{Gold83a}, and is thus appropriate to analyze \emph{the evolution aspect} of class comments.
    \item Smalltalk supports liveness since more than three decades; therefore, it can present interesting insights into code documentation in live programming environments.
    \item Class comments in Pharo neither use any annotations nor the same writing style as used in Javadocs or Pydocs, thus presenting a rather different aspect on commenting practices, and challenges for existing information identification approaches~\citep{Pasc17a,Zhan18a}.
    \item Pharo traditionally offers a concise template, consisting of commenting guidelines for class comments, to enter a class comment for newly-created classes, and this template has evolved over the years.
    Consequently, Pharo is appropriate as a case study to investigate to what extent developers follow the template in writing comments, and what additional information developers embed in them.
\end{itemize}
More details regarding the Pharo environment are discussed in \secref{Pharo-background}.

\textbf{Research Questions}.
To better understand class commenting practices in Pharo, we formulate the following research questions: \begin{itemize}
    \item \emph{\textbf{RQ1}: \rqI}
        \item \emph{\textbf{RQ2}: \rqII}
    \item  \emph{\textbf{RQ3}: \rqIII}
\end{itemize}

In this paper, we first study the class commenting practice trends of major Pharo releases over 11 years from 2008 to 2019, assessing whether developers do or do not change comments of old classes.
In addition, we quantitatively and qualitatively analyze the class comments of the latest version of Pharo to characterize the various types of information embedded in class comments, and we build a comment taxonomy, called \emph{Pharo-CTM} (Pharo Comment Type Model).
Finally, we evaluate how comments adhere to the template in terms of content and writing style.
For the content aspect, we observe how many information types in Pharo-CTM match the information types constituting the standard Pharo comment template (\ie a guideline template to write a class comment), and how many are not part of it.
For the writing style aspect, we compare the writing style of comments to the writing style guidelines suggested by the template.

Our work shows that the trend of writing class comments increased rapidly in the initial three Pharo versions and then was maintained over subsequent versions, and that developers tend to add comments to old classes in Pharo with or without code changes.
We observe that the current comment template substantially diverges from contemporary practices of developers, with 23 information types occurring in class comments by developers, while only seven of them are present in the Pharo class comment template.
Measuring the frequency of different information types, we find that the seven information types proposed by the template are present more often than others.
Additionally, while writing these information types, developers follow the writing style guidelines from the template, \eg using first-person pronouns in describing various information types, and mentioning the headers of different information types.
We find this behavior of comments adhering to the template throughout all Pharo versions.
Based on these insights we suggest adding commenting guidelines to the template to ensure consistent formatting of text, and enable highlighting of certain details, thus improving the quality of the template.

We argue that this work not only encourages stakeholders to revisit their commenting guidelines, but it also informs developers to comment on the essential details of a class in a more structured and complete way, and opens the way for research aimed at proposing tools for ensuring a high quality of code comments.
A direct implication of our work is that, in different programming languages, using the contemporary code comment template or guidelines is not always ideal when actual practices strongly diverge from it.
Thus, future research effort is needed to
(i) develop tools that are able to determine the extent to which the code comment template or guidelines diverge from actual practice,
(ii) establish language-independent approaches to automatically identify the information type from the comments, given the increasing usage of multi-programming languages in open source projects, 
and
(iii) automatically assess code comment quality in terms of both content and style.

In summary, this paper offers the following contributions:
\begin{enumerate}
    \item an overview of the Pharo commenting trends over all seven major releases till 2019,
    \item an empirically validated taxonomy, called \emph{Pharo-CTM}, characterizing the  information types embedded in class comments written by developers,
    \item a discussion of taxonomies available from the related work, and a mapping and discussion of these taxonomies compare to our taxonomy,
    \item an assessment of the extent to which developer commenting practices adhere to the standard Pharo template, and
    \item a publicly available dataset of manually dissected and categorized Pharo comments, including all versions of the data used for trend analysis in the replication package~\citep{RPackage}\footnote{\url{https://github.com/poojaruhal/CommentAnalysisInPharo/tree/master}}.
            \end{enumerate}

\textbf{Paper structure}.
The rest of the paper is organized as follows.
In \secref{Comment-trend-analysis} we analyze the trends in commenting activities for both old and new classes over the seven major Pharo releases (RQ1).
In \secref{Pharo-Commenting-Practices} we report on our study of Pharo commenting practices, in particular the types of information developers include in class comments (RQ2).
In \secref{Adherence-to-template} we compare the commenting practices of developers to the standard template, focusing on the types of information developers include in class comments, and the writing style they follow  (RQ3).
We highlight the possible threats to validity of our study in \secref{Threats-to-validity}.
Then \secref{Related-work} summarizes the related work, in relation to the formulated research questions.
Finally, \secref{conclusion} concludes our study, outlining future directions.


\section{Background}
\seclabel{Pharo-background}

\textbf{The Pharo environment}.
Pharo is a reflective programming language environment incorporating a Smalltalk dialect.
Smalltalk is one of the oldest object-oriented, dynamically-typed programming languages, still used extensively in various systems (Pharo, Squeak), and scored second place for \emph{most loved} \emph{programming language} in the Stack Overflow survey of 2017.\footnote{\url{https://insights.stackoverflow.com/survey/2017/} verified on 4 Feb 2020}
{Pharo is a fully open-source and live development environment with a large library integrating external packages.
The Pharo ecosystem has a significant number of projects used in research and industry~\citep{Pharo}, and code comments are a primary source of documentation in Pharo.
We computed the ratio of comment sentences to lines of code in the most recent Pharo release (\ie Pharo 7) and found that 15\% of the total lines are comments.

\begin{figure}[h]
    \centering
        \includegraphics[width=\linewidth]{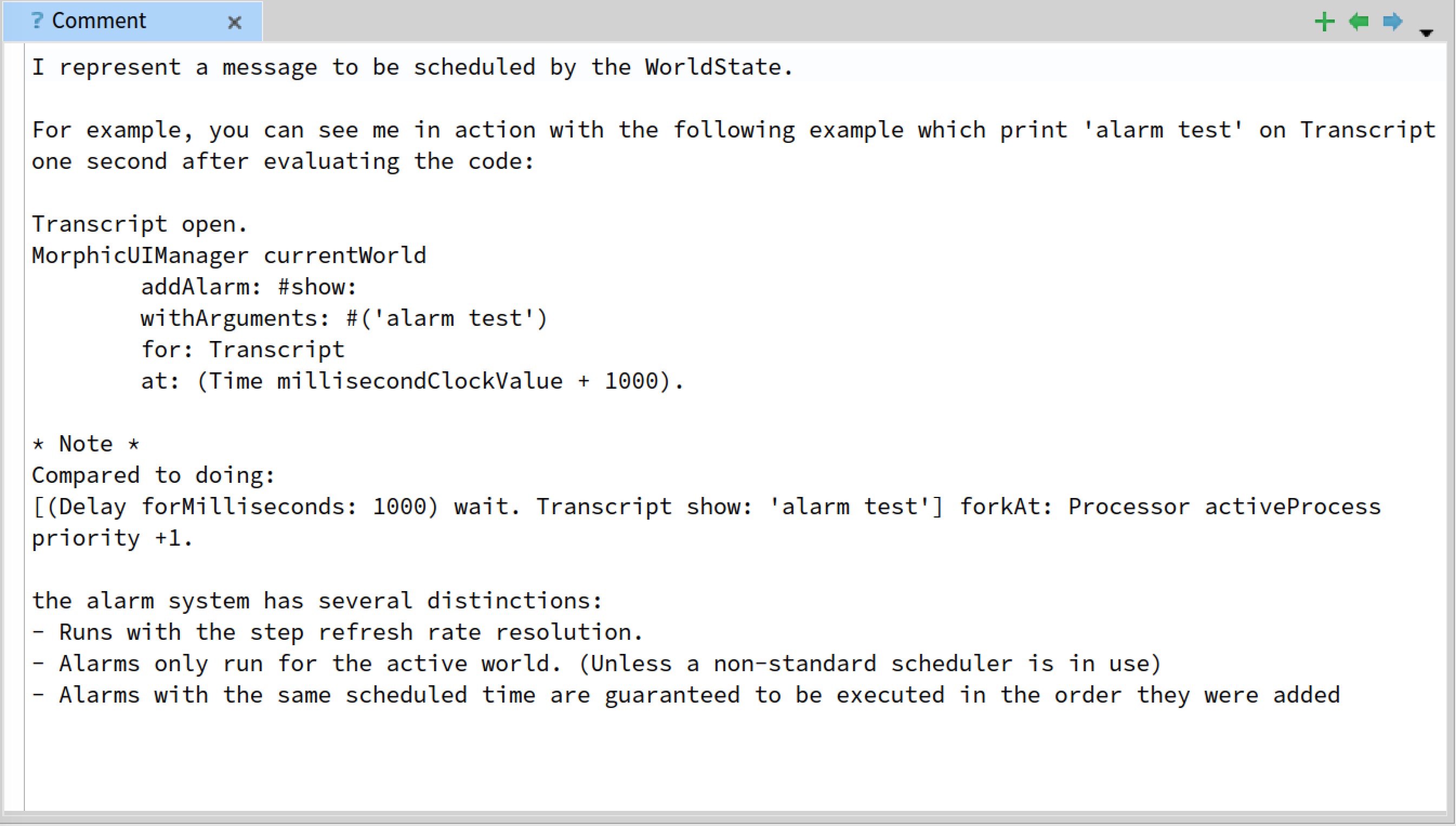}
        \caption{A class comment in Pharo
        }
        \figlabel{class-comment-pharo-7}
    \end{figure}%

\begin{figure}[h]
    \centering
    \includegraphics[width=\linewidth]{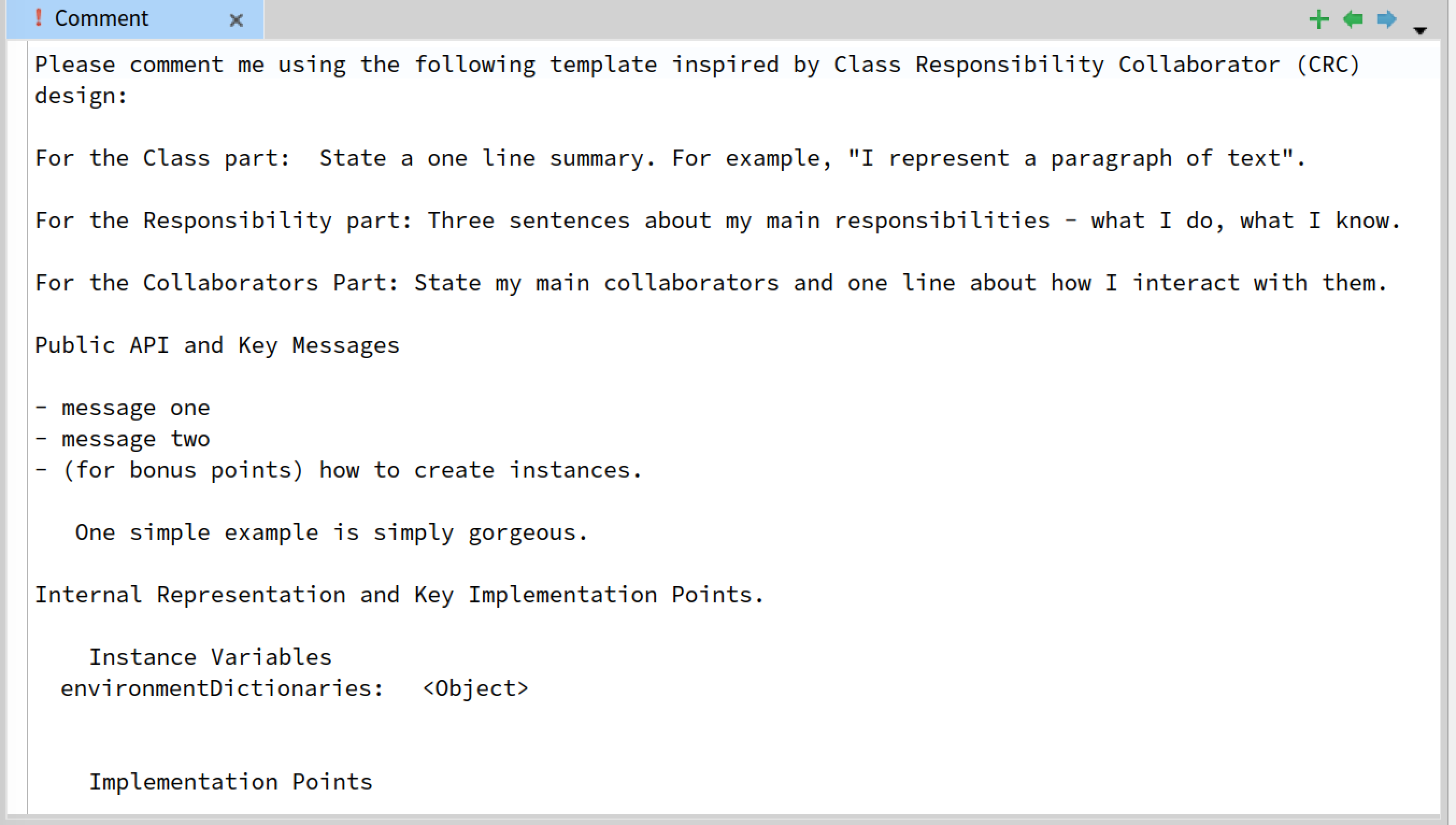}
    \caption{Class comment template in Pharo 7}
    \figlabel{template_pharo_7}
\end{figure}
   
According to our initial investigation into Pharo code comments, referred to as the pilot study later in this paper, a class comment in Pharo represents the main source of documentation for developers, as it provides detailed information about a class.
For instance, the class comment example of the class \emph{MorphicAlarm} in \figref{class-comment-pharo-7} shows the intent of the class mentioned in the first line (\emph{``I represent a message to be scheduled by the} \emph{WorldState''}), a code example to instantiate the class in the following two paragraphs, a note with the heading \emph{``* Note *''} to explain the corresponding comparison, and the features of the alarm system in the last paragraph.
The class comment appears in a separate pane instead of being woven into the source code of the class.
Within a class comment, complete sentences are used, but not annotations like \emph{@param, @see} to mark the type of information, as opposed to class comments in other languages.
However, the commenting patterns and practices in Pharo have not yet been studied or analyzed.

To guide developers in writing a class comment, Pharo offers a semi-structured default template, as shown in the Pharo 7 template in \figref{template_pharo_7}.
The template encourages developers to write different types of information like \emph{Intent}, \emph{Responsibilities}, \emph{Collaborators}, and \emph{Public API} to document important properties and implementation details of the class, but it is still unclear how frequently developers follow the template while writing class comments, and what additional information they actually add to the comments.


\section{RQ1: Comment trend analysis}
\seclabel{Comment-trend-analysis}

Classes are commented more frequently than other code entities, such as methods, variables, and control structures~\citep{Flur07b}.
As software evolves, changes to the source code of classes may invalidate the class comments~\citep{Wen19a}.
It is therefore important to understand how and when developers update classes and their comments.
This knowledge may be useful to inform developers when to update class comments to keep them in sync with the code.
Fluri \etal reported that developers rarely comment newly added classes in Java projects~\citep{Flur07b},  but whether developers have the same behavior in other programming languages or not, is unexplored.
With this investigation, our main aim is to understand developer class commenting behavior in Pharo, and how class documentation is updated over the years.
We therefore perform a trend analysis on developer class commenting practices.
In the commenting trend of class comments,  we specifically look at two main aspects: whether the number of commented classes increases or decreases over time, and whether developers change class comments of old classes over time.

\subsection{Study Setup}

 To better understand class commenting practices of Pharo and achieve reliable results, we analyzed the core libraries of Pharo.
We extracted the most recent revision of each major release of Pharo, from Pharo 1
to Pharo 7 (2008 to 2019), using a software analysis platform named Moose~\citep{Duca05f}.
For each version we used Moose~\citep{Moose} to extract the class comments and meta details of the classes in the standard image, known as the Pharo core.\footnote{
For Pharo 1 and Pharo 6, we only extracted Pharo 1.4 and 6.1 because we could not run Pharo 1 and 6 using Moose, due to the backward compatibility issues of Moose.}
This includes classes to work with files, collections, sockets, streams, exceptions, graphical interfaces, unit tests, \etc

\begin{table}[h]
    \centering
\caption{Overview of Pharo versions with the release dates and number of classes}
\tablabel{tab1}
\begin{tabular}{cccc}
\hline\noalign{\smallskip}
\textbf{Version} &\textbf{Release date} & \textbf{\# Classes} &\textbf{\# Classes with comments}  \\
\noalign{\smallskip}\hline\noalign{\smallskip}
    1.4& Apr, 2012& 2\,950 & 1\,486 \\
    2.0& Mar, 2013& 3\,248 & 1\,983 \\
    3.0& Apr, 2014& 4\,025 & 3\,264 \\
    4.0& Apr, 2015& 4\,923 & 3\,768 \\
    5.0& May, 2016& 5\,670 & 4\,493 \\
    6.1& Jun, 2017& 6\,484 & 5\,181 \\
    7.0& Jan, 2019& 7\,863 & 6\,324 \\
    \noalign{\smallskip}\hline
\end{tabular}
\end{table}

\tabref{tab1} shows the details of each version with version number, release date, the total number of classes and the total number of classes with comments.

\subsection{Methodology}
Using this dataset,\footnote{\repFolder{Dataset-for-Replication/Data/RQ1/Source-files}}
we measured the trend of commenting by calculating the ratio of commented classes to uncommented classes in each version.
To investigate whether \emph{developers change comments of old classes}, we tracked comment changes in already existing classes (old classes).
For comment changes, we compared each class in a given version to its previous version to assess added comments, removed comments and changed content.
Additionally, we tracked code changes of a class in comparison to the previous version to get an overall summary of the historical changes.
To compute code changes we extracted the class definition (instance side and class side), all methods of the class, and source code of all methods of each class for each version.\footnote{\repFolder{Dataset-for-Replication/Data/RQ1/Code-changes}}

\begin{figure}[t]
    \centering
    \includegraphics[width=\linewidth]{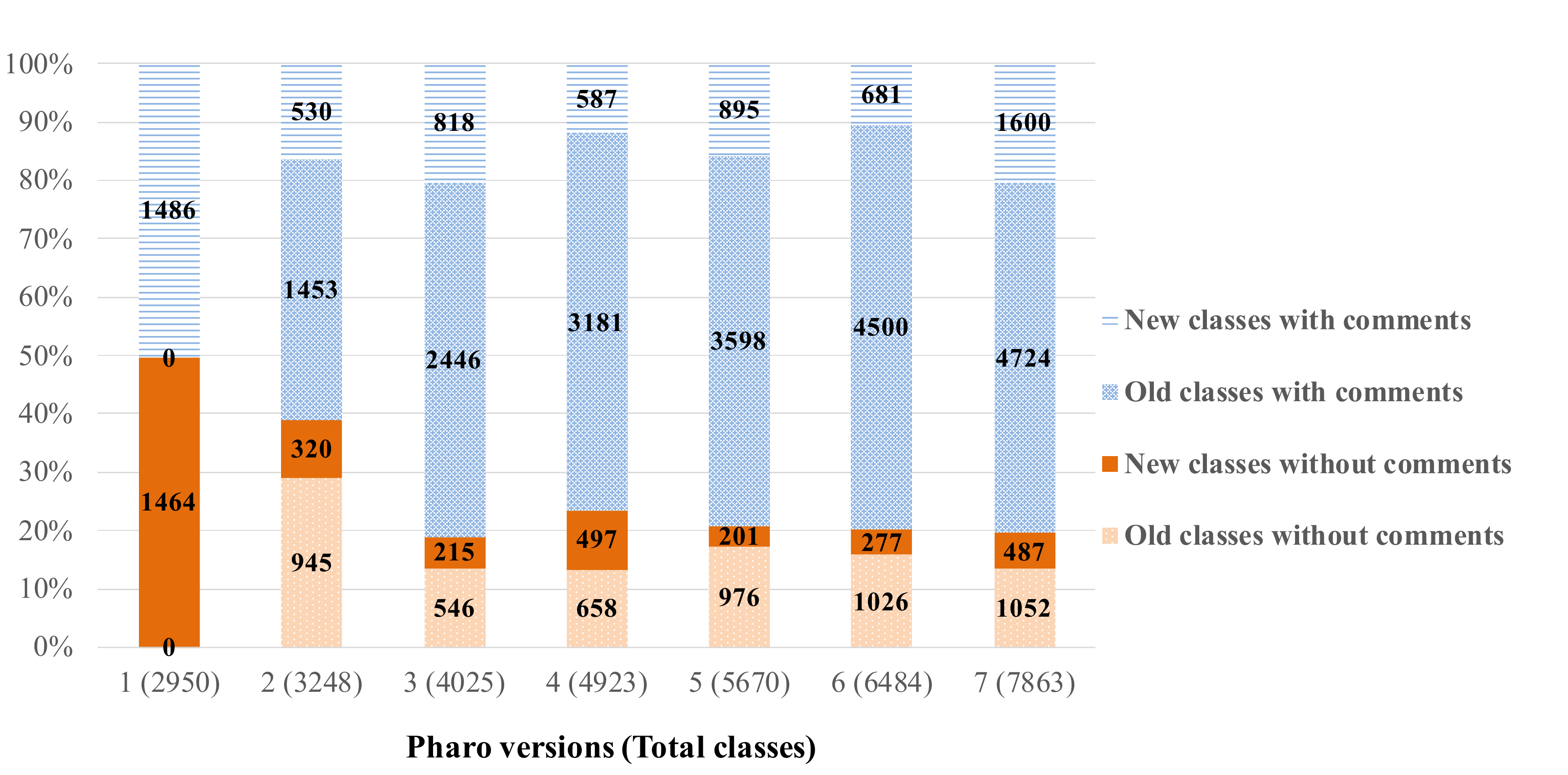}
    \caption{The trend of classes with and without comments in Pharo versions.}
    \figlabel{Percentage-classes-with-and-without-comments}
        \vspace{-4mm}
\end{figure}

\subsection{Result}
The result in \figref{Percentage-classes-with-and-without-comments} shows that the trend of commenting classes increases rapidly for initial Pharo versions, and is then maintained in subsequent versions.
Indeed, in the figure, we can see that the percentage of commented classes, in light and dark blue (for old and new classes), increased in initial versions, and then remained constant from the fourth version.

 \vspace*{2mm}
 \hspace*{-5mm}
 \begin{tikzpicture}
 \node [mybox] (box){%
 \centering
 \begin{minipage}{0.95\textwidth}  
 \fontsize{9.5}{9.5}\selectfont  
     \emph{\textbf{Finding 1:} The trend of commenting classes increases rapidly over the first three Pharo versions, from 50\% of commented classes in Pharo 1, to 80\% commented classes in Pharo 3 and subsequent versions}.
     
 \end{minipage}
 };
 \end{tikzpicture}%

 \figref{Percentage-classes-with-and-without-comments} also portrays the detailed aspect of classes that have survived from old versions and classes added in the current version.
 For instance, in version 3, we can see that the number of old classes without comments has decreased (height of the light orange bar segment decreased), and the number of old classes with comments has increased (height of light blue bar segment increased) implying that several old classes are commented in version 3, in addition to commenting new classes.
In version 7, we can see a major effort being put into commenting new classes (77\% of the new classes were commented) compared to old classes (12\% of old classes were commented).
In particular, 89\% 
 of the old classes from Pharo 6 survived to Pharo 7, of which 20\% 
were uncommented classes and only 12\% 
of the uncommented classes were commented in Pharo 7.


 \vspace*{2mm}
 \hspace*{-4mm}
 \begin{tikzpicture}
 \node [mybox] (box){%
 \centering
 \begin{minipage}{.95\textwidth}  
 \fontsize{9.5}{9.5}\selectfont  
     \emph{\textbf{Finding 2:}
     In later versions of Pharo, developers put effort into maintaining the code comment ratio, commenting new classes,  and adding comments to old classes}.
 \end{minipage}
 };
 \end{tikzpicture}%

In addition, we find that developers change comments of old classes as shown in \figref{changes-in-the-comments}.
Changing a class comment includes adding comments to an uncommented class, removing the comment, and updating the content of the comment.
 \begin{figure}[t]
    \centering
    \includegraphics[width=\linewidth]{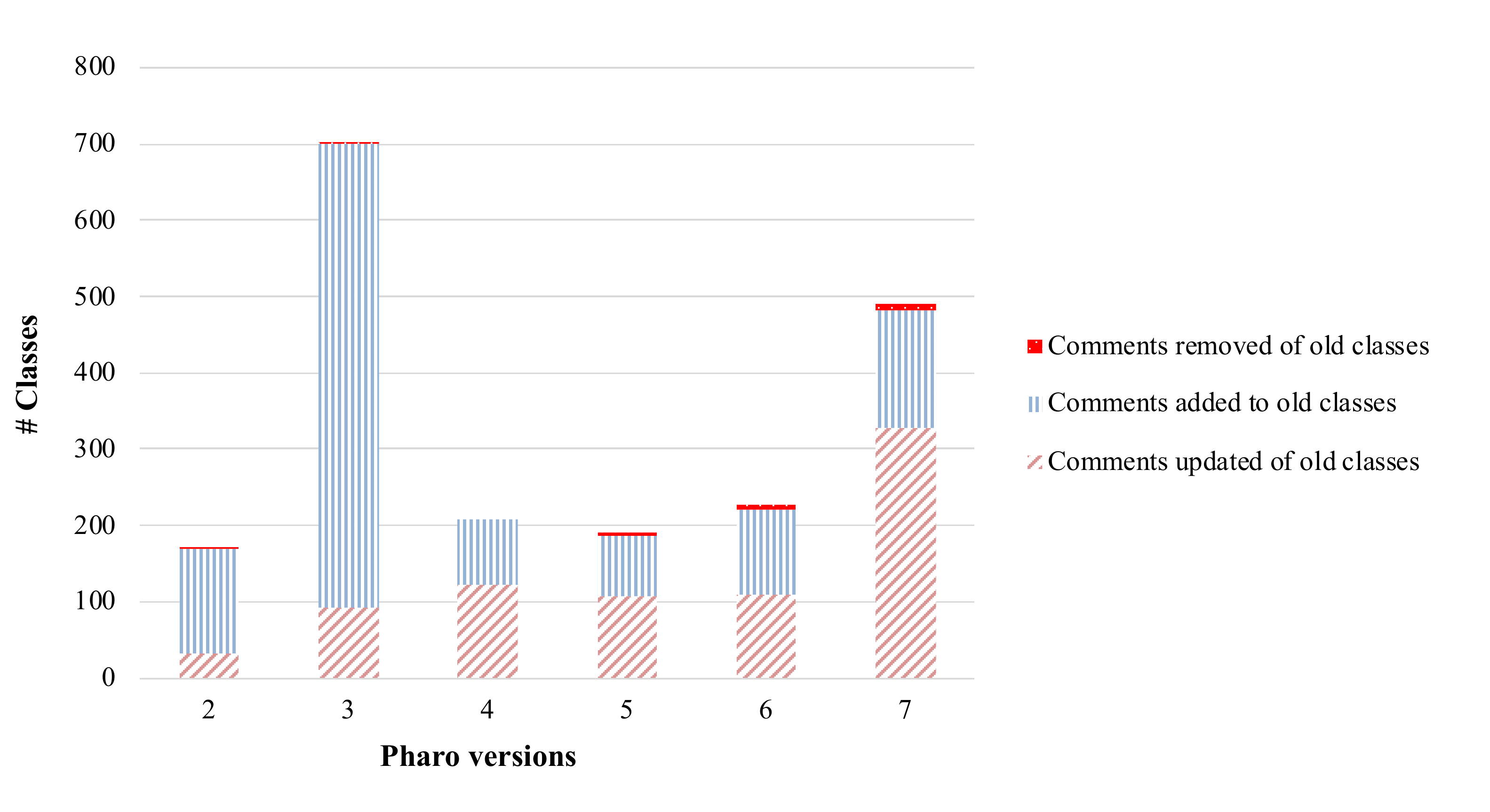}
    \caption{The trend of comment changes in old classes}
    \figlabel{changes-in-the-comments}
        \vspace{-4mm}
\end{figure}
Differentiating this change behavior in \figref{changes-in-the-comments} highlights that in versions 2 and 3, developers focused more on adding comments to old classes compared to updating or removing the comment content.
Since version 4, the focus of changing comments shifted to updating the content of class comments compared to adding comments to old classes.
For example, in Pharo 7, more class comments are changed compared to comments added to old classes.
To find the reason behind this behavior, we examine the code changes in old classes, and measure the extent to which developers update comments of old classes when changing their code.

\begin{figure}[t]
    \centering
    \includegraphics[width=\linewidth]{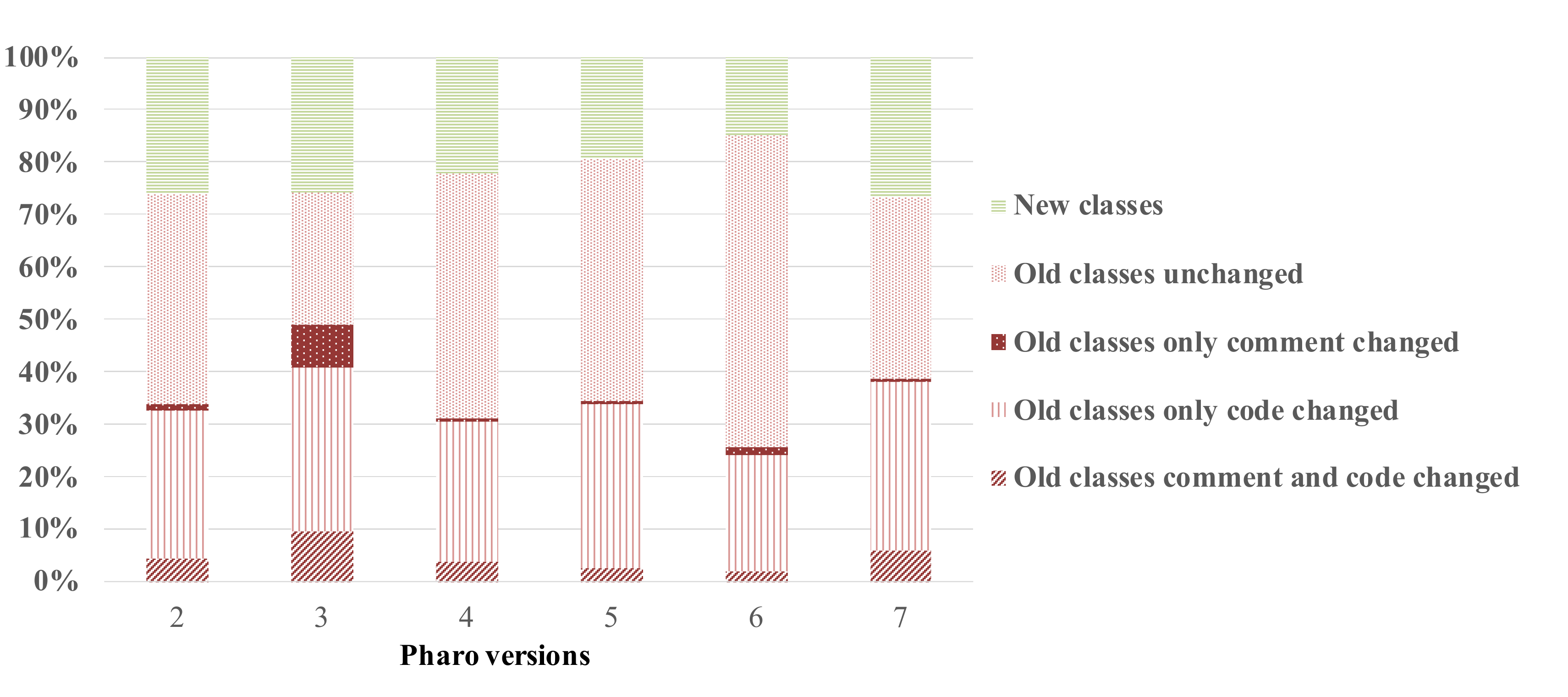}
    \caption{The trend of changing old classes in Pharo versions}
    \figlabel{percentage-changing-old-classes-trend}
        \vspace{-4mm}
\end{figure}

From \figref{percentage-changing-old-classes-trend}, we find that in Pharo 7, 52\% %
 of the old classes are changed either by changing code, comments, or both, indicating a major refactoring of the old classes.
Nearly 44\% 
 of old classes are changed without updating their class comment.
Specifically, 75\% of these changes were related to adding, removing, or updating methods, but we found no corresponding changes in the class comments.
We expected these changes to affect the class comments, due to the dedicated section in the class comment template for instance variables and key messages.
In contrast, the changes such as renaming a package or changing a method category carry a lower tendency to affect the class comment.
Only 7.9\% 
of the old class comments are changed together with the code in Pharo 7, as shown by the dark red bar segment at the bottom of version 7 in \figref{percentage-changing-old-classes-trend}.
We further explored this segment by analyzing a sample of 15\% of the 327 classes where both comments and code changed.
We find that 50\% of the changes in class comments are related to code changes, confirming the finding from earlier work~\citep{Flur09a}.
In our analysis, the most specific types of code changes that triggered comment changes were the deprecation of a class and the addition of new methods.
The rest of the code changes \eg updating a method or class definition changes, triggered comment changes less frequently.
In one particular case of code changes where a method is removed from the class, the method code is added to the class comment as an example.
In contrast, in another similar case, the method comment is added to the class comment as implementation details.
The reason for such a behavior can be the intent to keep the information about the removed method in the system for future tasks even though it is deleted.
The remaining 50\% of the comment changes are not related to code changes, even though 73\% of the code changes in these classes are adding new methods, updating methods, or removing existing methods which can potentially trigger the comment changes, according to previous work~\citep{Flur09a}.
These unrelated comment changes are about clarifying details of the class by changing the information types or formatting, improving the grammar, or changing the writing style from third person to first person or vice versa.\footnote{\repFile{Results/RQ1/trend-analysis/class-comment-code-changes-analysis.xlsx}}
We further analyzed which information types are frequently changed in comment changes irrespective of the code changes to find out the importance of specific information types.
We found that most specific information changes in the class comments were about adding and updating the intent of the class, warnings, usage examples, and implementation details of the class, thus indicating the importance of these information types.
On the other hand, in test classes (10\% of the classes where code and class comments changed) the most specific information changes were about removing the bug-related details from the class comments in the next version.
Analyzing what factors motivate developers to make such comment changes and at what stages of the project they change is the subject of future work.

\vspace*{2mm}
\hspace*{-4mm}
\begin{tikzpicture}
\node [mybox] (box){%
\centering
\begin{minipage}{.95\textwidth}   
\fontsize{9.5}{9.5}\selectfont 
    \emph{\textbf{Finding 3: }
    In 50\% of the cases, the code and class comments of old classes change together, with developers updating comments of the classes to keep them synchronized with the implementation}.
\end{minipage}
};
\end{tikzpicture}%

Until now we separated the old classes from the new classes, but did not distinguish between the originating versions of old classes and those that survived from a specific version.
For example, in Pharo 7, what portion of the classes survived from Pharo 1 or Pharo 2?
This information is crucial to gain insight into comment coverage of a particular version in each version, and which class comments developers considered important to refactor in the current version.
Furthermore, it helped us to analyze what happened to the old classes in the current version.
For example, if the system went through a major refactoring, then which old version's classes were deleted, re-introduced or modified?
We therefore need to keep track of the history of a class, from Pharo 1 to latest version, to get an overall view of the evolution of the system.
To answer all these questions, we track the origins of old classes and their survival history to the current version in \figref{comment-evolution}.

\begin{figure}[ht]
    \centering
    \includegraphics[width=\linewidth]{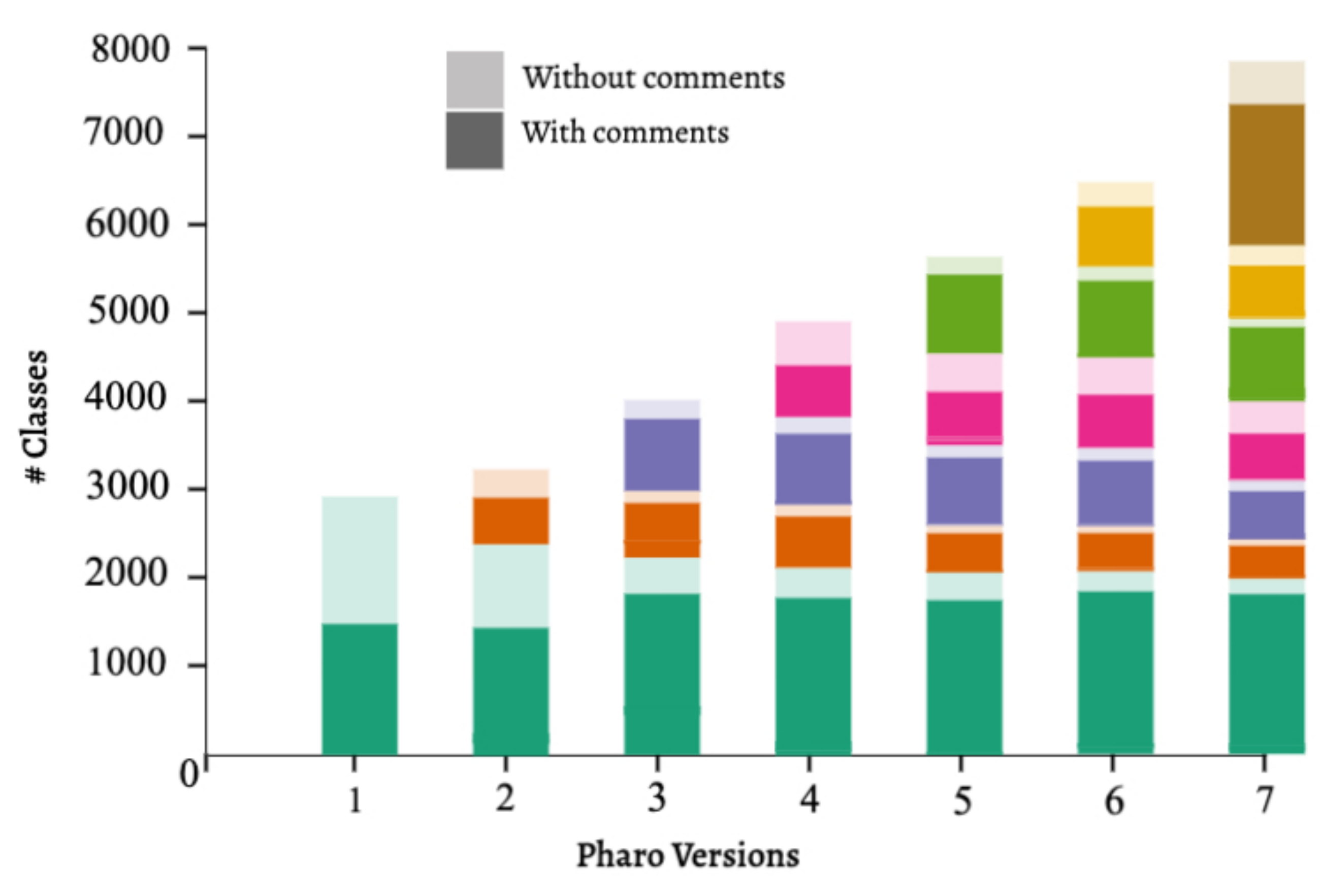}
    \caption{Survival analysis of Pharo versions.}
    \figlabel{comment-evolution}
\end{figure}

In \figref{comment-evolution}, each Pharo version is assigned a unique color.
The shading indicates the distribution of classes with and without comments.
The height of a bar segment in one color represents the classes surviving from a previous version to the new version.
The original versions of each class are ordered by age with the oldest version at the bottom and the newest version at the top.
Tracking the color of a version allows us to know how long classes are kept in the system.
For example, in Pharo 1 the dark shade of green shows the classes with comments and a lighter shade of green shows the classes without comments.
We find that until Pharo 6 the classes originating in version 1 still constitute the largest group of all older classes.
In Pharo 3, major efforts were devoted to refactoring and re-documenting classes from older versions, 1 and 2.
In Pharo 4, we observe that the ratio of adding comments to the new classes is less compared to preceding and succeeding versions except Pharo 1.
Pharo 7 shows the effort of documenting old classes and new classes, thus achieving maximum coverage \ie 80\% of classes with comments among past versions.

\begin{table}[h]
     \centering
    \caption{Overview of major projects added, removed and re-documented in each version}
    \tablabel{tab2}
    \begin{threeparttable} 

    \begin{tabular}{lp{3.4cm}p{3.0cm}p{3.4cm}}
        \toprule
        \textbf{Version} & \textbf{Added} & \textbf{Removed} & \textbf{Re-documented}   \\
        \midrule
        1& Ring metamodel   &  Squeak classes & Code simulator, Zinc, Refactoring, Monticello \\
        \noalign{\smallskip}\hline  
        2& QA tools, Spec, Fuel, Native Boost, Nautilius &  OmniBrowser, TrueType& Zinc, Refactoring, Monticello \\
        \noalign{\smallskip}\hline 
        3& Versioner, Opal, Athens, Debugger& & Kernel tests, Zinc, Monticello, Collection tests\\
        \noalign{\smallskip}\hline 
        4& GLM, Rubric, TxText, OSWindow, MetaLink & Slot tests  & Refactoring, AST, Athens, Zinc, Delay scheduler \\
        \noalign{\smallskip}\hline     
        5& Spur VM, UFFI, Renraku, STON & NativeBoost & Rubric, Refactoring, TxText, Nautilius, Komitter \\
        \noalign{\smallskip}\hline 
        6& Iceberg, Epicea, Tonel, Ombu&   &Refactoring, AST, UFFI, Spec, Renraku \\ 
        \noalign{\smallskip}\hline 
        7& Bootstrapping,  Traits2, Refactoring2, Calypso &TxText, Versioner, Nautilius, Kommitter, Traits  & UFFI, System tests, Tool, Kernel, STON, System, Iceberg \\
        \bottomrule
\end{tabular}
\end{threeparttable} 
\end{table}

In addition to showing the overview of a version, we also summarized the major projects that were added, removed and re-documented in each version in \tabref{tab2}.
We observed that the documentation of few projects such as \emph{Zinc}, and \emph{Refactoring} were actively updated, but whether it was due to their importance, or discipline of their developers, or both, is the subject of future work.
We summarized the projects by grouping the added, removed and recommented classes by their package in each version.
To verify our calculated list, we compare our project list to Pharo change logs.\footnote{\url{https://github.com/pharo-project/pharo-changelogs}}
From the aforementioned analysis we collected several observations about Pharo commenting patterns:

\begin{itemize}
    \item In Pharo 2, significant effort has been put into refactoring and removing classes from the old version, Pharo 1.
    The old system browser, \emph{OmniBrowser}, is replaced with \emph{Nautilus}.
        \item In Pharo 3, a major effort is put into commenting old classes, as shown in \figref{changes-in-the-comments}.
    \item In Pharo 4, developers focus less on commenting old classes but more on adding new classes.
    New projects added in the version are shown in \tabref{tab2}.
    \item In Pharo 5, the focus seems more on refactoring classes from old versions, specifically Pharo 1, 2 and 3 but not Pharo 4, as shown in \figref{comment-evolution}.
    The ratio of classes with comments to classes without comments is also higher compared to the previous Pharo 4.
    \item In Pharo 6, the effort is put into adding new classes and making sure that comments are also added to new classes.
    One of the main projects added in this version is for git support.
    \item In Pharo 7, we find that many new classes are added.
    After investigating further we found that new versions of \emph{Refactoring} and \emph{Traits}, and a new system browser \emph{Calypso} are added.
    Refactoring old projects is the primary focus of this version.
    A substantial number of old class comments are updated, in particular, the projects \emph{UFFI, Tool,} and \emph{System tests}.
    \item Analyzing \figref{comment-evolution}, we observe that Pharo 4 classes were rarely refactored in succeeding versions except Pharo 7 as the height of the Pharo 4 magenta bar remains the same through Pharo 6.
    We believe this is due to the importance of the project GLM (Glamorous toolkit), and 
    the general interest of developers to keep this project in the current, already stable, status.
\end{itemize}

  \vspace*{2mm}
\hspace*{-5mm}
\begin{tikzpicture}
\node [mybox] (box){%
\centering
\begin{minipage}{.95\textwidth}
\fontsize{9.5}{9.5}\selectfont   
\emph{\textbf{Finding 4:} In Pharo 3, a major effort is put into adding comments to old classes whereas in subsequent versions, more effort is put into updating comments of old classes.
Both cases show developers adding and updating comments of old classes.}
\end{minipage}
};
\end{tikzpicture}%

\subsection{Implications}
\seclabel{rq1-implications}
The investigation performed on commenting trends presents important insights into the commenting habits of Pharo developers.
These insights can assist developers and researchers in the following aspects:\\
\begin{itemize}
  \item \emph{Tool support to analyze the co-evolution of code and comments}: 
  Understanding software evolution is crucial to ease various software development tasks such as understanding a program, its software elements, finding the actual change that introduced a bug, or detecting change propagation patterns among software artifacts.
  Our comment evolution results show that developers tend to add class comments to old classes, however, once the ratio of class comments to the total classes reached a particular level (at least 75\%), developers do not allocate the same effort, thus indicating the stability of the system. 
  Also, we observe that developers put considerable effort into adding comments to classes newly added to the Pharo core, which is in contrast to previous results involving commenting practices of Java external systems~\citep{Flur09a}. 
  Whether such commenting behaviour is due to the expectation of better commenting practices from core systems compared to the external systems or due to Pharo developer habits requires further analysis.
  Fluri \etal showed that the Eclipse core system has a better commenting ratio compared to non-core systems such as  Eclipse JDT and Eclipse PDE~\citep{Flur09a}. 
  We observe similar behaviour in the Pharo core compared to external projects.
  Still, these systems lack appropriate tools to analyze the co-evolution of code and comments.
We suggest that further research needs to be devoted to developing tools providing co-evolution views of code and comments to monitor better the relative growth and quality of comments over time as well as the actual code comment coverage~\citep{Zaid08c}.

\item \emph{More accurate tools to automate the detection of comment changes}:
Soetens \etal envision that future IDEs will use the notion of changes as first-class entities (AKA change reification approaches).
These change-based approaches can help in communicating changes between IDEs and their architectures, and to produce accurate recommendations to boost complex modular and dynamic systems~\citep{Soet17a}.
Analyzing and detecting change patterns of comments can enable the vision of Soetens \etal of integrating code comments easily in such change-oriented IDEs.
Additionally, detecting which types of information in the comments tend to change more often can help researchers in generating comments automatically.
For example, we found a code change due to a class deprecation which triggered a comment change by adding the deprecation notice in the class comment to inform other developers. This effort of updating the class comment whenever a class deprecation code change is detected can be reduced by generating the notice information automatically in the class comment.
These comment change patterns are not only helpful for developers to reduce their commenting effort but can also help researchers to improve their bug-prediction models.
For instance, Ibrahim \etal showed statistically significant improvements in their bug-prediction models using comment update patterns; similarly, our comment update patterns can be used for future work~\citep{Ibra12a}.

\item \emph{Leveraging change data}:
Previous studies have leveraged the historical change data in various ways, such as in designing new applications in the IDE~\citep{Soet17a}, evaluating code completion algorithms~\citep{Robb10a}, and recommending future changes in specific code parts~\citep{Flur09a}.
In the context of comments, Fluri \etal implemented a tool named ChangeCommander, which recommends comment changes when a new method invocation is introduced in the system, based on the collected code-comment change patterns~\citep{Flur09a}.
However, the approach of Fluri \etal to detect comment changes does not work entirely for the Pharo system due to its dynamic nature, and its different comment structure and scope.
Based on our code-comment change analysis, we identified patterns of code changes in a class such as deprecating a class, or adding a new method which triggers comment changes more frequently than other code changes. 
Future tools can utilize these patterns for recommending developers when to update class comments. 
From a technological point of view, Epicea (a tool to log code changes in Pharo) supports source code changes on the class level.
Integrating the type of comment changes we identified in our study, such as formatting changes, typo fixes, instance variable changes, and code-comment change patterns, can help to answer particular developer questions such as \emph{``What specific type of the code change led to this comment change?} or \emph{``Which specific comment changes does a commit consist of?''}~\citep{Dias14a}.
\end{itemize}

This investigation helped us to gather the general practices developers follow towards class commenting but does not characterize the content of the comments, nor does it describe how comments adhere to the commenting guidelines of Pharo.
We cover these aspects in the rest of this paper.


\section{RQ2: Comment information types}
\seclabel{Pharo-Commenting-Practices}
With class comments being a primary source of detailed design and implementation documentation, developers add different types of information they deem important for the class.
The class comment in Pharo does not make use of any kind of annotation (\eg \emph{@param,@return}) as in other languages, and no fixed structure is followed to place the information in the class comment.
A few comments we found are written using the Pillar markup language,\footnote{\url{https://github.com/pillar-markup/pillar}} but the majority of comments do not adopt it, and instead are written in a free-text style.
The way of writing the same information thus varies among developers, so extracting and analyzing a certain type of information from comments is non-trivial.
Consequently, to answer RQ2 (\emph{\rqII}), we investigated the class comments manually.
We performed a pilot study and formed an initial taxonomy of comment information types.
We then conducted a three-iteration-based analysis on a sample set of 363 comments to finalize the taxonomy.
Following the same methodology, we analyzed 351 comments from external projects (not part of the Pharo core) to verify the commenting practices of other developers.

\subsection{Study Setup}

To investigate the commenting practices, we studied the latest stable version of Pharo, namely Pharo 7.
Since each class has one class comment, all the classes with class comments participated in the analysis dataset, resulting in a dataset of 6\,324 classes.
However, due to the semi-structured nature of comments and the lack of content headers or annotations,  
a content-wise investigation of comments requires manual effort, making the investigation of the whole dataset a non-trivial task.

We therefore selected a representative subset of comments for manual analysis by defining the required minimum sample size \emph{n} with the following standard formula~\citep{Trio06a}:

\[sample size( n) = \frac{\frac{z^2 \times p(1-p)}{e^2}}{1+(\frac{z^2 \times p (1-p)}{e^2 N})}\]
\emph{N} is the size of the dataset, \emph{e} is the margin of error, \emph{p} is the percentage of picking a comment and the \emph{z} is selected according to the desired confidence level.
 We calculated the required sample size from the finite population of 6\,324 to reach a confidence level of 95\% and error \emph{e} of 5\%.
The z-score is 1.96 according to the confidence level and \emph{p} is 0.5 used for the sample size needed.
The resulting dataset should therefore contain a subset of 363 class comments in total.
In order to choose 363 representative comments from the dataset, we investigated the distribution of comments based on the number of sentences present in a comment shown in \figref{comment-distribution-scatter-plot}.
The sentences were separated using a custom-built Pharo sentence splitter.
We found that the number of sentences in the comments varies from 1 to 272.
Therefore,
we used stratified random sampling approach to ensure that all kinds (or size) of comments are represented in the manual analysis dataset in case of skewed population.
This approach divides the whole dataset into smaller strata based on the comment distribution, and allows random samples to be drawn from each stratum.

\begin{figure}[ht]
    \centering
    \subfloat[Frequency of comments w.r.t comment length]{
    \figlabel{comment-distribution-scatter-plot}
     \includegraphics[width=0.48\linewidth,height=.45\linewidth]{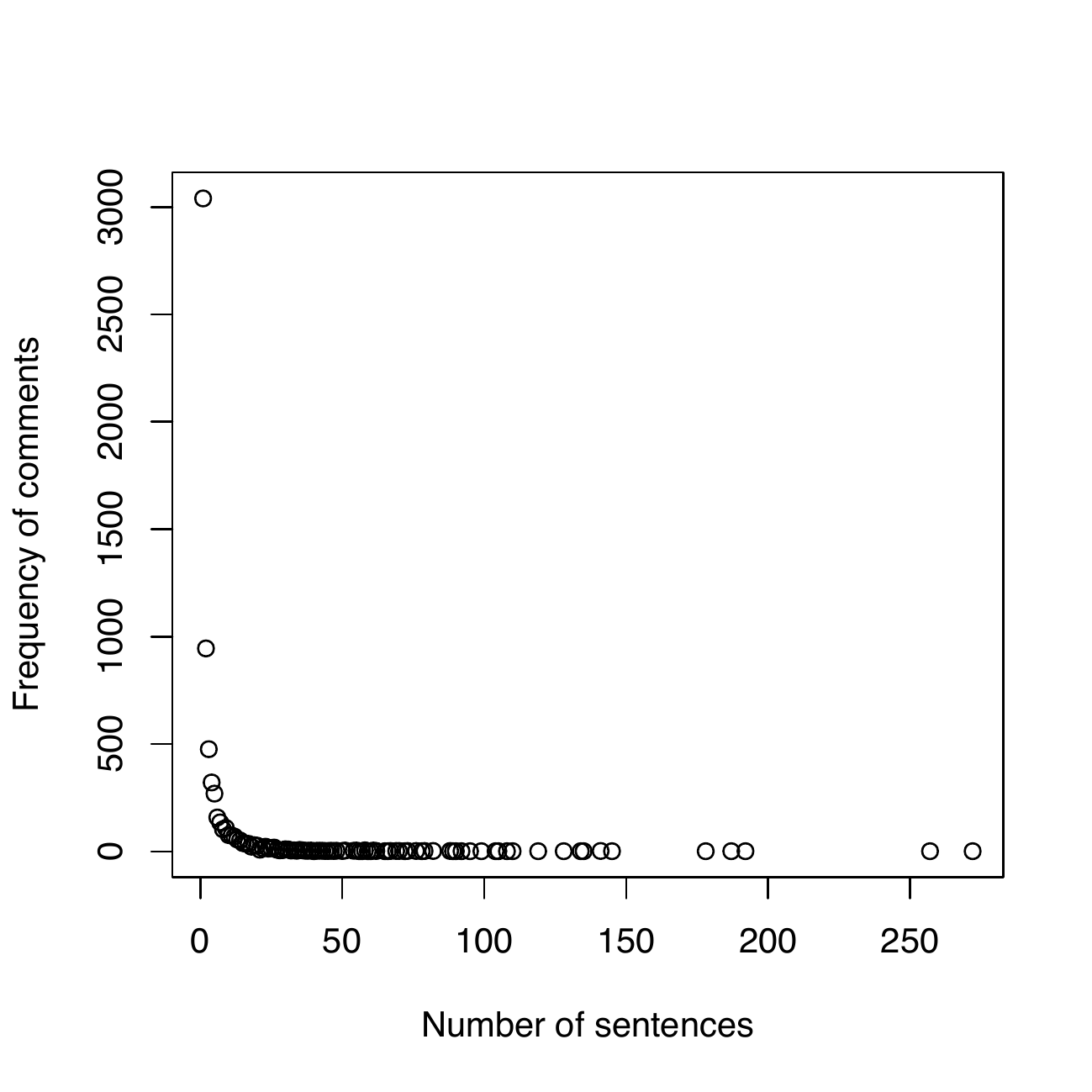}}
    \subfloat[Comments subgroup w.r.t number of sentences]{
    \figlabel{comment-distribution-box-plot}
      \includegraphics[width=0.48\linewidth,height=.40\linewidth]{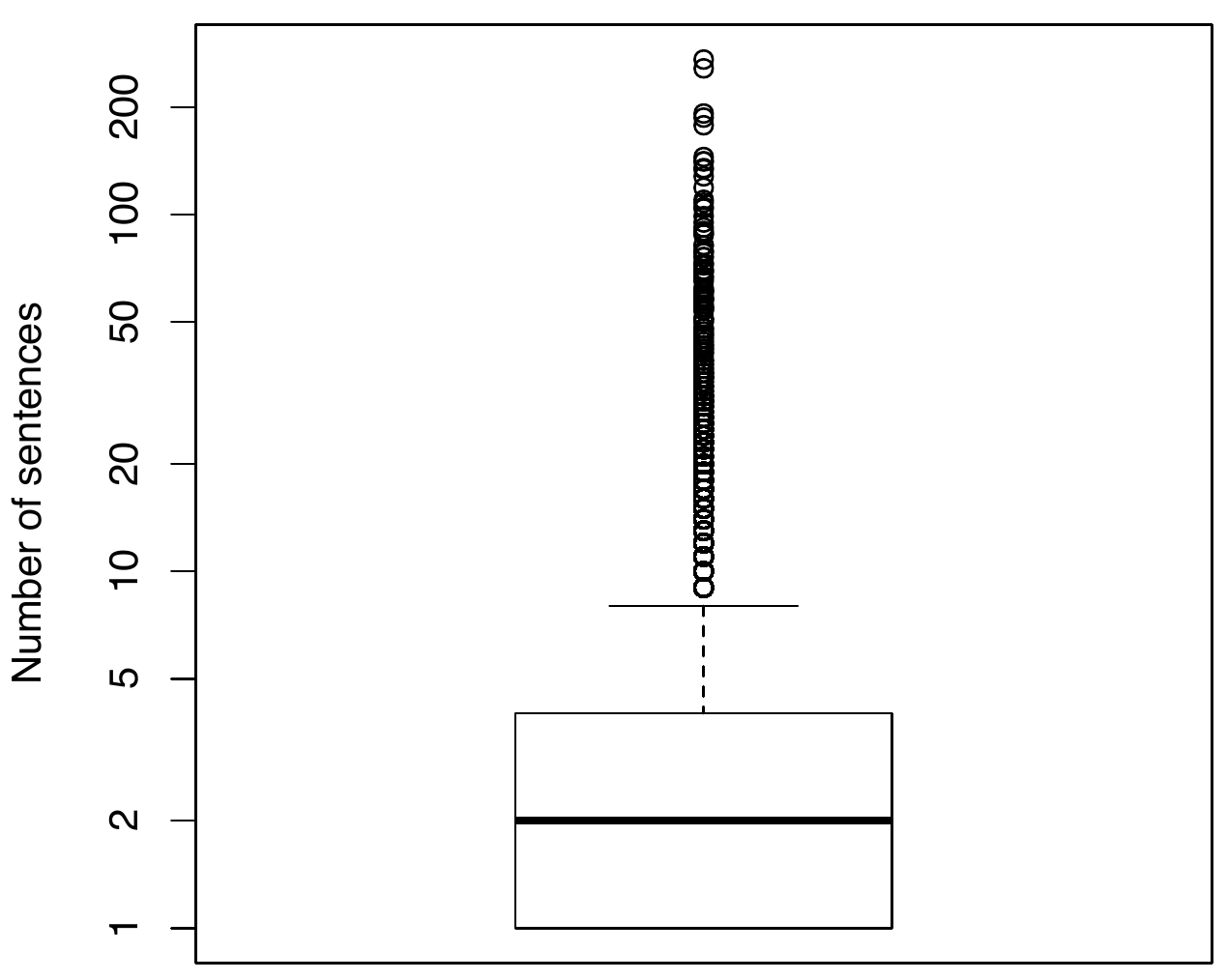}}
\end{figure}
    
In order to select the 363 sample comments according to the approach, we used quintiles from the logarithmic distribution based on the number of sentences in each class comment shown in \figref{comment-distribution-box-plot}.
Accordingly, we obtained five quintiles as follows 1, 1, 2, 6, and 272.
Based on the quintile values, we obtained comment strata, and calculated the comment proportion of each stratum shown in \tabref{comment-rate-per-group}.
We selected from each stratum a number of comments that correspond to the proportion of such comments in the entire dataset, following a random sampling approach without replacement.
For example, from a total of 3\,040 comments of comment stratum ``1-1'', we selected 175 comments \ie 48\% of 363 comments using a random sampling approach without replacement.
As the approach facilitates the selection of a random sample from a stratum and not all strata are entirely homogeneous (such as `3-6' compared to `1-1'), we observed that the margin of error varies from 7\% to 9\% within strata (measured using the formula \emph{samplesize(n)} for each stratum).
On the other hand, this approach is known to increase the overall precision instead of that of the individual strata, thus helping us to better select representative comments.

\begin{table}[h]
    \centering
    \caption{Comment proportion per stratum for the whole dataset, and the resulting sample dataset.}
    \tablabel{comment-rate-per-group}
\begin{tabular}{crcc}
\hline\noalign{\smallskip}
        \textbf{stratum} & \textbf{\#comments}& \textbf{comment rate} & \textbf{\#selected for study}\\
\noalign{\smallskip}\hline\noalign{\smallskip}        
        1-1 & 3\,040& 48\%   & 175  \\ 
        2-2 & 945& 15\%  & 54 \\
        3-6 & 1\,224& 19\%  & 69\\
        7-272 & 1\,115& 18\% & 65 \\
\noalign{\smallskip}\hline
\textbf{Total} & 6\,324 & 100\% & 363\\
\hline
    \end{tabular}
\end{table}

To verify the practices of Smalltalk developers in other projects than the Pharo core, we analyzed the selected comments from seven external projects.
We filtered the external projects from GitHub\footnote{\url{https://github.com/topics/pharo?o=desc\&s=stars}} based on several criteria:
(i) the project is not part of the Pharo core,
(ii) it has an active project activity since 2019, and the project history spans at least two years with at least 600 commits,
(iii) it is not a repository for books, an article, or documentation,
(iv) it has more than five contributors,
(v) the project does not contain more than 20\% code from other programming languages to avoid polyglot projects, \eg opensmalltalk-vm contains 89\% code from C, and SmalltalkCI contains 35\% shell scripts,\footnote{\url{https://github.com/OpenSmalltalk/opensmalltalk-vm}} and (vi) it contains more than 20\,000 lines of Smalltalk code, to remove small projects thus the projects MaterialDesignLite,\footnote{\url{https://github.com/DuneSt/MaterialDesignLite}} Kendrick,\footnote{\url{https://github.com/UNU-Macau/kendrick}} and PharoLauncher\footnote{\url{https://github.com/pharo-project/pharo-launcher}} were removed.

We sorted the projects based on commits and 
size (based on lines of code), and selected the top seven projects.
The projects consequently vary in size, domain, and contributors.
For each project we followed the same methodology used for selecting representative Pharo core comments.
Depending on the proportion of each project's comments with respect to the comments of all projects, we selected the sample comments.
 We extracted 351 comments in total from the selected external projects and analyzed their information types.\footnote{\repFolder{Dataset-for-Replication/Data/RQ2/external-projects}}

\subsection{Methodology}

We conducted a pilot study to construct initial categories of the content of comments.
We selected a sample of 100 classes from Pharo 7 classes with comments (6\,324) using a random sampling approach.
We used an open card-sorting approach and established the categorization procedure for the next larger-scale study.
The study was performed by the first author, and the classification granularity was set to sentence-level.
She manually analyzed the selected 100 classes, constructed new categories, and placed the comment sentences into appropriate categories according to the intent of the sentence.
Thus, she formed 21 categories, among them seven categories being inspired by the recent Pharo template.

She constructed the category names by looking at the intent of the sentence and type of information, resulting in an initial draft of the Pharo-CTM.\footnote{\repFile{Results/RQ2/pilot-study/Pilot-study-result.xlsx}} 
Once an initial taxonomy was elicited from the pilot study, we started the taxonomy study on 363 further comments to verify the completeness of the initial taxonomy, and to mitigate the chances of bias due to analysis by a single evaluator.

\subsubsection{Taxonomy study}
In this study, three evaluators (two Ph.D.\ candidates, one of whom was involved in the pilot study, and one faculty member, all authors of this paper) having at least four years of programming experience, participated in the study.
We divided our sample dataset (363 comments) equally among the three evaluators so that each subset (of size 121) had an equal number of comments selected randomly from each of the groups identified (see column \emph{selected for study} of \tabref{comment-rate-per-group} according to the distribution shown in \figref{comment-distribution-box-plot}).
This ensured that each evaluator's dataset included comments of all lengths and projects.
Then, we used a two-step validation approach to validate the content classification of the comment and the category name assigned to the content type.
This way, all the categories were discussed by all the evaluators for the better naming convention, and whenever required, unnecessary categories were removed and duplicates were merged.

\emph{Execution:} The evaluators analyzed the assigned comments by applying a hybrid card-sorting technique \ie assigning class comments to the initial taxonomy, and adding new categories whenever existing categories were found to be unsuitable for classifying the content.
This step was performed to verify if the taxonomy was exhaustive, or if potential categories were missing.

Once we finished the assigned individual evaluation of the comments, 
we started the collaborative validation explained next.

\emph{Validation:}
After analyzing all the comments, we validated the content classification of the comments over three iterations.
In the first iteration, each evaluator reviewed a random 50\% of the comments categorized by the other two evaluators.
This way, each comment categorization was reviewed by at least one of the other evaluators.
The reviewer (the evaluator who reviewed the comment's classification) marked his or her opinion by \emph{agreeing} or \emph{disagreeing} with each comment.
In case of disagreement, the reviewer highlighted the disputed categories and suggested changes.
In the second iteration, the evaluator studied the changes suggested by the reviewers and marked his or her agreement or disagreement for the changes.
In case of agreement, the classification was simply confirmed, otherwise the disagreements were carried to the next (third) iteration where the third evaluator who had not yet seen the comment reviewed it, and a decision was made based on majority voting.
In case all evaluators disagreed about a categorization, a discussion was started, and all three then discussed it to agree on a final classification.
Thus, only the marked discrepancies were resolved by reviewing each case with the involvement of all three evaluators.
The evaluators used pair-sorting~\citep{Guzz13a} to discuss discrepancies in their thoughts for each card during the card sorting itself.

\begin{figure}[h]
    \centering
    \includegraphics[width=\linewidth]{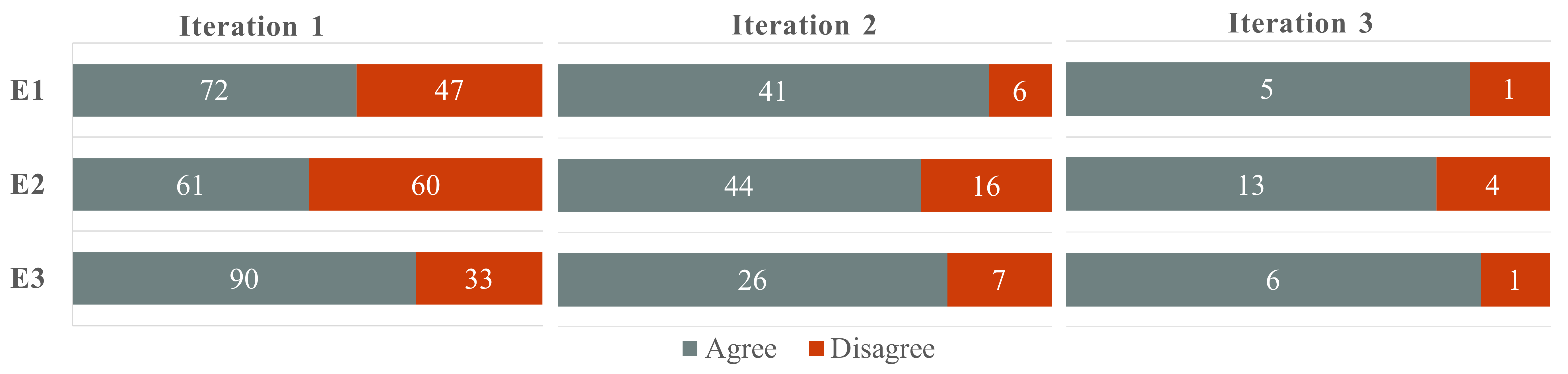}
    \caption{The status of comment classification discrepancies by reviewers in each iteration per evaluator}
    \figlabel{comment-classification-process}
\end{figure}

Levels of agreement and disagreement among the evaluators are reported in \figref{comment-classification-process}.
Specifically, in the first iteration, the reviewers reviewed the classification by the first evaluator, (E1) and agreed on the classifications of 72 comments and disagreed with 47 ones, suggesting changes for the disputed categories of 47 comments.
In the second iteration, the evaluator E1 agreed with suggested changes on 41 comments and disagreed with six.
In the third iteration, the cases where the reviewer and the evaluator disagreed were reviewed by the third reviewer who had not yet seen the comment.
The third reviewer agreed with the classification of five comments, but disagreed with one suggesting a different classification.
Finally,
for such a case, we discussed the conflict among all the evaluators and used a majority voting mechanism to finalize the classification.

After reaching a final agreement on the comment classification, we validated the category names.
We gathered all categories, and merged some redundant categories or renamed them using a majority voting mechanism, thus generating a final version of the taxonomy \ie Pharo-CTM.

\subsection{Results}
\seclabel{sub:rq2-results}
Our taxonomy study led to the finalization of Pharo-CTM, identifying 23 types of information (categories) present in the class comments
 The majority of these types, \ie 21 categories, are taken from the pilot study even though several categories of the pilot study underwent the refinement process (renaming, merging) for the final Pharo-CTM. From these 21 categories, seven belong to the Pharo template while six categories were merged to three categories in the taxonomy study.\footnote{\repFile{Results/RQ2/pilot-study/pilot-study-categories.pdf}}
The rest of the types, such as \emph{Subclasses Explanation}, \emph{TODO comments}, and \emph{Others}, were added during the taxonomy study.

\begin{table}[htp]
    \centering
    \caption[23 information types]{The 23 identified information types}
    \tablabel{taxonomy-table}
    \begin{adjustbox}{width=\textwidth}   
        \begin{tabular}{lp{0.40\textwidth}lp{0.20\textwidth}}
        \noalign{\smallskip}\hline\noalign{\smallskip}
        \textbf{Category} & \textbf{Description} & \textbf{Implicitness level} & \textbf{Keywords} \\
        \noalign{\smallskip}\hline\noalign{\smallskip}
        Intent & Describe purpose of the class & {Often Implicit} & {I represent, I am, I'm, This class is, A *Class* is}  \\
        \noalign{\smallskip}\hline\noalign{\smallskip}
        {Responsibility} & { List responsibilities of the class} & {Often Implicit}  &{ provide, implement, I do, I know, responsible} \\ 
        \noalign{\smallskip}\hline\noalign{\smallskip}
        {Collaborator} & { List interactions of the class with other classes} & { Implicit} &{use, interact, provide, collaborate} \\
        \noalign{\smallskip}\hline\noalign{\smallskip}
        {Public API} & { List key methods and public APIs of the class}   & {Sometimes Implicit} & {Key Messages, Public API} \\
        \noalign{\smallskip}\hline\noalign{\smallskip}
        {Example} & {Provide code examples to instantiate the class and to use API of the class} & {Often Explicit} & {Usage,  Example, For example, code examples}\\
        \noalign{\smallskip}\hline\noalign{\smallskip}
        {Implementation Points} & {Provide internal details referring to the internal representation of the objects, particular implementation logic, conditions about the object state, and settings important to understand the class} & {Often Implicit} & {Internal representations, Implementation points:}\\
        \noalign{\smallskip}\hline\noalign{\smallskip}
        {Instance Variables} & {List state variables of the object}  & {Often Explicit} & {instance variables:} \\
        \noalign{\smallskip}\hline\noalign{\smallskip}
        {Class references} & {Overlaps with Collaborator category but includes extra cases when developers refer to other classes in the class comment to explain the context of the class}  & {Implicit} & {} \\
        \noalign{\smallskip}\hline\noalign{\smallskip}
        {Warnings} & {Warn readers about using various implementation details of the class} & {Often Implicit} & {Note, do not, remarks, should} \\
        \noalign{\smallskip}\hline\noalign{\smallskip}
        {Contracts} & {Inform readers about potential conditions before or after using a class/method/component of the class} & {Often Implicit} & {Precondition:, do..when..} \\
        \noalign{\smallskip}\hline\noalign{\smallskip}
        {Dependencies} & {Describe the dependency of the class on other classes/methods/components}  & { Implicit} & {used by} \\
        \noalign{\smallskip}\hline\noalign{\smallskip}
        {Reference to other resources} & {Refer reader to extra internal or external resources}  & {Often Explicit} & { {See, Look}} \\
        \noalign{\smallskip}\hline\noalign{\smallskip}
        {Discourse} & {Inform the readers about a few class details in an informal manner} & { Implicit} & {developers use conversational language} \\
        \noalign{\smallskip}\hline\noalign{\smallskip}
        {Recommendation} & {Recommend the ways to improve the class implementation} & { Implicit} & {recommended, see, should be} \\
        \noalign{\smallskip}\hline\noalign{\smallskip}
        {Subclasses explanation} & {Describe details about its subclasses, the intent of creating the subclasses, and when to use which subclass} & {Implicit} & {My subclasses} \\
        \noalign{\smallskip}\hline\noalign{\smallskip}
        {Observations} & {Record developer observations while working with the class} & {Often Implicit} & {} \\
        \noalign{\smallskip}\hline\noalign{\smallskip}
        {License} & {Store license information of the code} & {Often Implicit} & {} \\
        \noalign{\smallskip}\hline\noalign{\smallskip}
        {Extension} & {Describe how to extend the class}  & {Often Implicit} & {extend, extension} \\
        \noalign{\smallskip}\hline\noalign{\smallskip}
        {Naming conventions} & {Record the different naming convention such as acronyms used in the code} & { Implicit} & {} \\
        \noalign{\smallskip}\hline\noalign{\smallskip}
        {Coding Guideline} & {Describe rules to be followed for coding by the developer while writing the class}  & {Often Implicit} & {} \\
        \noalign{\smallskip}\hline\noalign{\smallskip}
        {Link} & {Refer to a web link for extra or detailed information}  & {Sometimes Implicit} & {} \\
        \noalign{\smallskip}\hline\noalign{\smallskip}
        {TODO comments} & {Record actions to be done or remarks for developers} & {Explicit} & {todo} \\
        \noalign{\smallskip}\hline\noalign{\smallskip}
        {Other} & {Include the comments from other programming languages} & {Explicit} & {JavaDoc comments} \\
        \noalign{\smallskip}\hline\noalign{\smallskip}
    \end{tabular}
    \end{adjustbox}
    \end{table}

\tabref{taxonomy-table} presents an overview of this taxonomy.
The list of 23 identified information types, with full details and examples is available online.\footnote{\repFile{Results/RQ2/taxonomy-study/All-categories-with-examples.pdf}}
The column \emph{Description} describes the category, \emph{Implicitness level} defines the degree to which information is hidden in the text, and \emph{keywords} lists the keywords and patterns observed during manual analysis for each category.
The implicitness level is taken from a five-level Likert scale with items \emph{Implicit}, \emph{Often Implicit}, \emph{Sometimes Implicit}, \emph{Often Explicit}, and \emph{Explicit}.
A category is marked \emph{Implicit} when it is either in the same line or paragraph with other categories or without a header in the comment, making it difficult to identify.
For example, the category \emph{todo} is always mentioned in a separate paragraph with a header Todo, which makes it \emph{Explicit}.
On the other hand, a majority of the time the category \emph{Intent} is combined with \emph{Responsibility}  in one line thus making them \emph{Often Implicit}, but \emph{Collaborator} is always combined with other categories in the same paragraph without a header.
Based on the formulated criteria, one author evaluated the \emph{Implicitness level} of each category, and other authors reviewed them and possibly proposed changes. 
All authors resolved the disagreements by the majority voting mechanism and refined the measurement criteria by mutual discussions.
The examples for the categories are present in the respective category of classified comments.\footnote{\repFile{Results/RQ2/taxonomy-study/Taxonomy-study-results.xlsx}} 
We found that in one-line comments developers usually describe the \emph{Intent} of the class, and a very few times \emph{Responsibilities}.
A substantial number of comments contain warning information of some type (\eg a note about the code, or behavior of the class, an important point to keep in mind while extending the class).
In \emph{Others}, we observed a few comments having the source code from other languages and following the commenting style of other languages, such as C and Java.

\begin{figure}[t]
    \centering
    \includegraphics[width=\linewidth]{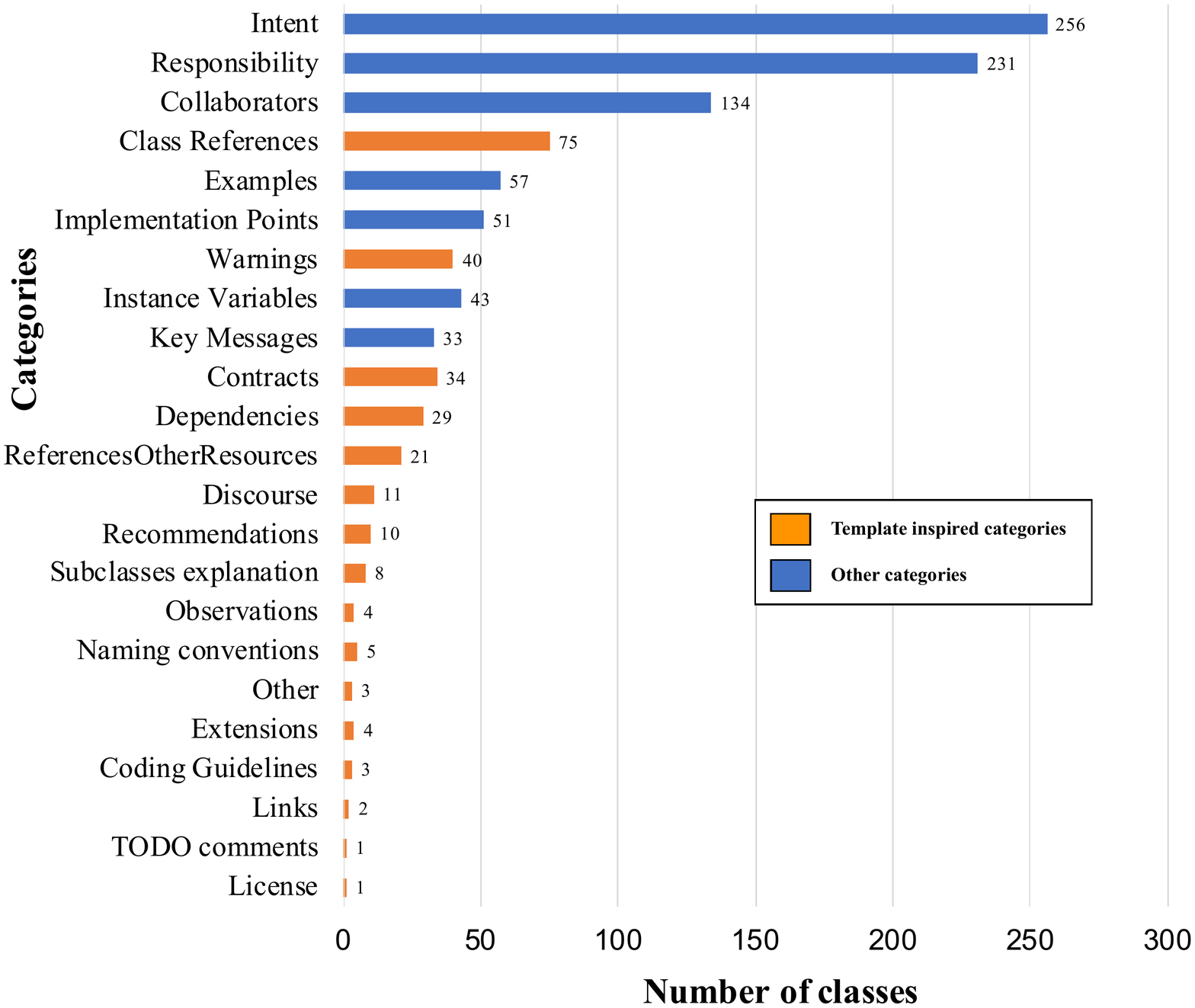}
    \caption{Information categories of class comments formed during manual analysis of the Pharo core (internal projects)}
    \figlabel{categories-from-manual-analysis}
\end{figure}

\figref{categories-from-manual-analysis} presents the distribution of the comments across all 23 categories.
There are seven template-inspired categories, which are colored in blue and the remaining categories are colored in orange.
The template-inspired categories contain the details proposed by the recent template.
Other categories, composed of 16 definitions, contain comment details that developers deem important to understand their class and therefore mention in the class documentation.

 \vspace*{2mm}
 \hspace*{-5mm}
 \begin{tikzpicture}
 \node [mybox] (box){%
 \centering
 \begin{minipage}{.95\textwidth}
\fontsize{9.5}{9.5}\selectfont 
 \emph{\textbf{Finding 5: } The most recent Pharo class comment template suggests writing seven different types of details, namely Intent, Responsibility, Public API, Example, Instance Variable, Collaborators, and Internal details.
Interestingly, developers frequently add other types of details such as Warnings, References to other classes and external docs, Dependencies, and Contracts in the class comments.}
\end{minipage}
};
\end{tikzpicture}%

\begin{figure}[t]
    \centering
    \includegraphics[width=\linewidth]{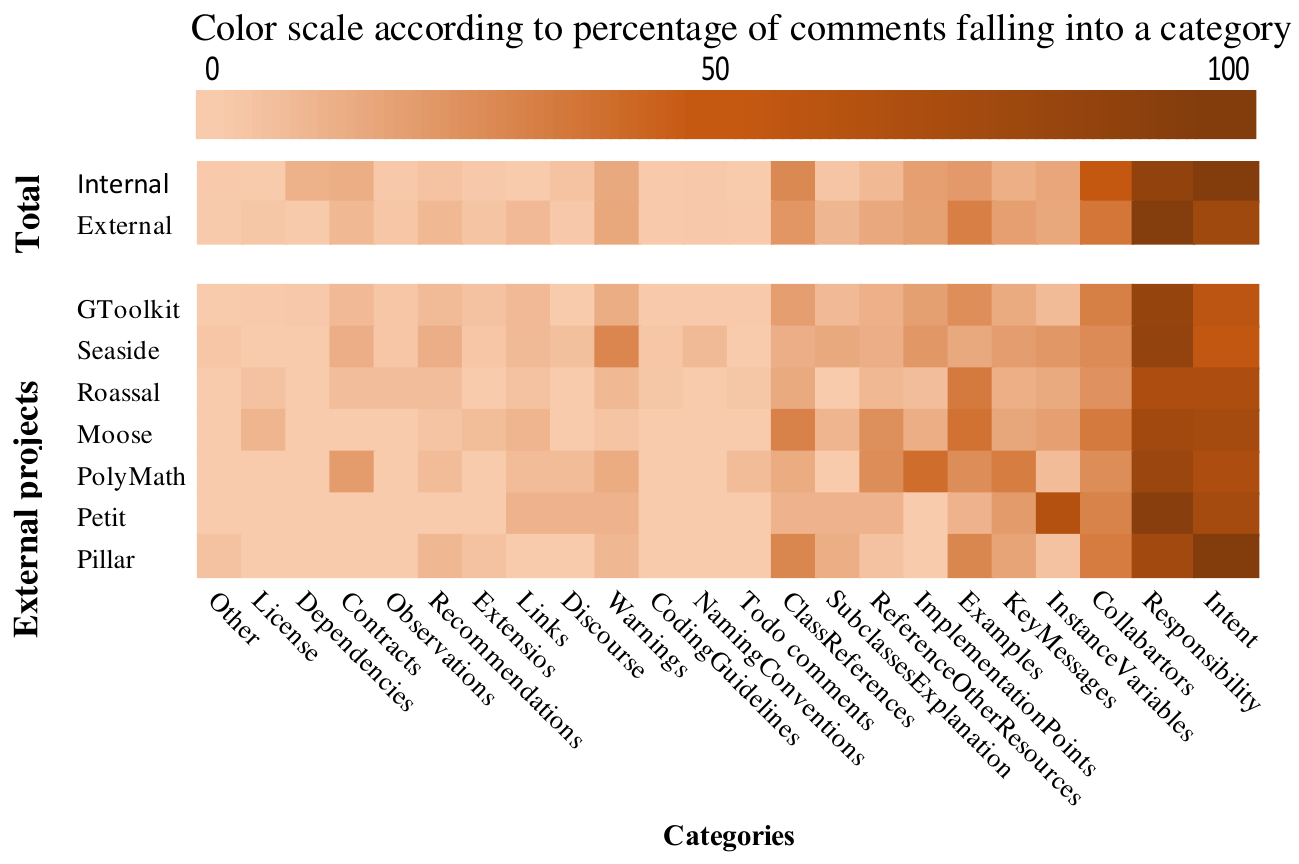}
    \caption{The trend of information types in external Pharo projects and comparison of total comments from Pharo external projects with Pharo internal (core) projects}
    \figlabel{external-projects-categories-existence}
\end{figure} 

In external projects, we found all 23 types of information embedded by developers as shown in \figref{external-projects-categories-existence}, though the frequency of some information types in comments is not as high as in Pharo core comments.
For example, \emph{Collaborators}, \emph{Implementation Points}, \emph{Contracts}, and \emph{Dependencies} are not found so often in the external projects as in the Pharo core.
Interestingly, we found that the project domain plays an important role in having a particular type of information.
For instance, Roassal, a visualization engine project, 
contains a large number of \emph{Examples} in the comments.
Most of the examples are small code snippets to create different visualizations using the class.
In contrast, we found detailed code examples (tutorials) in GToolkit class comments to explain how the project works.
Additionally, we found that template-inspired categories are not used so often as in the Pharo core.
On the other hand, some additional information types (not inspired by the template) are used more often than in the Pharo core.
A few such information types are \emph{Links, Recommendation, Subclasses explanation}, and \emph{References to other resources}.
Specifically, we found \emph{Links} in less than 1\% of Pharo core comments whereas nearly 6\% of comments from external projects contain \emph{Links}.
This suggests that the Pharo core and external projects contain similar information types (23) but with different frequencies.
Padioleau \etal analyzed operating system (OS) and non-OS projects and found similarities and differences in the kinds of details in project comments.
However, whether these similarities are due to common developers or coding guidelines, if any, is not investigated~\citep{Padi09a}.
On the other hand, our preliminary investigation found few common developers from the external projects \emph{Moose}, \emph{GToolkit} who contributed to the Pharo core projects as well.
Whether developers change their commenting practices in core and external projects would be an interesting topic to explore in the future.
Additionally, investigating the impact of the template on external projects in addition to Pharo core comments can also highlight the differences in developer commenting practices across projects.
In the future, we plan to investigate the impact of the template on external projects.

\vspace*{2mm}
\hspace*{-5mm}
\begin{tikzpicture}
\node [mybox] (box){%
\centering
\begin{minipage}{0.95\textwidth}  
\fontsize{9.5}{9.5}\selectfont  
    \emph{\textbf{Finding 6:} External projects in Pharo also contain 23 types of information as found in the Pharo core (internal projects).
    However, the frequencies of certain information types vary.}
\end{minipage}
};
\end{tikzpicture}%

\emph{Discussion:}
A very few categories are explicit, such as \emph{Examples}, and \emph{Instance variables}, and they are generally indicated by a header, such as \emph{Usage}, and \emph{Instance variables} respectively.
Most of the categories we found are implicit in the text and thus pose a challenge for the automated identification and extraction.
However, we observed various patterns for them.
Such patterns can help the researchers in designing approaches and heuristics to extract the specific information automatically.
For implicit categories mentioned more frequently, we observed that developers mostly use common keywords to indicate the specific types of information in their comments.
For instance, developers use a keyword \emph{Note} while describing any kind of warning, sometimes as a header as shown in \lstref{Explicit-warning-AthensCairoSurface-class}, or in the first line of the warning shown in \lstref{explicit-warning-BitBlt-class} whereas in some cases the information is implicit in the text as shown in \lstref{Implicit-warning-RBAbstractClass-class}.
Similarly to the implicit warnings, instructions for using a class as in \lstref{Implicit-Contracts-RPackageRenamed-class} are implicit, without any header or specific pattern.

\begin{minipage}{0.9\linewidth}
\begin{lstlisting}[caption= {Explicit warning given in the ``AthensCairoSurface'' class}, label={lst:Explicit-warning-AthensCairoSurface-class}]
**NOTE**
As a workaround of bitblt bug, the actual Cairo surfaces, created internally is with 1 extra pixel higher than requested.
This is, however completely hidden from users.
\end{lstlisting}
\end{minipage}

\begin{minipage}{0.9\linewidth}  
\begin{lstlisting}[caption= {Warning mentioned in the ``BitBlt'' class}, label={lst:explicit-warning-BitBlt-class}]
(Note also that a Form can be copied to itself, and transformed in the process, if a non-nil colorMap is supplied.)
\end{lstlisting}
\end{minipage}

\begin{minipage}{0.9\linewidth}
\begin{lstlisting}[caption= {Implicit warning given in the ``RBAbstractClass'' class}, label={lst:Implicit-warning-RBAbstractClass-class}]
They shouldn't be directly used and always be a part of a refactoring namespace - the model.
\end{lstlisting}
\end{minipage}

\begin{minipage}{0.9\linewidth}
    \begin{lstlisting}[caption= {Implicit Contracts given in the ``RPackageRenamed'' class}, label={lst:Implicit-Contracts-RPackageRenamed-class}]
        I am a public annoucement, sent when a new package is renamed.
    \end{lstlisting}
    \end{minipage}

For categories like \emph{Intent}, we observed that developers mostly mention the intent of the class in the first line of a comment.
For \emph{Class references}, we observed that class names are broken into words and not capitalized, thus making it hard to recognize the class name from the text.
Pharo does not provide any language mechanism to support private or public scope for APIs, therefore APIs used by other services are generally marked \emph{Public} by grouping such APIs in a protocol (interface) named \emph{Public}, and documenting these in the class comment as a recommended practice.
Additionally, we found that not all classes describe their public APIs in the class comments, and not all public APIs of the class are mentioned.
The APIs mentioned are those that are considered to be important by the developer who is writing the comment \eg the class ``FTAllItemsStrategy'' has eight methods, three of which are public APIs, but not all three are mentioned in the comment, and only one API ``realSearch'' is mentioned in the comment under the \emph{Public API and Key Messages} section.
Similarly, for other information types, developers follow different commenting practices, and the writing style shown in \tabref{taxonomy-table}.

\vspace*{2mm}
 \hspace*{-5mm}
 \begin{tikzpicture}
 \node [mybox] (box){%
 \centering
 \begin{minipage}{.95\textwidth}  
 \fontsize{9.5}{9.5}\selectfont
    \emph{\textbf{Finding 7:} The top three types of information found in comments are template-inspired categories and these categories are implicitly present in the text, but developers mostly use common patterns or keywords in mentioning them.}
\end{minipage}
};
\end{tikzpicture}%

All of these information types answer different developer questions in understanding the program, and assist them in various software development activities.
LaToza \etal surveyed  179 developers during coding activities and collected the questions perceived as being hard-to-answer by developers~\citep{LaTo10b}.
Questions about rationale, intent, and implementation are the topmost categories of those marked hard-to-answer by developers.
In our study, we also found that developers mention intent, rationale, and implementation information in their comments with high frequency, indicating that developers find such pieces of information important.
However, these information types are implicit in the text, which makes them hard to extract and present to the developers.
Better tool support and more studies are needed to address the general problem of identifying information types and highlighting them to assist developers.

\textbf{Code commenting practices in other systems}.
Several works in the past have explored the idea of identifying the information embedded in code comments to leverage them in various development tasks.
We attempt to summarize these related works based on the development systems, programming language, comment entity (\eg class comments, inline comments), and when possible, mapping their taxonomies to our taxonomy, as shown in \tabref{Related-work-comment-info-category-table}.
\begin{table}[h]
    \centering
    \caption[Comparison of related works on comment information categorization]{Comparison of related works on comment information categorization\\
    Note: M in the column \emph{Mapping to our taxonomy} represents mapping of one category from the related work's taxonomy to multiple categories in our taxonomy
    }
    \tablabel{Related-work-comment-info-category-table}
    \begin{adjustbox}{width=\textwidth}   
        \begin{tabular}{p{0.20\textwidth}lp{0.30\textwidth}p{0.40\textwidth}p{0.40\textwidth}}
        \noalign{\smallskip}\hline\noalign{\smallskip}
        \textbf{Study} & \textbf{Comment types analyzed} & \textbf{System analyzed} & \textbf{Categories proposed} & \textbf{Mapping to our taxonomy (M)} \\
        \noalign{\smallskip}\hline\noalign{\smallskip}
        \citep{Ying05a}  & Task comments &  \textbf{[Java]}: Eclipse Architect's Workbench (AWB) project & \textbf{7 categories}: communication, pointer to a change request, bookmark, current task, future task, location marker, concern tag & \textbf{1 category}: Task comments \\
        \noalign{\smallskip}\hline\noalign{\smallskip}
        \citep{Padi09a}  & Source code comments & \textbf{[C]}: Linux, FreeBSD, OpenSolaris 
        \textbf{[Java]}: Eclipse, \textbf{[C/C++]}: MySQL and Firefox & \textbf{6 categories (comment content)}: type, interface, code relationship, past future, meta, explanation & \textbf{5 categories}: type (M), code relationship (M), past future (Todo), meta (copyright), explanation (M) \\    
        \noalign{\smallskip}\hline\noalign{\smallskip}
        \citep{Haou11a} & Source code comments & \textbf{[Java]}: DrJava, SHome3D, jPlayMan & \textbf{3 categories (comment type)}: explanation comments, working comments, commented code, other &   \textbf{3 categories}: explanation comments (M), working comments (Todo), other (M) \\
        \noalign{\smallskip}\hline\noalign{\smallskip}
        \citep{Stei13b} & Source code comments & \textbf{[Java]}: CSLessons, EMF, Jung, ConQAT, jBoss, voTUM, mylun, pdfsam, jMol, jEdit, Eclipse, jabref, C++ & \textbf{7 categories}: Copyright comments, header comments, member comments, inline comments, section comments, code comments (commented code), task comments &  \textbf{5 categories}: copyright comments (license), header comments, member comments (M), section comments (M), and task comments (Todo) \\
        \noalign{\smallskip}\hline\noalign{\smallskip}
        \citep{Pasc17a} & Source code comments & \textbf{[Java]}: Apache (Spark, Hadoop), Google (Guava, Guice), Vaadin, Eclipse & \textbf{16 categories}: summary, expand, rational (intent), deprecation (warning), usage, exception, TODO, incomplete, commented code, directive, formatter, license, pointer, auto-generated, noise &  \textbf{9 categories}: summary (M), expand (M), rational, deprecation, usage (M), TODO, license, pointer (M), noise\\
        \noalign{\smallskip}\hline\noalign{\smallskip}
        \citep{Zhan18a} & Source code comments & \textbf{[Python]}: Pandas, Django, Pipenv, Pytorch, Ipython, Mailpile, Requests & \textbf{11 categories}: metadata, summary, usage, parameters, expand, version, development notes, todo, exception, links, noise & \textbf{8 categories}: metadata (M), summary (M), usage, expand (M), parameters, development notes(M), todo, links (M), noise (other) \\
        \noalign{\smallskip}\hline\noalign{\smallskip}
        \citep{Shin18a} & Local comments (inside methods) & \textbf{[Java]}: 1\,000 projects \textbf{[Python]}: 990 projects & \textbf{11 categories}: Preconditions, post conditions, value description, instructions, guide, interface, meta information, comment out, directive, visual cue, uncategorized & \textbf{7 categories}: pre conditions (contracts), post conditions (contracts), value description (instance variables), guide (examples), interface (key message), meta (license), uncategorized (other) \\
        \noalign{\smallskip}\hline\noalign{\smallskip}
        \citep{Hata19a} & Links in comments &\textbf{[C]}, \textbf{[C++]}, \textbf{[Java]}, \textbf{[JavaScript]}, \textbf{[Python]}, \textbf{[PHP]}, \textbf{[Ruby]}:  Projects from GitHub & - & Links\\

        \noalign{\smallskip}\hline\noalign{\smallskip}
    \end{tabular}
    \end{adjustbox}
    \end{table}
Based on our comparison analysis, code commenting practices vary across programming languages and systems.
For common information types present in the comments across systems such as summary, links, code examples, we observed that they differ in the way they are located in the system and the way they are written.
Hata \etal investigated the \emph{Links} embedded in the comments and found top three links \emph{github.com}, \emph{stackoverflow.com}, and \emph{en.wikipedia.com}\citep{Hata19a}. 
In our analysis, none of the links from Pharo core comments or external project comments point to \emph{github.com} or \emph{stackoverflow.com}. 
We did, however, find instances of \emph{Links} pointing to \emph{en.wikipedia.com} in Pharo external projects.

Padioleau \etal explored comments in different programming languages by focusing on Eclipse (IDE) written in Java, MySQL (a database server) and Firefox (a web browser) written in C and C++.~\citep{Padi09a}.
We observed similar information types with our taxonomy, such as code relationship, TODO, and deprecated code. 
In our work, we also observed these information types in both internal and external projects, though with lower frequency compared to Java, C and C++.
Indeed, Padioleau \etal found that several projects embed often these specific concerns, which can vary among different domains.
For example, OS-related projects contain a higher number of memory management, and lock/synchronization related concerns. 
In contrast, Eclipse comments include null references, error management, or links to issue tracker services (\eg Bugzilla).
Similar results have been reported by Pascarella \etal and Zhang \etal for code comments in Java and Python~\citep{Pasc17a,Zhan18a}.
In our study, we find that class comments of \emph{Roassal} contain a large number of code examples, with \emph{PolyMath} containing more implementation details compared to other external projects.
However, we did not find any error management related information, or links to issue tracker services in Pharo class comments.
Similarly to other languages, Pharo class comments contain object-oriented programming guidelines or design pattern details.
Hence, our results show a high diversity in commenting practices across various systems and languages.
In future we plan to systematically and more precisely compare class commenting practices in other popular languages.

\subsection{Implications}
\seclabel{rq2-implications}
Finding different types of information embedded in class comment can assist developers to quickly find and access information required for various development tasks.
In this section, we discuss the need of identifying information types in code comments of various application domains and languages.
We then discuss language-independent approaches to organize and identify such information type automatically:
\begin{itemize}
  \item
\emph{Need to analyze class commenting practices in other systems}:
Previous studies, as shown in \tabref{Related-work-comment-info-category-table}, have focused on classifying code comments, or specific types of information on these comments (\eg links and task comments). 
However, we observed that such studies do not classify code comment information according to specific comment types (\eg package comments, class comments, function comments).
According to standard coding style guidelines, different comment types report various kinds of information.
For example, the Java Oracle style guideline suggests adding author information to the class comments but not to the method comments. In contrast, Python PEP8 suggest to place this information after the module \emph{docstring}, and before the relevant statement.
On the other hand, in Pharo, the guidelines (and the class comment template) do not mention author information but we found instances of author information in the class comments.
This shows that class commenting guidelines vary across languages but to what extent developer class commenting practices vary is still unclear and it requires a systematic investigation.

  \item
\emph{Identify information types automatically}: 
The task of accessing the type of information embedded in comments depends on the kind of information (warning, rationale), level of detail (design level or implementation level) developers seek, the type of development activities they are performing, and the type of audience (user or developers) accessing them. 
Tools to automatically identify these information types can reduce the effort developers and other stakeholders invest in reading code comments when gathering particular types of information.
In addition,  on top of these automated tools, visualization strategies could be implemented to highlight and organize the content embedded in the comments, to further ease the process of obtaining the required information.
For example, identifying warnings from the comments can help turn them into executable test cases, so developers can automatically check that the mentioned warnings are respected.
Similarly, automatically identifying code examples from the comments and executing them can ensure that code examples are up to date.
In recent work by Pascarella \etal the authors build a machine learning-based tool to identify information types for Java automatically~\citep{Pasc17a}.
Similarly, Wang \etal developed such an approach for Python~\citep{Zhan18a}.
However, given the increasing trend of open-source systems written in multiple programming languages, these approaches can be of limited use for developers contributing to these projects~\citep{Toma14a}.
Our work has the aim to foster the building of language-independent tools based on comprehensive taxonomies for comments analysis of multi-language projects.
Future studies can leverage our labelled data as a starting point to build language-independent tools, and verify the correctness of their tools.

  \item
\emph{Designing an annotation language}:
Annotation languages have proven to improve the reliability of software.\footnote{\url{https://docs.microsoft.com/en-us/cpp/c-runtime-library/sal-annotations?redirectedfrom=MSDN&view=vs-2019}} 
They can help the community in labelling and organizing a specific type of information, and to convert particular information types into formal specification which can further help in synchronizing comments with the code~\citep{Padi09a}.
Even though Pharo comments do not follow any annotation, they do have hidden patterns for different information types such as instance variables denoted by \emph{Instance variables} or main methods of a class are indicated by \emph{Key Messages}.
We identified various such patterns in constructing our taxonomy highlighted in \emph{Keywords} in \tabref{taxonomy-table}. 
Pharo community can use such patterns in developing an annotation language for Pharo comments.
In our study, we find some information types express properties (according to implicitness level in \tabref{taxonomy-table}) which can be described via annotations such as \emph{Examples}, \emph{public APIs}, \emph{Links}.
Tool/language designers can utilize the identified patterns to design information headers and annotations.
\end{itemize}


\section{ RQ3: Adherence of commenting practices to the template }
\seclabel{Adherence-to-template}

Programming languages and communities not only provide guidelines to maintain uniform coding styles, they also provide documentation guidelines for writing comments to have a uniform commenting style across projects.
Java has JavaDoc,\footnote{\url{https://www.oracle.com/technetwork/java/javase/documentation} verified on 28 Jan 2020} Python follows a standard documentation style,\footnote{\url{https://www.python.org/doc/} verified on 28 Jan 2020} and Google suggests style guidelines.\footnote{\url{https://developers.google.com/style/api-reference-comments} verified on 28 Jan 2020}
JavaDoc provides certain guidelines such as ``Class descriptions can omit the subject, and simply state the object, use third person rather than second person.''\footnote{\url{https://www.oracle.com/technetwork/java/javase/documentation/index-137868.html}}
In Pharo, developers are guided by a template, shown in \figref{template_pharo_7}, which recommends the use of first-person pronouns, writing complete sentences, following CRC style, and providing extra information sections like \emph{Public API and Key Message}, \emph{Example}, and \emph{Internal Representation}.
However, it is not known
how the template has evolved, what sections of the template are used more often than others, and to what degree developer commenting practices conform to the template.
We investigate these aspects in our third research question:
\emph{\textbf{RQ3}: \rqIII}

After expanding our understanding of the templates gathered from all versions, we investigate the adherence of comments to the template.
We define adherence by focusing on two main aspects: adherence to the content type, and to the writing style.
We elaborate these two aspects as:
\begin{itemize}
    \item \emph{Content adherence: } If the comments contain information types as mentioned in the respective template, then we say the comments adhere to the template in the content aspect.
    \item \emph{Writing style adherence: } If the comments follow the writing style conventions of the template, then we say the comments adhere to the template in the writing style aspect.
    The writing style conventions are composed of various constraints formulated for each template information type.
    If the comments containing specific information fulfill the corresponding constraints, we say the comments adhere to the writing style.
\end{itemize}

We measure the content adherence of the comments in \secref{Pharo-Commenting-Practices} by analyzing the content of the selected comments manually.

To measure adherence to writing style, we first extract the guidelines from the template regarding how a comment should be written.
We convert the guidelines into writing style constraints to identify the writing style influence of the template on the comments.
Then we manually analyze the 364 comments selected using stratified sampling, according to the writing style constraints of corresponding template version.
With the manual analysis study, we verify our definition and uncover other patterns of writing style.
Once we calculate both aspects of comment adherence, we answer RQ3.

We argue that this analysis will help researchers in evaluating the usage and importance of a comment template, and highlighting potential aspects to improve it.

\subsection{Study Setup}
To study the evolution of the template, we extracted the template from each Pharo version since Pharo 1 and compared all template versions to record the differences.

In order to measure the adherence of commenting practices to the template, we extracted the class comment template and a sample of an equal number of classes from each version, then identified the information types they contain.
The classes chosen for the study should be the newly added classes of each version, to make sure that the developer got a chance to look at the default template.
This is because, in Pharo, the template appears only when developers add a class comment to the class for the first time.
For each comment in the sample set (363) used in the RQ2, we therefore identified the original Pharo version when the comment was first added to the class.
We then extracted the class comment of that version to compare the comment to the corresponding template in content and writing style aspects.
For example, for a class comment added in Pharo 2, we compared the comment to the Pharo 2 template.

This partitioning of 363 comments according to the original Pharo version led to an unequal number of comments for each Pharo version \eg out of 363 comments version 2 has fewer than 40 comments whereas version 7 has more than 60 comments.
Furthermore, to compare the class commenting practices of all versions across each other, we selected an equal number of comments from each version.
To balance the equal sample comments from each version, we set a lower threshold of 52 comments for each Pharo version, summing to a total of 364 comments.
We extracted more comments from the Pharo versions where there were fewer than 52 comments, mainly Pharo 2 and Pharo 4.
For each such version, we selected the sample classes from newly added classes with comments shown in the top dark blue segment of \figref{Percentage-classes-with-and-without-comments} according to the distribution of comments based on the number of sentences present in a comment.
Similarly, we removed the classes from Pharo versions where there were more than 52 comments, mainly Pharo 1, Pharo 6, and Pharo 7, based on the distribution of comments of each version.
We followed the same approach to choose representative comments as used in 363 comments from the earlier study (taxonomy study).

\subsection{Methodology}
\subsubsection{Template Evolution}
We analyzed the template of each Pharo version and created a template meta-model for each version.
When a class is created, a default class comment template is added to the class, \eg the recent template is shown in \figref{template_pharo_7}.
We created a class with one instance variable and then observed the changes in the default class comment template.
According to the available details in the comment template, each author of the paper prepared their own interpretation of the template model for each Pharo version.
Once we prepared the template models for all versions, we compared and discussed them to reconstruct and establish one template model for each version.
There were few intermediate Pharo versions where the template had not changed; in such cases we used the same template model from the earlier version.
Thus each template model captures the differences from preceding and succeeding versions and presents the evolution of the template (models of the various template versions are reported in \figref{template-meta-models}).

\subsubsection{Adherence of comments to the template}
We grouped all 364 comments according to their original Pharo versions (when the comment was first added to the class) so that we could differentiate the comments of one version from another version, analyze their evolution, and compare them to the corresponding template of that version.
Then we identified the comment information types of 364 comments following the methodology used for the taxonomy study.
Once we identified the comment information types of all comments, we identified the information types and writing style guidelines from the templates by studying the content of each template corresponding to the Pharo version.
Three authors of the paper participated in the study and analyzed each version's template independently.
Then, we used a two-step validation approach, thus validating the content classification of the template and the name assigned to the classified content.
Specifically, the content classification was validated by an iterative evaluation process where each evaluator reviewed the other's content classification.
This way, all the information types were discussed by all the evaluators for the better naming convention and classification.

Similarly, we extracted the writing style guidelines hinted by each information type of each version's template, discussed among ourselves and formulated several constraints for each information type.
For instance, \emph{For the Class part} section of the Pharo 7 template in \figref{template_pharo_7} is identified as \emph{Intent} information type.
For this type, we extracted the guidelines from the keywords \emph{State one line, I represent} and converted them into rules such as \emph{description should be one line, subject should be first person}, and have a pattern of <\emph{subject}>, <\emph{verb}> from \emph{I represent}.
The process of finalizing the constraints for all information types of the Pharo 7 template is shown in the replication package.\footnote{\repFile{Results/RQ3/Constraints-definition-for-template-writing-style.xlsx}}

\begin{figure}[ht]
    \centering
    \includegraphics[width=\linewidth]{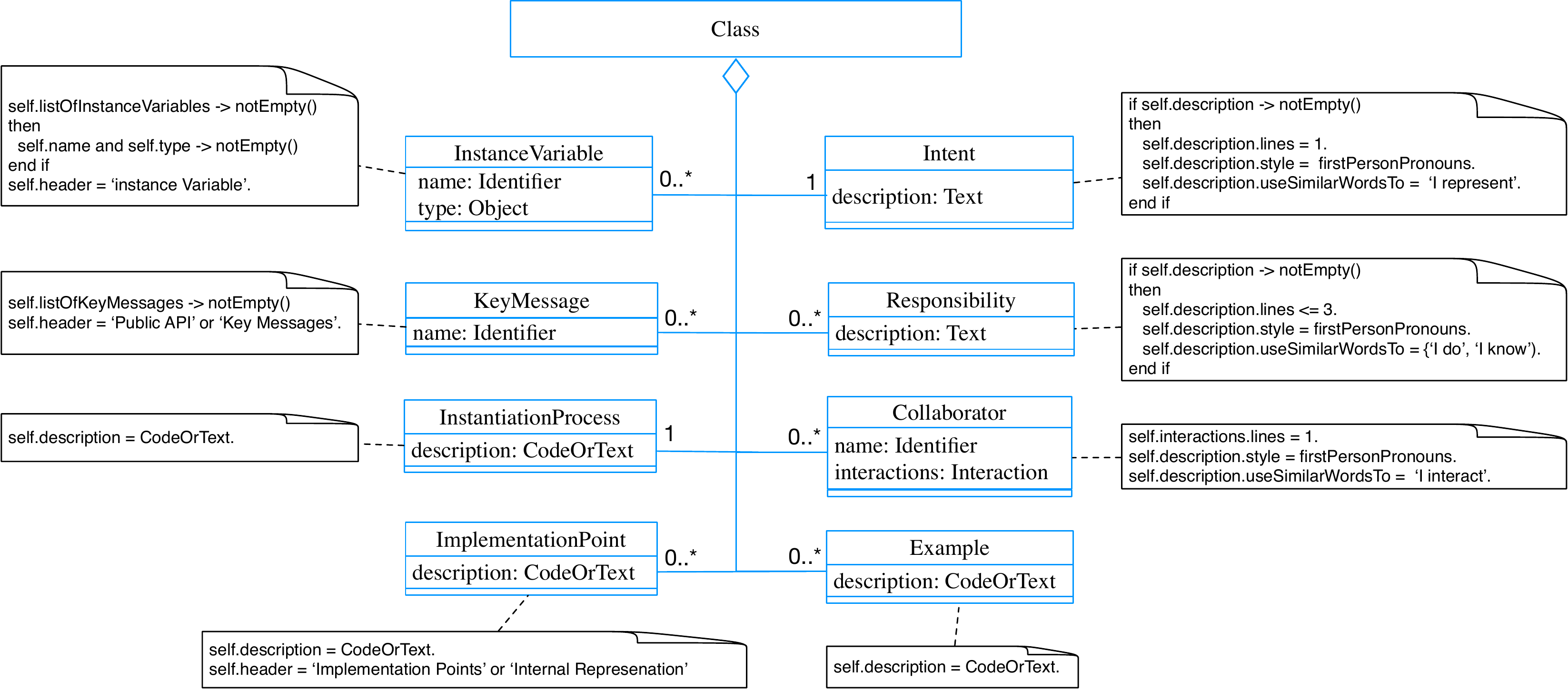}
    \caption{Writing style constraints formulated for Pharo 7 template}
    \figlabel{Pharo-7-writing-style-constraints}
\end{figure}
 The final constraints for the Pharo 7 template are shown in \figref{Pharo-7-writing-style-constraints}.
A complete list of all constraints and their examples for each Pharo version can be found in \autoref{appendix:template-models}.
There were few intermediate Pharo versions where the template had not changed; in such cases we used the same information types and writing style guidelines from the earlier template.

\textbf{Content adherence:}
After identifying all the information types from each template version, we compared them to each version's information types identified via Pharo-CTM.
For example, for a class comment added to the class in Pharo 2, we compared the information types of the comment to the information types identified from the template that existed in Pharo 2, thus comparing what developers typically write in their comments to the information proposed by the template.

\textbf{Writing style adherence:}
Some of the constraints identified from a template can be verified automatically in the comments and do not require manual intervention but could lead to less reliable results due to the freedom of writing free text in the class comments, non-availability of formatting standards, and limited patterns available in the template.
Additionally, there are chances to miss the cases where selected patterns are not present, and instead developers use synonyms to describe the same detail or do not describe the detail under a specific section header, say \emph{Instance variables}, and just write the instance variable details without any header.
We therefore manually analyzed the 364 comments (52 comments from each version), using the same setup as that of our studies of manual analysis performed in RQ2 and RQ3 for identifying the information types.
We followed the same iterative approach for evaluating the writing style constraints and the same validation approach as used in the taxonomy study.
We used the pair sorting approach to decide whether a sentence in the comment fulfills the constraints, and was influenced by the template or not.

After collecting all the data, we used statistical tests to verify whether there is a statistically significant difference between the scores (\eg the number of classes that adhere to the Pharo template style) when observing different Pharo versions.
 We employed non-parametric tests since the Shapiro-Wilk 
test revealed that the numbers of commented classes among Pharo versions do not follow a normal distribution ($p \ll 0.01$).
Hence, we used the non-parametric Wilcoxon Rank Sum test with a $p$-value threshold of $0.05$.
Significant $p$-values indicate that there is a statistically significant difference between the scores.
In addition, we computed the effect-size of the observed differences using the Vargha-Delaney \^{A}$_{12}$ statistic~\citep{Varg00a}.
The Vargha-Delaney \^{A}$_{12}$  statistic also classifies the obtained effect size values into four different levels (\emph{negligible}, \emph{small}, \emph{medium} and \emph{large}) that are easier to interpret.

\subsection{Results}
\subsubsection{Template Evolution}

\begin{figure}[ht]%
    \centering
    \subfloat[Template model of Pharo 1]{
    \figlabel{Template-Model-Pharo-1}
     \includegraphics[width=0.5\linewidth]{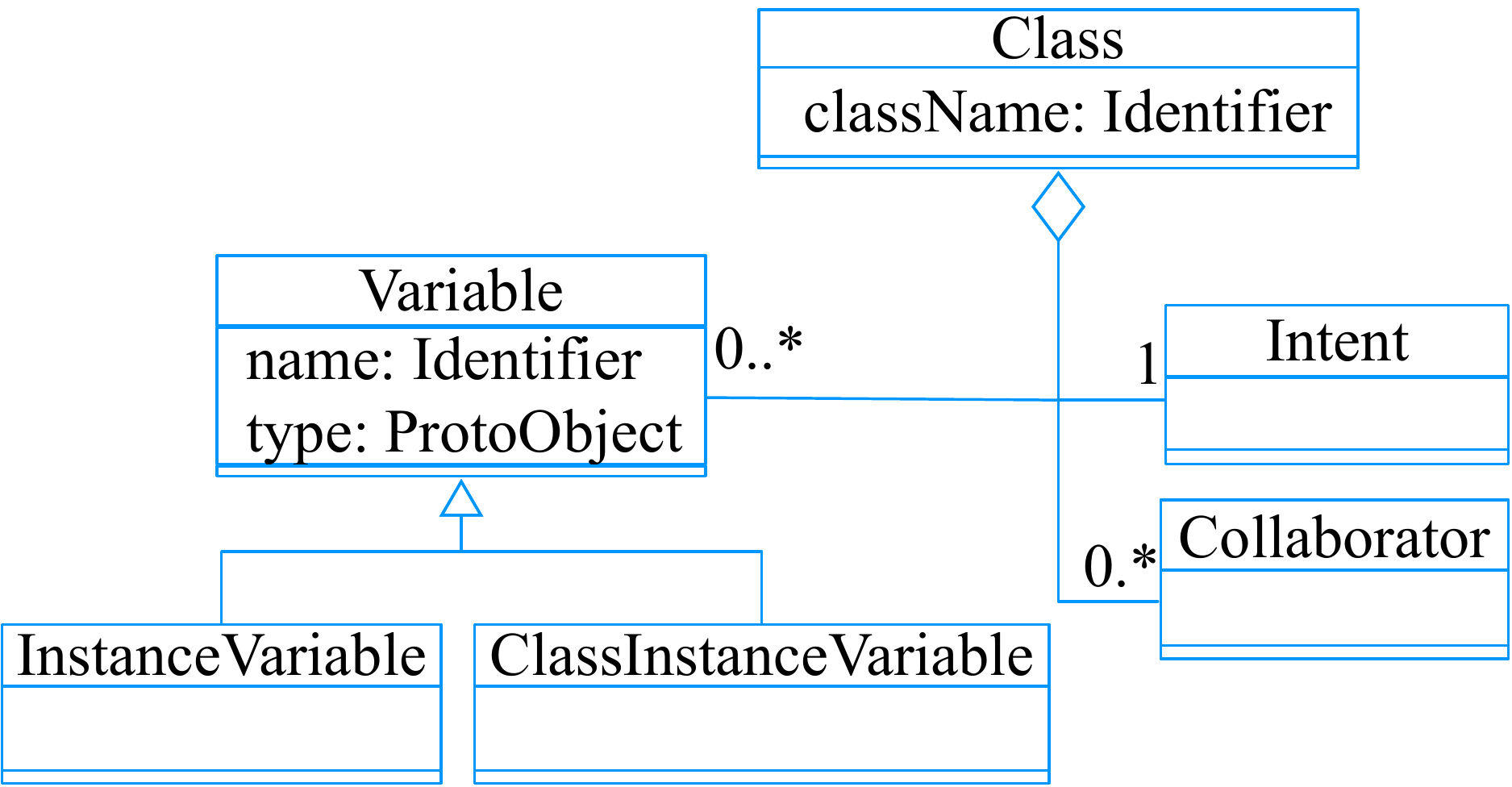}}%
     \qquad
    \subfloat[Template model of Pharo 2 and 3]{
    \figlabel{Template-Model-Pharo-2-3}
      \includegraphics[width=0.5\linewidth]{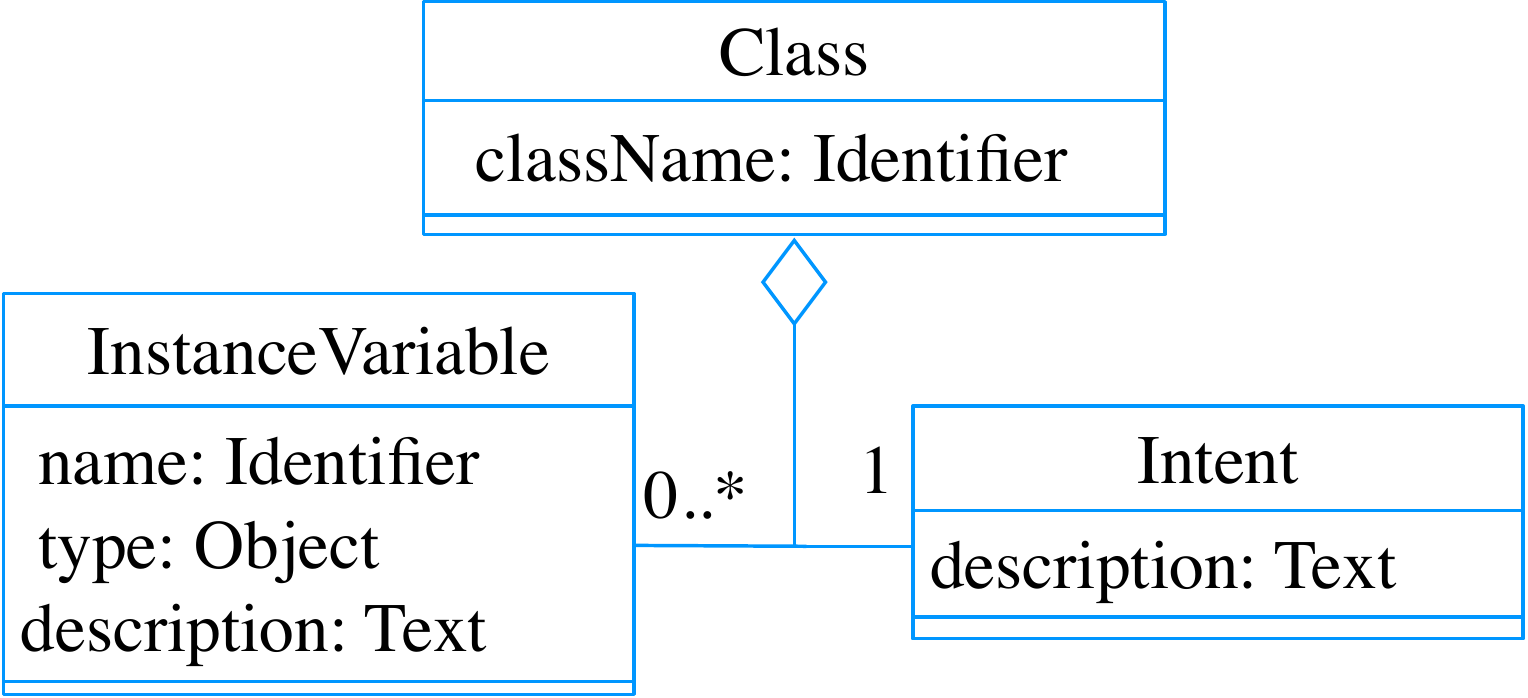}}%
     \qquad 
    \subfloat[Template model of Pharo 4]{
    \figlabel{Template-Model-Pharo-4}
      \includegraphics[width=0.5\linewidth]{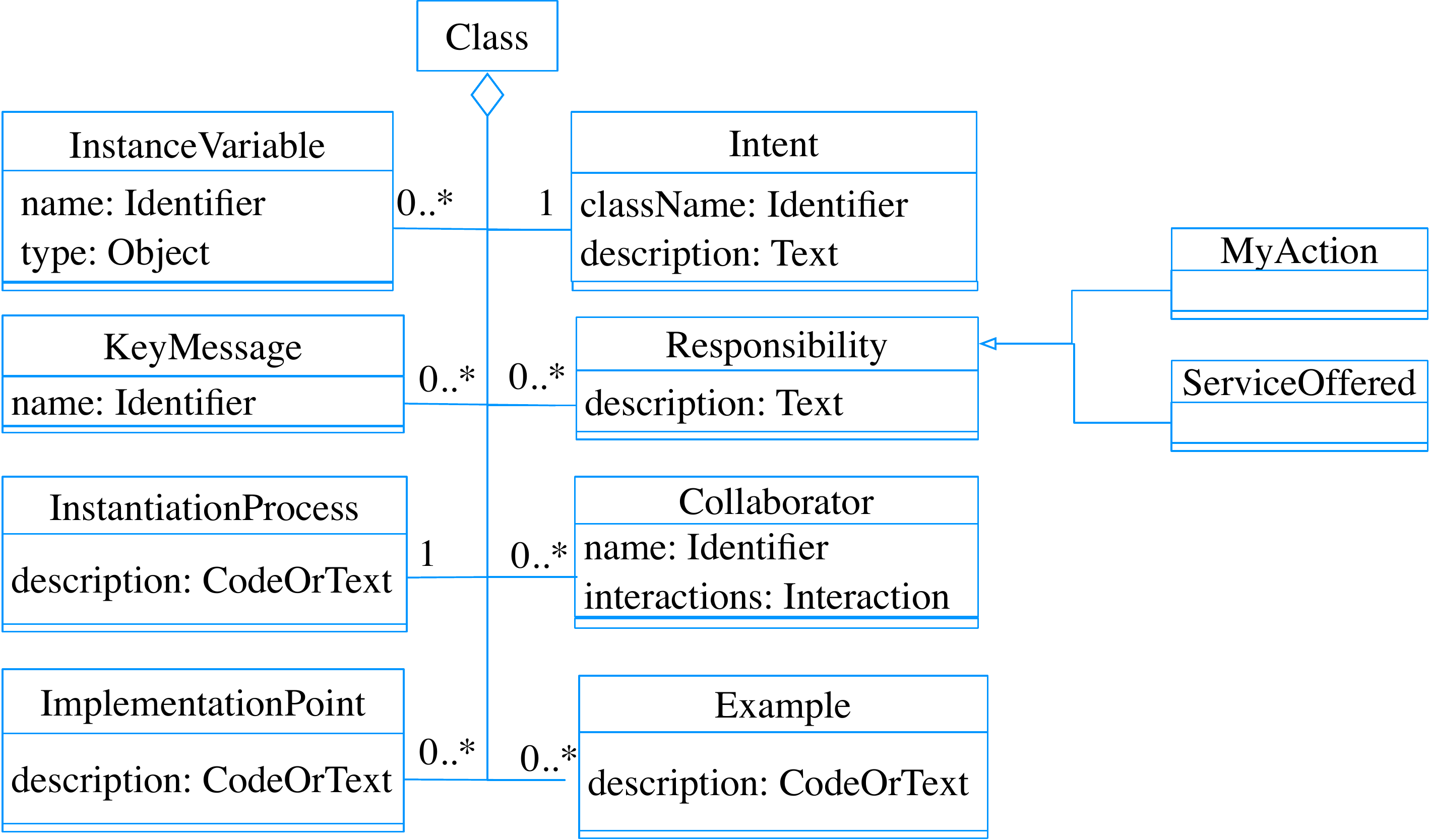}}%
     \qquad 
    \subfloat[Template model of Pharo 5, 6 and 7]{
    \figlabel{Template-Model-Pharo-5-6-7}
      \includegraphics[width=0.5\linewidth]{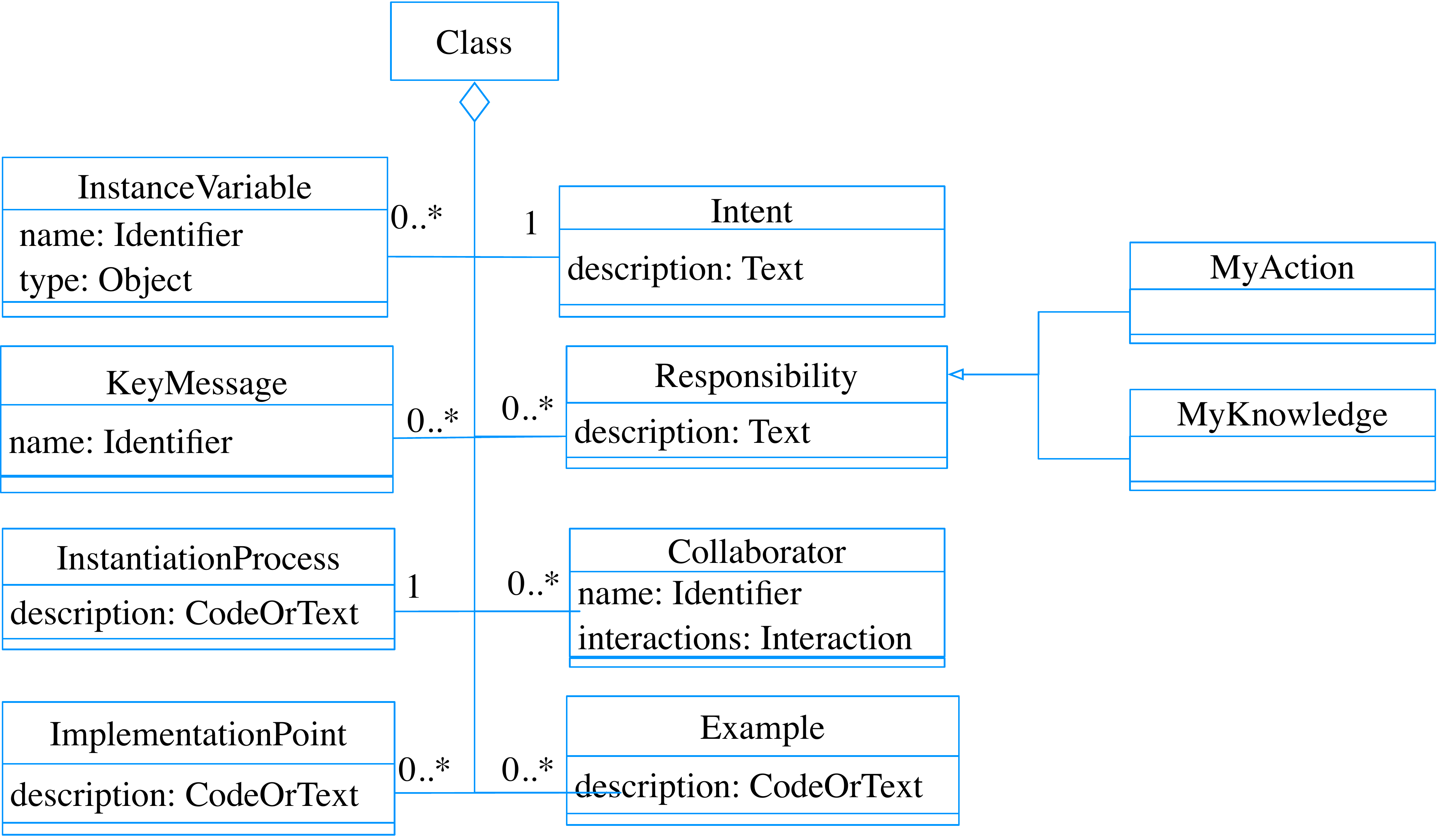}}%
    \caption{Template models for Pharo versions}%
    \figlabel{template-meta-models} 
\end{figure}

    Analyzing the template meta-models in \figref{template-meta-models}, we found that in the first Pharo template version shown in \figref{Template-Model-Pharo-1}, the template includes class side and instance side variables, and adds the class name and instance variable names by default.
    In later Pharo versions, class side variable information is omitted, and is shifted to the class side template.
    In the second and third Pharo versions in \figref{Template-Model-Pharo-2-3}, the template adds a description line for each instance variable to encourage developers to explain each instance variable.
    Additionally, the first line of the template refers to the intent of the class.
    In Pharo version 4 in \figref{Template-Model-Pharo-4}, the template underwent major changes and incorporated the CRC design to encourage the developers to describe the class intent, its responsibilities and its collaborators.
    The template presents different types of details to include in the class comment, and also gives examples to show developers how to write a comment.
    Since Pharo version 5 shown in \figref{Template-Model-Pharo-5-6-7}, the template remains the same.
    Compared to the previous Pharo version 4, the template asks developers to document ``what I know'' rather than ``what services do I offer'' in the responsibility section.
    
    We also observed that in Pharo version 1, there is a common template for the class side and the instance side.
    Then in later versions (from version 2 to 6), different default templates exist for the class side and the instance side.
    In recent version (7), again a single template is introduced for both the class side and the instance side.
    The reason for removing such a feature can be to simplify the template behavior, but this loses the facility of documenting the class side instance variables automatically in the template.

    \subsubsection{Adherence of comments to the template}
    This section aims at understanding the template of each Pharo version, finding the differences among templates, and comparing the commenting practices of developers with the class comment template.
    For each part of the question, we present our results and discussion.

\textbf{Content Adherence:}
Analyzing the information embedded in the comments shows that developers document different kinds of information in the class comments to make their classes more understandable and maintainable.
However, whether the practice of embedding various information types in the class comments is recent or present from initial Pharo versions, is unexplored and unknown.
\begin{figure}[h]
    \centering
    \includegraphics[width=\linewidth]{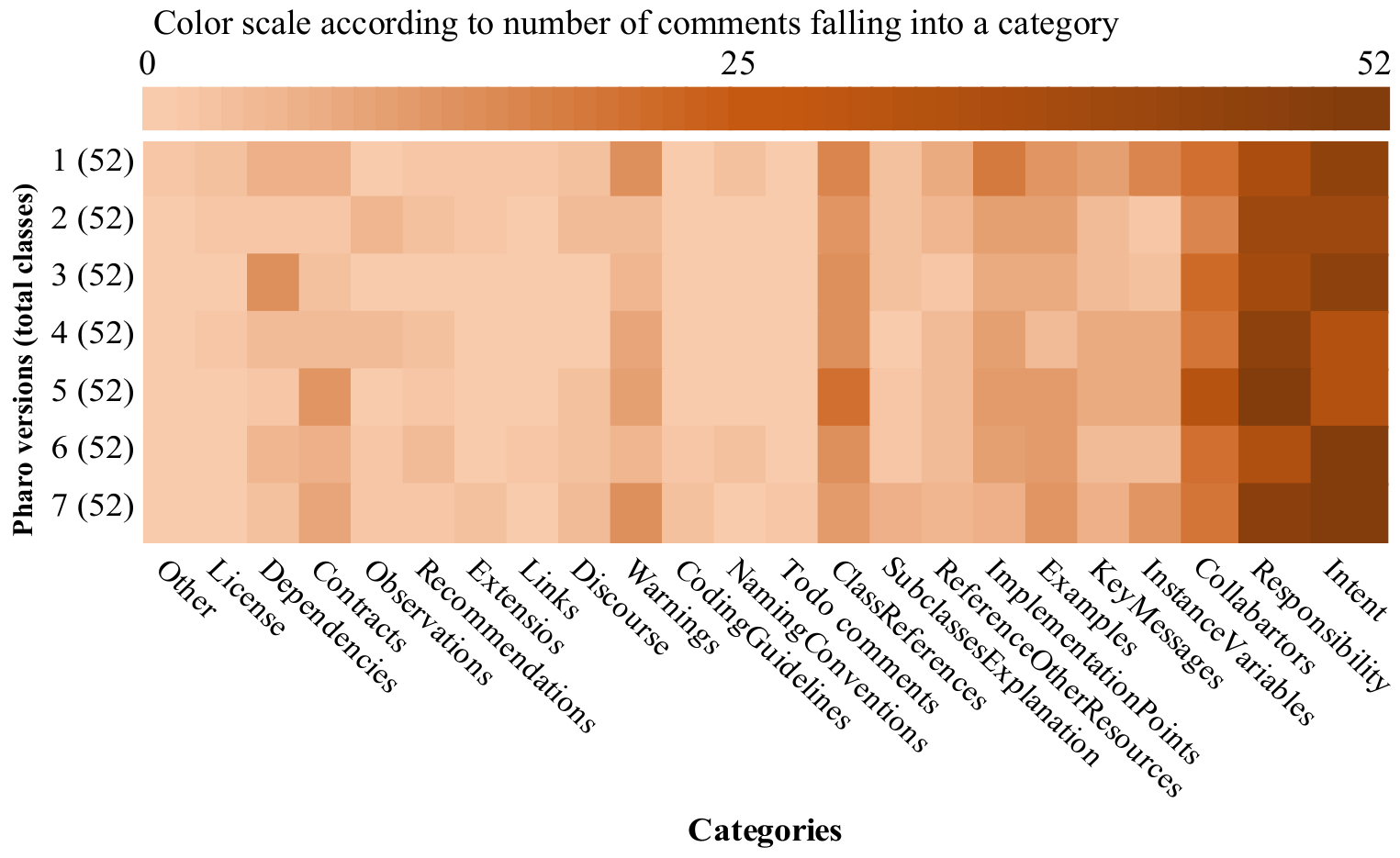}
    \caption{The trend of information types in Pharo versions}
    \figlabel{categories-existence}
\end{figure} 

In \figref{categories-existence}, the x-axis lists the information types, and the y-axis shows the Pharo versions with a number of classes considered for each Pharo version.
A darker shade of orange indicates a large number of comments having a particular type of information, and a lighter shade indicates a smaller number of comments falling into the information type.
 From our analysis, we found that most of the information types are present in the comments since Pharo 1 except \emph{Todo comments}, \emph{Coding Guidelines}, and \emph{Observations}.
A few information types like \emph{Intent}, \emph{Responsibility}, \emph{Collaborators}, and \emph{Examples} are highly frequent in all versions of Pharo.

\begin{table}[t]
    \centering
    \caption{The trend of information types in Pharo Template versions}
    \tablabel{template-information-types-trends}
    \begin{tabular}{lp{5.6cm}}
        \noalign{\smallskip}\hline\noalign{\smallskip}
    \textbf{version} & \textbf{categories}\\
\noalign{\smallskip}\hline\noalign{\smallskip}        
        1 & Intent, Collaborator, Instance Variables \\
        2-3 & Intent, Instance Variables \\
        4-7 & Intent, Responsibility, Collaborator, Instance Variables, Key Messages, Example, Implementation Points \\
        \noalign{\smallskip}\hline\noalign{\smallskip}
    \end{tabular}
\end{table}

Looking at \tabref{template-information-types-trends}, we see that the template suggests only a few information types to write in the class comment, especially in the initial three Pharo versions.
Later on, the template suggested seven types of information.
However, there are other information types mentioned by developers than those suggested by the template.
For example, the Pharo 1 template mentions three types of information shown in \tabref{template-information-types-trends}, but developers mention 20 other types of information shown in \figref{categories-existence}.
In the most recent template, among 23 types found in the comments only seven are present in the template.
Analyzing the developer practices of writing information seen in \figref{categories-from-manual-analysis}, we found that the information types suggested by the template are mentioned more frequently in the comments than other information types found in comments.
For instance, \emph{Intent} and \emph{Responsibility} are present in 65\% of sample class comments, while \emph{Warnings} is present in 12\% of the sample class comments, indicating the relevance of the template in terms of its information types.

 \vspace*{2mm}
 \hspace*{-5mm}
 \begin{tikzpicture}
 \node [mybox] (box){%
 \centering
 \begin{minipage}{.95\textwidth}  
 \fontsize{9.5}{9.5}\selectfont 
     \emph{\textbf{Finding 8:} Most of the information types are available in the comments since Pharo version 1.
A few information types like \emph{To do comments, Coding guidelines}, and \emph{Observations} are not found in the initial version.}
 \end{minipage}
 };
 \end{tikzpicture}%

 \vspace*{2mm}
 \hspace*{-5mm}
 \begin{tikzpicture}
 \node [mybox] (box){%
 \centering
 \begin{minipage}{.95\textwidth}  
 \fontsize{9.5}{9.5}\selectfont 
     \emph{\textbf{Finding 9:} The template-suggested information types are mentioned more frequently in the comments than other types of information.}
 \end{minipage}
 };
 \end{tikzpicture}%

\begin{figure}[ht]%
    \centering
    \subfloat[Distribution of the differences between Pharo versions]{
    \figlabel{writng-style-rules-boxplot}
     \includegraphics[width=0.45\linewidth,height=.44\linewidth]{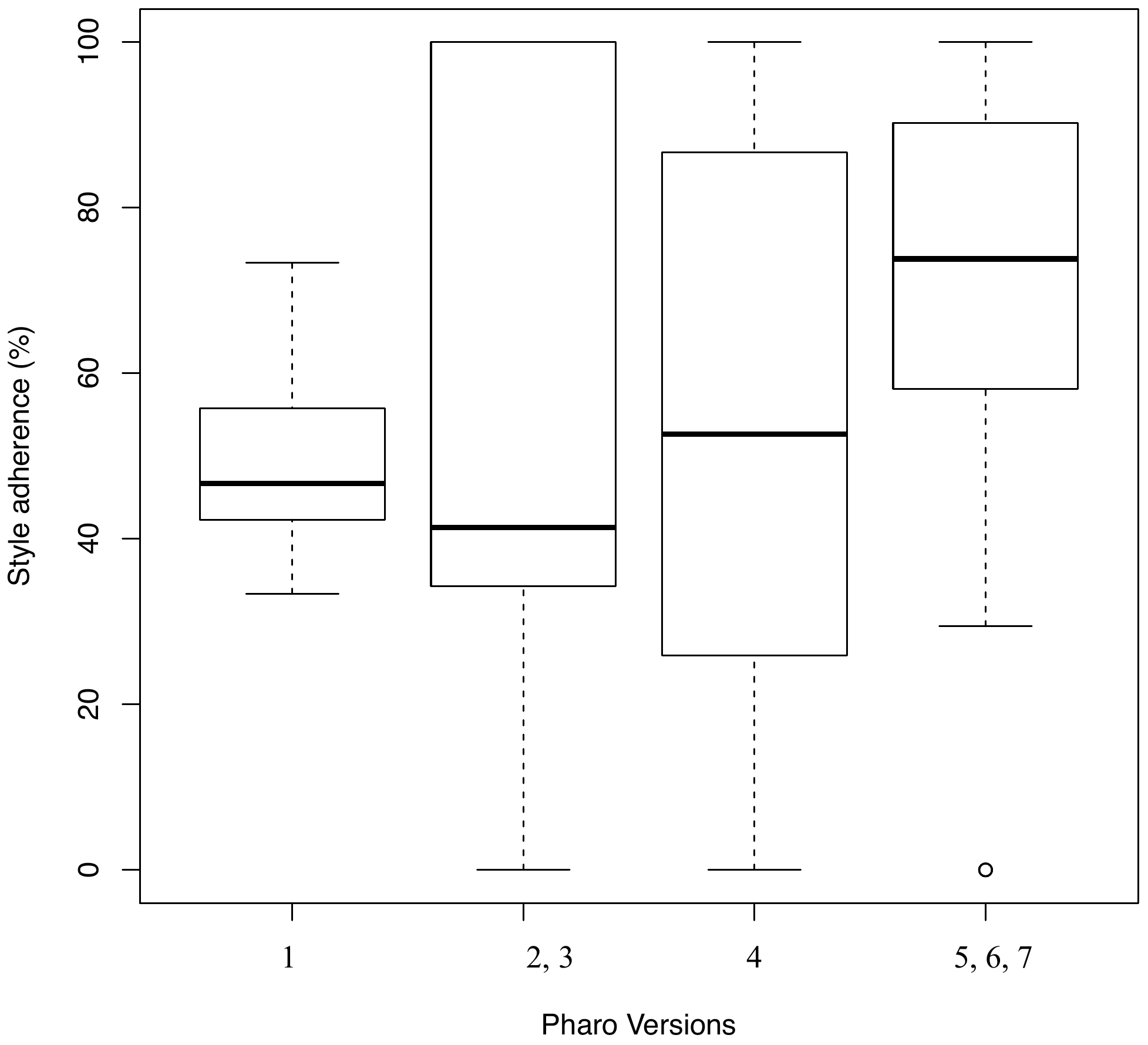}}%
    \qquad
    \subfloat[The trend of following the writing style rules in Pharo versions]{
    \figlabel{writing-style-rules-evolution}
      \includegraphics[width=0.45\linewidth,height=.47\linewidth]{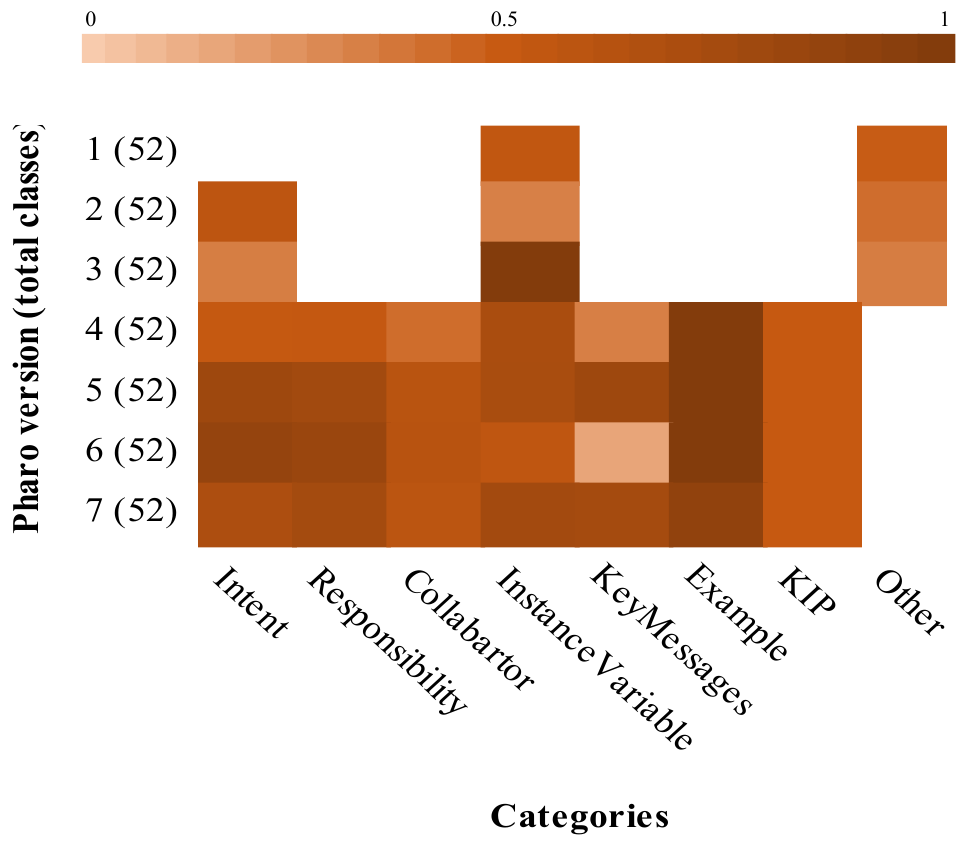}}%
    \caption{Comments following the writing guidelines over Pharo versions}
\end{figure}

\textbf{Writing Style Adherence:}
Analyzing \figref{writng-style-rules-boxplot}, we observe that Pharo 1 comments follow the rules 50\% of the time whereas, since Pharo 4, the trend of comments adhering to the style rules increased to 75\%.
To understand these differences between Pharo versions, we grouped  comments according to the changes in the template \eg the template in Pharo 2 and Pharo 3 has been the same, therefore, we grouped the comments from Pharo 2 and Pharo 3 and measured the percentage of comments adhering to the writing style rules.
After grouping the comments according to the version, we use the Wilcoxon test as well as the Vargha-Delaney \^{A}$_{12}$ statistic to observe potential statistical significant differences in the results achieved by classes of the grouped versions.
The results of the Wilcoxon test highlight a marginal significant difference (\ie $p$-values of $0.0673$ ) is observed between Pharo 1 and the Pharo 4, 5, 6 groups.
For these groups, the Vargha-Delaney statistic also reveals that this difference is large.

\vspace*{2mm}
\hspace*{-5mm}
\begin{tikzpicture}
\node [mybox] (box){%
\centering
\begin{minipage}{.95\textwidth}  
\fontsize{9.5}{9.5}\selectfont 
    \emph{\textbf{Finding 10:} Developer commenting practices adhere more to the writing style guidelines since Pharo 4 especially in describing the \emph{Intent, Responsibilities}, and \emph{Instance Variables} of the class.}
\end{minipage}
};
\end{tikzpicture}%

We further explored the differences between Pharo versions by measuring the adherence of comments to specific information types of each template version shown in \figref{writing-style-rules-evolution}.
We found that \emph{Example} and \emph{KIP} (Key Implementation Points) are always inconsistent due to unavailability of strict guidelines to write them.
The rule in the \emph{Example} section mostly checks the presence of an example in the comment written either in natural language or a code snippet, but the templates do not suggest any guidelines to write and format it.
Developers therefore follow various conventions to mention examples, such as using dedicated headers \emph{Usage, Examples, Code examples}.
Similarly, for \emph{KIP}, one of the rules just checks the presence of the implementation details in the comment.
Another rule in \emph{KIP} section suggests to write the header \emph{Internal representation and Implementation points} while mentioning the implementation details, but this is rarely followed by developers.

\begin{minipage}{0.9\linewidth}
    \begin{lstlisting}[
            showstringspaces=false,
            basicstyle=\footnotesize\ttfamily,
            % background
            frame=single,
            framerule=0.5pt,
            backgroundcolor=\color{source},
            caption= {Implementation points header present in the ``SycMethodCommand'' class}, label={lst:implementation-header-SycMethodCommand-class}]
    I am a base class for commands which perform operations with collection of methods.

    Internal Representation and Key Implementation Points.
            
    Instance Variables
    methods:		<Collection of<CompiledMethod>>
    \end{lstlisting}
\end{minipage}

In our analysis, we found several comments where only the header is present, but no further details are mentioned below the header.
We believe this is due to a lack of attention from developers in deleting unused section headers.
One of the cases we encountered is in the class ``SycMethodCommand'', shown in \lstref{implementation-header-SycMethodCommand-class}, where the developers have not provided any details under \emph{Internal representation and Implementation points} section, but the header is still present.
In the case of writing the \emph{Instance Variable} information, its header is mentioned in most of the cases with the instance variables.
One of the reasons for such a behavior can be the feature of Pharo of adding an instance variables section automatically to the class comment template if the class is created with instance variables.

\begin{figure}
    \centering
    \subfloat[Comments following subject-form guidelines]{
    \figlabel{subject-form-analysis}
     \includegraphics[width=0.45\linewidth,height=.44\linewidth]{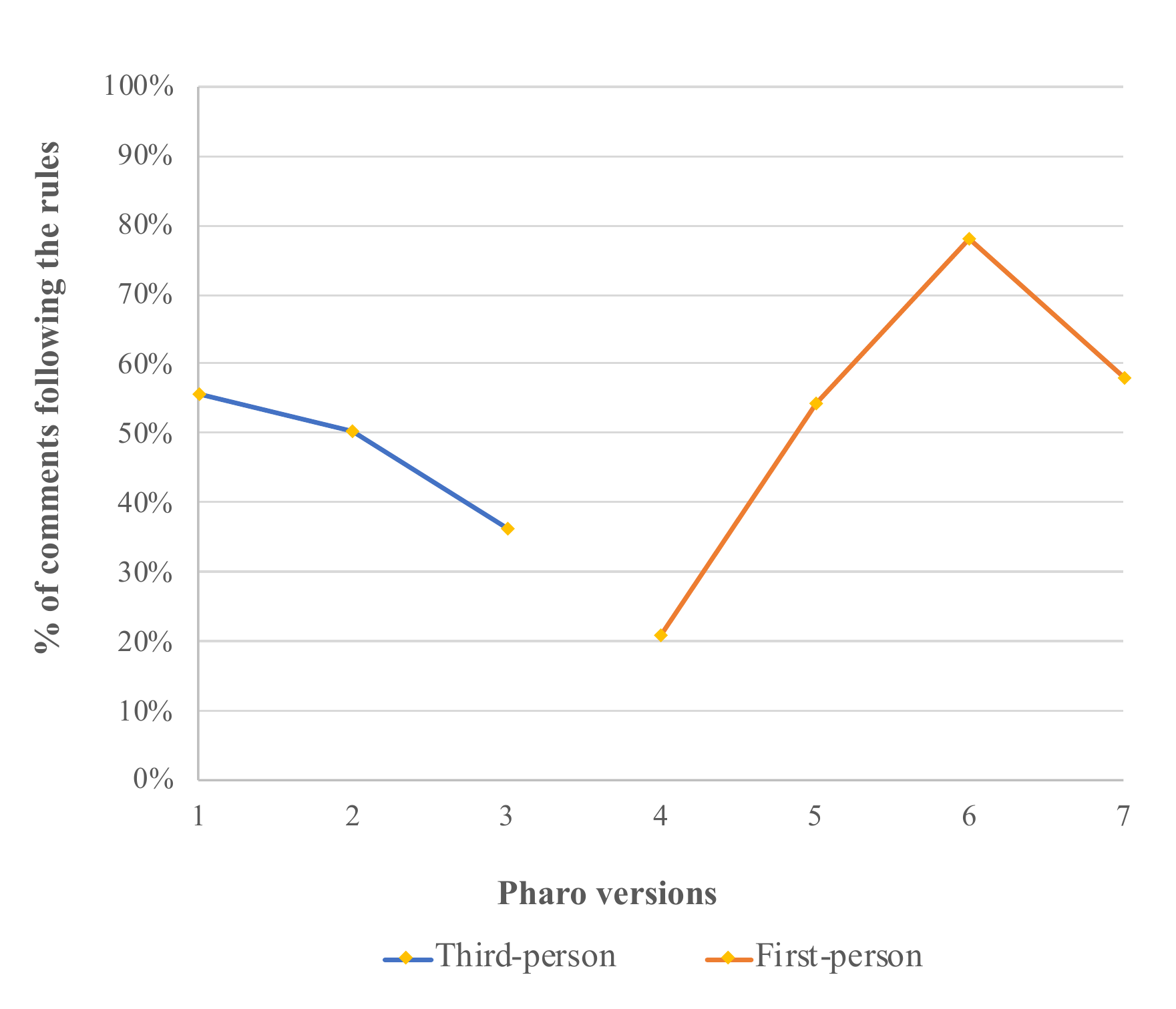}}%
    \qquad
    \subfloat[Comments following formatting guidelines]{
    \figlabel{formatting-rules-analysis}
      \includegraphics[width=0.45\linewidth,height=.47\linewidth]{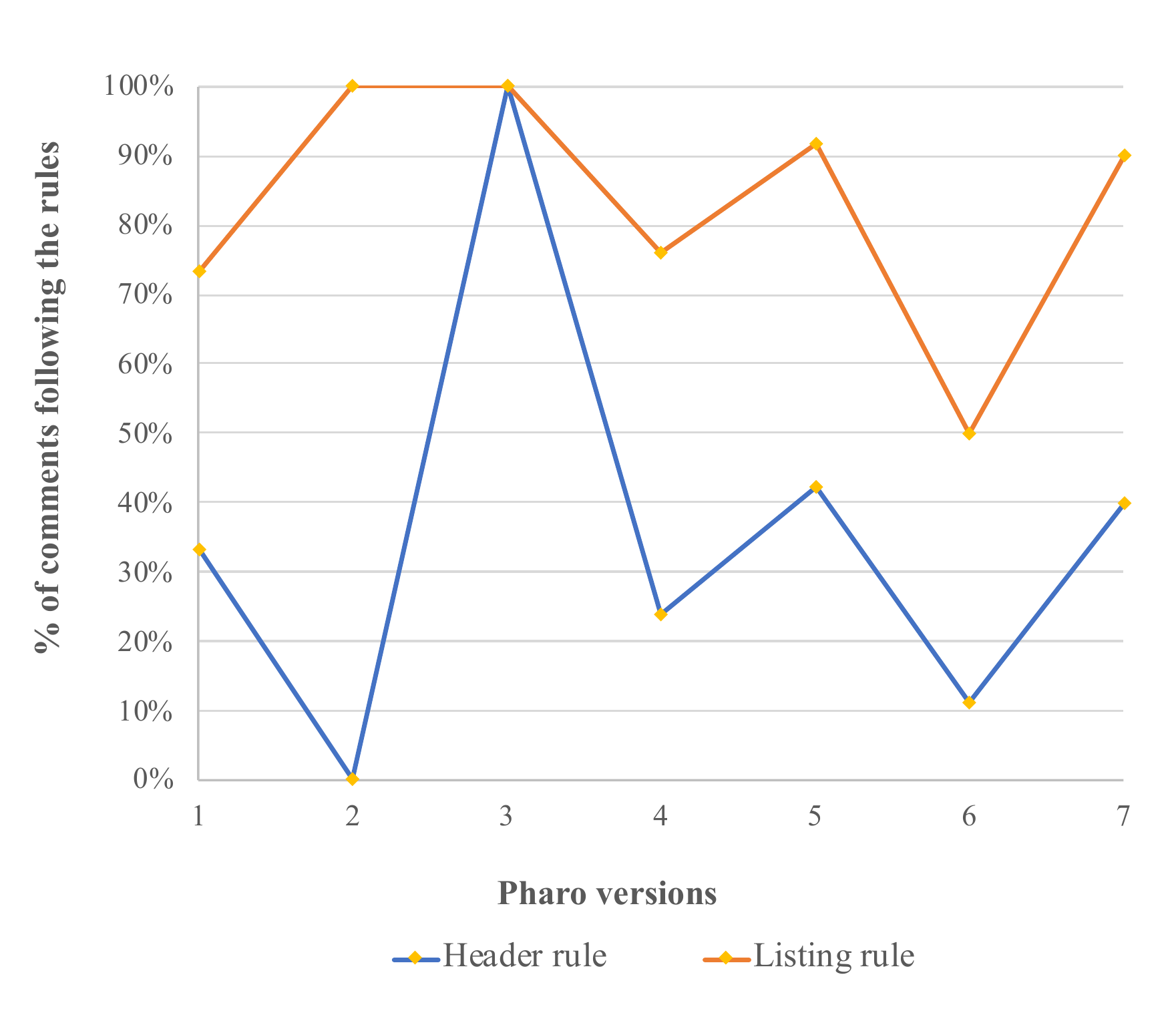}}%
    \caption{Comments following different guidelines over Pharo versions}
\end{figure}

We observe a high degree of inconsistency in using or not using headers to delimit different information types in class comments.
In \figref{formatting-rules-analysis} (Header rule) we see that the use of headers fluctuates significantly across all Pharo versions.
We note a similar fluctuation in the adherence to the rules to document instance variables and Key APIs as lists (\figref{formatting-rules-analysis}, Listing rule).
This indicates the need to have a better and consistent standard for formatting and providing headers for different information types.

\vspace*{2mm}
\hspace*{-5mm}
\begin{tikzpicture}
\node [mybox] (box){%
\centering
\begin{minipage}{.95\textwidth}  
\fontsize{9.5}{9.5}\selectfont 
\emph{\textbf{Finding 11:}
In the majority of Pharo versions, fewer than 40\% of the comments make use of the headers suggested by the comment template.
Where headers are used, developers often use different and inconsistent headers for the same information types.}
\end{minipage}
};
\end{tikzpicture}%

On the other hand, for a few rules, we notice the consistent declining rate of following them.
For instance, in Pharo 1, the rules ask developers to write specific information types in the third person.
Instead, developers often write this information in the first person.
Since Pharo version 5, such rules are respected more than 50\% of the time, showing the increasing usage of first person.
We confirm our observation by mining the rules related to first person and third person from all information types in all versions as shown in \figref{subject-form-analysis}, and find that the usage of third person started declining in the initial versions even though the template proposed to use it.
In later versions since Pharo 4 the usage of first person and active-voice rules is increasing, however, it is still not entirely followed, showing the inconsistency of the writing style in comments.

\vspace*{2mm}
\hspace*{-5mm}
\begin{tikzpicture}
\node [mybox] (box){%
\centering
\begin{minipage}{.95\textwidth}  
\fontsize{9.5}{9.5}\selectfont 
\emph{\textbf{Finding 12:}
Developers use various verb forms to describe the top three information types \emph{Intent}, \emph{Responsibilities}, and \emph{Collaborators} of a class but mainly adhere to the template's use of the first-person pronouns.}
\end{minipage}
};
\end{tikzpicture}%

\emph{Discussion.}
Examining the information types suggested in the template (seven categories), we found that a few information types like \emph{Intent}, and \emph{Responsibility} are found more frequently in the comments than other details, indicating that developers follow the template in writing the template information types.
On the other hand, the availability of extra information types mentioned in different writing styles without a consistent header, like warnings, points out the need for adapting the template to the developer needs.
We believe that adding the commenting guidelines for  other frequent information types in the template will encourage developers to add such details uniformly to their class comments whenever necessary.
We specifically suggest to add headers and organization guidelines about the extra frequent categories to the template, which are not currently present: 
\emph{Reference to external resources}, \emph{Warnings}, \emph{Contracts}, \emph{Dependencies}, \emph{Observation}, and \emph{Todo}.

We additionally observed that Pharo class comments range from high-level design details to low-level implementation details.
This unique way of documenting can help developers and users to get all the information about the class from one place, but poses a challenge at the same time in identifying the specifically required information from such an interwoven text.
Not all developers need to know the low-level details of the class.
A study by Cioch \etal~\citep{Cioc96a} proposes different documents for each stage, \eg interns require task-oriented documentation such as process description, examples, and step-by-step instructions, whereas experts require low-level documentation as well as a design specification.
In the current state of Pharo comments, developers seeking a specific type of information have to go through the whole comment due to the lack of annotations, the non-uniform way of placing information, and the relaxed style conventions.
Similarly, users looking for design details have to go through the implementation details.
Building tools to automatically identify and highlight information from the class comment, according to the desired level of detail and the targeted users of the information, could help developers to search more efficiently within documentation.
At the same time, such tools could also be used to identify the parts of the code that are poorly documented, thus generating documentation fixes.

Analyzing the writing style aspect, we find that developers follow a mix of the first person and third person to express the same information about the class.
Although more than 75\% of the comments of recent versions follow the writing style conventions of the template, there is a substantial proportion of comments that are written differently, creating an inconsistent style across projects.
This suggests a need for better structure conventions, as the template does not follow any strict structural guidelines to organize the content, thus making developers look through the whole comment to find a piece of information.
Encouraging developers to follow structural guidelines in the text, and writing comments with standard headers will allow other developers to extract information from them more easily.
We suggest that the Pharo comment template should impose a formatting and markup style to structure the details in comments.

\subsection{Implications}
\seclabel{rq3-implications}

Assessing the adherence of comments to the suggested guidelines provides important directions on how to maintain comments and keep them consistent with such guidelines.
Based on our study insights, we provide implications for developers and researchers to address the comment quality and consistency with commenting guidelines:\\
\begin{itemize}
	\item 
\emph{Verifying comments adherence in other languages}.
To write useful and consistent comments, numerous programming languages such as Java and Python, and communities such as Google and Oracle, provide coding guidelines~\citep{Goog20a,Orac20a}.
For example, Oracle's guidelines suggest ``using third person (descriptive) style and second person (prescriptive)'' while writing documentation comments, but it is not known whether developers actually follow this guideline in their comments or not.
To ensure developers follow such guidelines, various automated style checkers or linters \eg Checkstyle,\footnote{\url{https://checkstyle.org/checks.html}, accessed on 10 Sep, 2020} Pylint,\footnote{\url{https://www.pylint.org/}, accessed on 10 Sep, 2020}{ ESLint}\footnote{\url{https://eslint.org/}, accessed on 10 Sep, 2020} turn such guidelines into rules and then evaluate the rules against comments automatically.
However, these style checkers are not available for all programming languages, and for the supported ones, they provide limited rules for addressing code commenting guidelines.
The majority are limited to detecting missing comments and verifying formatting guidelines, but not adherences to guidelines concerning the content of comments .
Our results for Pharo show that developers embed template-inspired information types in the class comments.
Moreover, they also follow various syntactic guidelines to add such information types.
Whether developers follow similar commenting guidelines (suggested by the coding guidelines) in other programming languages is not yet explored.
Our dataset provides relevant data in which one can observe which commenting guidelines developers frequently follow in their comments and which they do not. 
Thus, it can help in conducting similar studies for other languages.

\item
\emph{Comment quality tools}:
Researchers have provided various heuristics-based approaches to evaluate comment quality~\citep{Kham10a,Stei13b,Scal16a}.
However, these approaches provide limited checks, they focus on particular programming languages (mainly Java), and they are not designed to be used for other domains and languages~\citep{Kham10a,Stei13b,Scal16a}. 
In particular, most approaches are based on language-specific heuristics such as comment syntax, common keywords used in the comments, and the supported annotations for comments~\citep{Kham10a,Stei13b}, which cannot be directly applied to other languages.
For instance, in Pharo code comments follow a different comment structure and writing style, and do not rely on annotations, which makes these approaches not suitable for this language.
In addition, Tan \etal also showed that previous approaches concerning the detection of inconsistencies in the comments require adaptation to new domains and languages~\citep{Tan07c}.
Hence, our study insights about Pharo commenting practices
provide further data to help researchers in designing tools for assessing comment quality across other languages and domains.

\item
\emph{Template-based comment generation and code summarization approaches}: 
Comment templates not only provide developers with concrete examples on how to write comments, but can also employed by researchers to enable automated generation of code comments for various code entities.
In recent work, Moreno \etal proposed a template-based approach to automatically generate comments for Java classes~\cite{More13c}.
Their template includes certain types of information which they deem essential for understanding a Java class.
However, the information types included in the template were not derived from class comments written by developers, which could make them potentially out of date with current Java commenting practices.
In Pharo, class comments are guided by a default template which includes seven types of information considered important to document a class.
We observed in our study that developers write template-inspired information types more often compared to other information types found in comments. 
We compared the information types included in the class comment template by Moreno \etal and Pharo class comment template.
We observed that their template does not include information types such as \emph{related classes}, \emph{algorithmic implementation details}, or an \emph{example} to show the usage of the class. 
In contrast, the Pharo template includes these information types and Pharo developers frequently refer them and with headers \emph{Collaborators}, \emph{Implementation points}, and \emph{Example} respectively.
On the other hand, both templates suggest describing the intent of the class, responsibilities of the class and the main important methods, which are again frequently reported by Pharo developers.
Thus, our study insights suggest that further information, typically embedded by developers in code comments developers, need to be included in template-based comment generation or code summarization approaches.
\end{itemize}


\section{Threats to validity}
\seclabel{Threats-to-validity}
We now outline potential threats to the validity of
our study.

\emph{Threats to construct validity}
mainly concern the measurements used in the evaluation.
First, we are aware that, to answer research questions RQ2 and RQ3, we sampled only a subset of the extracted class comments.
However, (i) the sample size limits the estimation imprecision to 5\% of error for a confidence level of
95\%, and (ii) to limit the subjectiveness and the bias in the evaluation, three evaluators (three authors of this work) manually analyzed the resulting sample.

Another threat to construct validity concerns the definition of the taxonomy, information types, and writing rules from the template, which are performed on data analyzed by three subjects.
Indeed, there is a level of subjectivity in deciding whether a Pharo comment type belongs to a specific category of the taxonomy or not.
To counteract this issue, we performed a two-level validation step.
This validation step involved further discussion among the evaluators, whenever they had divergent opinions, until they reached a final decision.

\emph{Threats to internal validity} concern confounding factors that could influence our results.
To analyze the commenting trend of old and new classes, we map the classes by their name.
This implies that a renamed class will be considered to be a new class, underestimating the tendency to comment old classes.
The main threat to internal validity in our study is that the assessment is
performed on data provided by human subjects, hence it
could be biased.
To counteract this issue, the evaluators of this work were two Ph.D.\ candidates and one faculty member, each having at least four years of programming experience.
To make transparent all decisions drawn during the evaluation process, all  results of the various validation steps are shared in the replication package (to provide evidence of the non-biased evaluation) and described in detail in the paper.

A second threat involves the taxonomy definition since some of the categories could overlap or be missing in the  \emph{Pharo-CTM}.
To alleviate these issues one of the authors performed a pilot study involving a validation task on a smaller set of Pharo comments.
Then a wider validation was performed involving three authors of this work.
A final threat to the internal validity is represented by the possibility that the chosen sample comments are not representative of the whole population.
To handle this problem we used a stratified sampling approach to choose the sample comments from the dataset, thus considering the quintiles of the comment distribution shown in \figref{comment-distribution-box-plot}.

\emph{Threats to  external validity} concern the generalization of results.
The main aim of this paper is to investigate the class comments and commenting practice evolution characterizing the Pharo core system.
Programmers developing an end-user application might have entirely different commenting practices.
To alleviate this concern to some extent, we analyzed a sample set of comments from a combination of external projects from the Pharo ecosystem.
The projects vary in terms of size, contributors and popularity.
Thus, our empirical investigation is limited to the Pharo ecosystem, and not generalizable to other programming languages.
On the other hand, our results highlight how previous findings on other programming languages\,---\,such as Java~\citep{Stei13b,Pasc17a}, showing that comments contain information like exceptions, IDE directives, bug references, formatters to separate code into logical section, and author ownership\,---\,are not applicable to the Pharo Smalltalk environment.
However, it is important to point out that variables such as developer experience (\eg more experienced developers could be more prone or be more aware of the actual Pharo commenting practices) could have influenced the results and findings of this work.

Finally, during the definition of our taxonomy (\ie  \emph{Pharo-CTM}) we mainly rely on a quantitative analysis of class comments of Pharo, without directly involving the actual Pharo developers.
Thus, for future work, we plan to involve developers in the loop, via surveys and (face-to-face or conference call) interviews.
This step is particularly important for proposing and evaluating automated approaches that can help them achieve a high quality of comments.

\emph{Conclusion Threats}.
We support our findings by using appropriate statistical tests,
such as the Wilk-Shapiro normality test to verify whether the non-parametric test could be applied to our data.
Finally, we used the Vargha and Delaney \^{A}$_{12}$ statistical test to measure the magnitude of the differences between the studied distributions.


\section{Related Work}
\seclabel{Related-work}

\subsection{Comment Evolution}
Considering the importance of code comments, several researchers have analyzed comments quantitatively and qualitatively.
Woodfield \etal study the usefulness of comments quantitatively, and measure the effects of comments on program comprehension~\citep{Wood81a}.
They find that the groups of programmers who were given a program with comments were able to answer more questions about a program in a quiz than the programmers who were given the program without comments.
A few studies focus on the evolution of comments.
Schreck \etal qualitatively analyze the evolution of comments over time in the Eclipse project~\citep{Schr07a},
whereas Jiang \etal~\citep{Jian06a} quantitatively examine the evolution of source code comments in PostgreSQL.
Their focus is on comments associated with functions while we study the comments associated with classes in Pharo and focus on analyzing the comments quantitatively over Pharo versions.

Fluri \etal analyze the co-evolution of code and comments in Java and discover that changes in comments are triggered by a change in source code~\citep{Flur07b}.
They find that newly-added code is rarely commented.
Interestingly, in contrast to their results, we find that the commenting behavior of developers in Pharo is different.
Developers comment newly-added code, as well as commenting old classes.
In another study, Fluri \etal claim that the investigation of commenting behavior of a software system is independent of the object-oriented language under the assumption that common object-oriented languages follow similar language constructs to add comments~\citep{Flur09a}.
We investigate the assumption with another object-oriented programming language and discover that Pharo follows a different comment convention for class comments.
Pharo separates the class comment from the source code and supports different kinds of information like warnings, pre-conditions, and examples in class comments.

\subsection{Comment information categorization}
Comments contain useful information to support various tasks in software development cycle.
Previous literature has explored this idea and analyzed various systems to find the information contained in comments.
We mapped taxonomies of other related work to our work to establish which systems have been analyzed, which kinds of comments are frequently analyzed, and which categories from these works are available in our taxonomy in \tabref{Related-work-comment-info-category-table}.
Several categories from their taxonomy mapped to multiple information types in our taxonomy. 
We highlighted such categories with the symbol (M) in \emph{Mapping to our taxonomy} in \tabref{Related-work-comment-info-category-table}.
In the next paragraphs, we discuss all these related works.

Ying \etal categorize a specific type of comment, namely Eclipse task comments, to see what information they contain. They categorize them on the basis of the various uses of the task comments, such as for communication, or to bookmark current and future tasks~\citep{Ying05a}.
Similarly Hata~\citep{Hata19a} categorized the links found in comments.
Padioleau \etal use multiple dimensions to analyze comments and propose comment categories based on the meaning of a comment.
They use W questions such as ``What is in a comment?'', ``Who can benefit?'', ``Where is the comment located?'', and ``When was the comment written?''
Our aim is to support developers to find important and different kinds of information from the class comment so we choose one specific dimension, namely \emph{``What is in a comment?''}, and classify Pharo class comments accordingly~\citep{Padi09a}.
Haouari \etal categorized the comments based on their position relative to code, \emph{comment type}, \emph{style}, and their \emph{quality.}\citep{Haou11a}
Similar to their work, we also categorized comments based on their content. 
They proposed three subcategories of \emph{comment type}, namely \emph{Explanation comments}, \emph{Working comments}, and \emph{Other}. 
However, due to the abstract nature of these categories, especially \emph{Explanation comments}, most of our categories can fit into it. We categorized the comments based on what specific types of information developers provide.

Steidl \etal assess the quality of comments in Java and C/C++ programs based on different comment categories.
They proposed seven high-level categories based on the position and syntax of the comments, \eg inline comments, block comments \etc~\citep{Stei13b}.
We focus particularly on class comments, which map to their Header comments.
Additionally in Pharo, four other categories (task comments, copyright comments, member comments, and section comments) from their work are available inside Pharo class comments, but are not annotated with any specific tags, and do not have a fixed position as in Java and C/C++.
Farooq \etal compared comments of popular programming languages based on the types of symbols used to denote them, parsing rule, recursivity, and usage of the comments for various purposes such as documentation, and debugging~\citep{Faro15a}.
In our case, the position of Pharo class comments is fixed and does not contain commented code as Pharo class comments are presented in a separate region, therefore, the categorization based on position does not apply to this case.

Pascarella \etal propose a taxonomy of code comments for Java projects~\citep{Pasc17a}.
Five of our categories, namely \emph{Intent}, \emph{Examples}, \emph{Warnings}, \emph{License}, and \emph{References to external documentations}, are close to their taxonomy categories \emph{Rationale},
\emph{Usage}, \emph{Notice}, \emph{License}, \emph{Pointer} respectively.
However, our categorization is specific to class comments. 
We found a number of cases in which the categories from their work did not fit Pharo comments, such as \emph{Ownership}, \emph{Commented code}, \emph{Directive}, \emph{Formatter}, \emph{Discarded}, and \emph{Exception}, due to unavailability of such information in the Pharo class comments.
We found other, different types of information that developers write in Pharo class comments, such as warnings, observations, and contracts, that are not reported in their work.
Zhang \etal constructed a Python comment taxonomy based on the work of Pascarella \etal~\citep{Zhan18a}. 
Shinayam \etal identified the information embedded in local comments, as shown in \tabref{Related-work-comment-info-category-table}~\citep{Shin18a}.
Mapping to their work showed that Pharo class comments contain low-level information also in addition to high-level information.
Based on the mapping analysis, several categories from related work did not map to our taxonomy.
As the scope of comments we analyzed is different from other works \eg Pascarella \etal and Zhang \etal, it is still possible that other kinds of Pharo comments (method comments or inline comments) contain other missing information types.
Additionally, all of the previous classifications have been performed on external projects of a language rather than internal core libraries such as String, or Collection. We categorized the comments from Pharo internal (core) and external projects to identify if developers have different commenting practices in internal and external projects.
In future work, we plan to investigate the class comments of other popular languages and compare them to Pharo commenting practices.

\subsection{Template evolution and adherence}
Nurvitadhi studies the impact of class comments and method comments on program comprehension in Java, and creates a template for class comments in Java~\citep{Nurv03a}.
He suggests to include the purpose of the class, what the class does, and the collaboration between classes.
The Pharo class comment template covers similar aspects with CRC style for the class comment.
However, whether developers follow these aspects or not in their comments is unstudied.
We therefore evaluate the adherence of the template to developer commenting practices.
Jiang \etal study the source code comments in PostgreSQL.
Their focus is on the function comments \ie comments before the declaration of the function named header comments and comments within function body and trailing the functions named non-header comments.
They observe that there is an initial fluctuation in the ratio of header and non-header comments due to the introduction of a new commenting style, but they do not investigate further about the commenting style~\citep{Jian06a}.
Marin investigates the psychological factors that drive developers to comment~\citep{Mari05c}.
The study concludes that developers use different comment styles in their code depending on the programming language they have used earlier.
We also partially confirm this result as we find Java style block comments present in Pharo class comments.
To best of our knowledge, we are first to conduct a study to evaluate the commenting style of developers, and measure the extent of their adherence to the standard guidelines.


\section{Summary}
\seclabel{summary}
High-quality code comments facilitate developers in various development and maintenance tasks~\citep{Souz05a}.
However, their semi-structured or unstructured nature, freedom to adopt various conventions in writing comments, and lack of quality assessment tools make their quality evaluation a non-trivial problem.
Therefore, building tools to ensure their quality requires a good understanding of the system, and the content and style-related aspects that developers follow. 
As not all OOP languages support the same commenting conventions and not all types of comments (class, method, inline) are expected to provide information at the same abstraction level, the quality assessment tools need to be tailored by considering the comment type and practices associated with it.
In this study, we explored Pharo class comments that neither have similar annotations nor the same writing style of Javadocs and Pydocs, thus can provide insights into different code comment characteristics.
To understand Pharo developer commenting practices, we analyzed comments from various prescriptive in terms of when do developers add or change comments, what they write in comments, and whether they follow the commenting guidelines or not in their comments. 

In the context of RQ$_1$ (when do developers add or change comments), we investigated the practices of adding or changing comments in \secref{Comment-trend-analysis} and identified various patterns that trigger comment changes. 
Such patterns are relevant to help developers in building tools to prevent inconsistent comments.
Our results highlighted that developers are motivated to comment on new classes as well as old classes to maintain the overall code-comment ratio (at least 75\% in Pharo). 
Once a particular level is achieved, developers do not put in the same effort, thus indicating the stability of the system.
As discussed in detail in the  implications of RQ$_1$ (\secref{rq1-implications}), we demonstrated the need for tools that are able to support co-evolution analyses of code and comments, by determining specific code changes that (should) trigger comment changes and then updating such comments.

In the context of RQ$_2$ (what do developers write in comments), we qualitatively identified various kinds of information embedded in class comments, as reported in \secref{Pharo-Commenting-Practices}.
We found that developers embed 23 types of information in comments, ranging from high-level design details to implementation-specific details, showing class comments to be a rich source of documentation.
We observed that these information types are present also in Pharo external projects, indicating the Pharo community practice are not limited only to core libraries.
We compared our taxonomy to other similar works in Java, C/C++, and Python~\tabref{Related-work-comment-info-category-table}.
In contrast to Java or Python commenting conventions, 
we found instances of specific information types that are not reported in earlier studies.
Based on our insights, we discussed various implications in \secref{rq2-implications}, in which we highlighted the need for approaches that systematically analyze and compare class commenting practices across languages.
In our comment content investigation, we found several frequent information types that are only \textit{implicitly} present in the text.
As a consequence, identifying such information types from comments automatically is not straightforward due to the unavailability of standard headers or annotations, the inconsistent use of headers, and the lack of a fixed order of writing these information types.
However, our manual analysis highlighted various keywords and patterns to identify certain types of information. 
Such patterns represent an important starting point for researchers interested in designing machine-learning-based approaches and heuristics to identify comment information type automatically.

To investigate RQ$_3$ (whether developers follow the commenting guidelines or not in their comments), we compared comments to the guidelines extracted from the default comment template in \secref{Adherence-to-template}.
We observed that developers write information types mentioned by the comment template more frequently than other information types, but there are some other information types not included in the template that are frequently adopted in practice by developers.
We found that developers follow different conventions to write such information types, thus resulting in the same kinds of information being scattered throughout the comments in different styles.
However, in the majority of comments, developers do follow the writing style of the template in writing such information types.
Hence, while our findings in Pharo suggested that developers follow commenting guidelines, it is yet unknown if this is also the case in other languages.
This motivates the need to explore this aspect in other languages in \secref{rq3-implications}, which is a critical aspect to integrate into comment quality techniques.

Our results shed not only some light about the extent to which developers use comment templates, but suggested to leverage such information to improve the template-based approaches behind various automated comment generation and code summarization techniques.
In essence, our study presented important insights about Pharo commenting practices such as
what do comments contain, what is their writing style, and how do they change over time.
We provided further data to help researchers in designing tools for assessing comment quality across other languages and domains. 


\section{Conclusion}
\seclabel{conclusion}
Class comments can provide a high-level understanding of the program, and help one to understand a complex program.
We analyze the class comments of Pharo releases over 11 years (from 2008 to 2019), characterizing the evolution of commenting practices, identifying the information types from class comments across versions and projects, and assessing the adherence of comments to the commenting guidelines.
This study highlights, from a quantitative and qualitative point of view, 
important patterns concerning class commenting practices of developers.
A direct implication of our work is that, in different programming languages, using the contemporary code comment template is not always ideal when actual practices strongly diverge from it.
This suggests a need to standardize guidelines for formatting and writing headers of the new emerging information types, with the goal of better supporting developer information needs, and ensuring a consistent and higher quality of class comments.
For future work, we are interested in conducting further studies on other programming languages, to investigate potentially different commenting practices, program comprehension, and code documentation patterns.
Additionally, we want to use the identified patterns concerning the implicit information types for building efficient tools to extract the information automatically and (possibly) present the specific information to the developers in a more exhaustive form (\eg by auto-completion of missing comment types).
More in general, we envision as future work, further research effort into
(i) developing tools able to determine the extent to which the code comment template is diverging from current practice;
(ii) automatically identifying information types from comments;
(iii) automatically assessing code comment quality in terms of content, style, and consistency with the source code; and
(iv) automatically generating code comments for templates designed from language guidelines and developer practices.


\begin{acknowledgements}

We gratefully acknowledge the financial support of the Swiss National Science Foundation for the project
``Agile Software Assistance'' (SNSF project No.\ 200020-181973, Feb 1, 2019 - Apr 30, 2022).

\end{acknowledgements}

\bibliographystyle{spbasic}           
\bibliography{scg}

\begin{thebibliography}{56}
\providecommand{\natexlab}[1]{#1}
\providecommand{\url}[1]{{#1}}
\providecommand{\urlprefix}{URL }
\expandafter\ifx\csname urlstyle\endcsname\relax
  \providecommand{\doi}[1]{DOI~\discretionary{}{}{}#1}\else
  \providecommand{\doi}{DOI~\discretionary{}{}{}\begingroup
  \urlstyle{rm}\Url}\fi
\providecommand{\eprint}[2][]{\url{#2}}

\bibitem[{Bavota et~al.(2013)Bavota, Canfora, Di~Penta, Oliveto, and
  Panichella}]{Bavo13b}
Bavota G, Canfora G, Di~Penta M, Oliveto R, Panichella S (2013) An empirical
  investigation on documentation usage patterns in maintenance tasks. In: 2013
  IEEE International Conference on Software Maintenance, IEEE, pp 210--219

\bibitem[{Cioch et~al.(1996)Cioch, Palazzolo, and Lohrer}]{Cioc96a}
Cioch FA, Palazzolo M, Lohrer S (1996) A documentation suite for maintenance
  programmers. In: Proceedings of the 1996 International Conference on Software
  Maintenance, IEEE Computer Society, Washington, DC, USA, ICSM '96, pp
  286--295, \urlprefix\url{http://dl.acm.org/citation.cfm?id=645544.655870}

\bibitem[{Cline(2015)}]{Clin15a}
Cline A (2015) Testing thread. In: Agile Development in the Real World,
  Springer, pp 221--252

\bibitem[{Cornelissen et~al.(2009)Cornelissen, Zaidman, van Deursen, Moonen,
  and Koschke}]{Corn09a}
Cornelissen B, Zaidman A, van Deursen A, Moonen L, Koschke R (2009) A
  systematic survey of program comprehension through dynamic analysis. IEEE
  Transactions on Software Engineering 35(5):684--702,
  \doi{10.1109/TSE.2009.28},
  \urlprefix\url{http://swerl.tudelft.nl/twiki/pub/Main/TechnicalReports/TUD-SERG-2008-033.pdf}

\bibitem[{Dias et~al.(2014)Dias, Peck, Ducasse, and Ar\'evalo}]{Dias14a}
Dias M, Peck MM, Ducasse S, Ar\'evalo G (2014) Fuel: a fast general purpose
  object graph serializer. Software: Practice and Experience 44(4):433--453,
  \doi{10.1002/spe.2136}, \urlprefix\url{http://dx.doi.org/10.1002/spe.2136}

\bibitem[{Ducasse et~al.(2005)Ducasse, G\^irba, and Nierstrasz}]{Duca05f}
Ducasse S, G\^irba T, Nierstrasz O (2005) {Moose}: an agile reengineering
  environment. In: Proceedings of ESEC/FSE 2005, pp 99--102,
  \doi{10.1145/1081706.1081723},
  \urlprefix\url{http://scg.unibe.ch/archive/papers/Duca05fMooseDemo.pdf}, tool
  demo

\bibitem[{Farooq et~al.(2015)Farooq, Khan, Abid, Ahmad, Naeem, Shafiq, and
  Abid}]{Faro15a}
Farooq M, Khan S, Abid K, Ahmad F, Naeem M, Shafiq M, Abid A (2015) Taxonomy
  and design considerations for comments in programming languages: A quality
  perspective. Journal of Quality and Technology Management 10(2)

\bibitem[{Fluri et~al.(2007)Fluri, Wursch, and Gall}]{Flur07b}
Fluri B, Wursch M, Gall HC (2007) Do code and comments co-evolve? {On} the
  {Relation} between {Source Code} and{ Comment Changes}. In: Reverse
  Engineering, 2007. WCRE 2007. 14th Working Conference on, IEEE, pp 70--79

\bibitem[{Fluri et~al.(2009)Fluri, W{\"u}rsch, Giger, and Gall}]{Flur09a}
Fluri B, W{\"u}rsch M, Giger E, Gall HC (2009) Analyzing the co-evolution of
  comments and source code. Software Quality Journal 17(4):367--394

\bibitem[{Goldberg and Robson(1983)}]{Gold83a}
Goldberg A, Robson D (1983) {Smalltalk} 80: the Language and its
  Implementation. Addison Wesley, Reading, Mass.,
  \urlprefix\url{http://stephane.ducasse.free.fr/FreeBooks/BlueBook/Bluebook.pdf}

\bibitem[{Google Style Guidelines(2020)}]{Goog20a}
Google Style Guidelines (2020) Google style guidelines.
  \urlprefix\url{https://google.github.io/styleguide/}, verified on 10 Jan 2021

\bibitem[{Guzzi et~al.(2013)Guzzi, Bacchelli, Lanza, Pinzger, and van
  Deursen}]{Guzz13a}
Guzzi A, Bacchelli A, Lanza M, Pinzger M, van Deursen A (2013) Communication in
  open source software development mailing lists. In: Proceedings of the 10th
  Working Conference on Mining Software Repositories, IEEE Press, pp 277--286

\bibitem[{Haiduc et~al.(2010)Haiduc, Aponte, Moreno, and Marcus}]{Haid10a}
Haiduc S, Aponte J, Moreno L, Marcus A (2010) On the use of automated text
  summarization techniques for summarizing source code. In: 2010 17th Working
  Conference on Reverse Engineering, IEEE, pp 35--44

\bibitem[{Haouari et~al.(2011)Haouari, Sahraoui, and Langlais}]{Haou11a}
Haouari D, Sahraoui HA, Langlais P (2011) How good is your comment? {A} study
  of comments in java programs. In: Proceedings of the 5th International
  Symposium on Empirical Software Engineering and Measurement, {ESEM} 2011,
  Banff, AB, Canada, September 22-23, 2011, {IEEE} Computer Society, pp
  137--146, \doi{10.1109/ESEM.2011.22},
  \urlprefix\url{https://doi.org/10.1109/ESEM.2011.22}

\bibitem[{Hartzman and Austin(1993)}]{Hart93a}
Hartzman CS, Austin CF (1993) Maintenance productivity: Observations based on
  an experience in a large system environment. In: Proceedings of the 1993
  conference of the Centre for Advanced Studies on Collaborative research:
  software engineering-Volume 1, IBM Press, pp 138--170

\bibitem[{Hata et~al.(2019)Hata, Treude, Kula, and Ishio}]{Hata19a}
Hata H, Treude C, Kula RG, Ishio T (2019) 9.6 million links in source code
  comments: Purpose, evolution, and decay. In: Proceedings of the 41st
  International Conference on Software Engineering, IEEE Press, pp 1211--1221

\bibitem[{Ibrahim et~al.(2012)Ibrahim, Bettenburg, Adams, and Hassan}]{Ibra12a}
Ibrahim WM, Bettenburg N, Adams B, Hassan AE (2012) On the relationship between
  comment update practices and software bugs. Journal of Systems and Software
  85(10):2293--2304

\bibitem[{Jiang and Hassan(2006)}]{Jian06a}
Jiang ZM, Hassan AE (2006) Examining the evolution of code comments in
  {PostgreSQL}. In: Proceedings of the 2006 international workshop on Mining
  software repositories, ACM, pp 179--180

\bibitem[{Khamis et~al.(2010)Khamis, Witte, and Rilling}]{Kham10a}
Khamis N, Witte R, Rilling J (2010) Automatic quality assessment of source code
  comments: the {JavadocMiner}. In: International Conference on Application of
  Natural Language to Information Systems, Springer, pp 68--79

\bibitem[{LaToza and Myers(2010)}]{LaTo10b}
LaToza TD, Myers BA (2010) Hard-to-answer questions about code. In: Evaluation
  and Usability of Programming Languages and Tools, ACM, New York, NY, USA,
  PLATEAU '10, pp 8:1--8:6, \doi{10.1145/1937117.1937125},
  \urlprefix\url{http://doi.acm.org/10.1145/1937117.1937125}

\bibitem[{Lidwell et~al.(2010)Lidwell, Holden, and Butler}]{Lidw10a}
Lidwell W, Holden K, Butler J (2010) Universal {Principles} of {Design}.
  Rockport Publishers

\bibitem[{Liu et~al.(2015)Liu, Sun, and Duan}]{Liu15b}
Liu Y, Sun X, Duan Y (2015) Analyzing program readability based on {WordNet}.
  In: Proceedings of the 19th International Conference on Evaluation and
  Assessment in Software Engineering, ACM, p~27

\bibitem[{Maalej et~al.(2014)Maalej, Tiarks, Roehm, and Koschke}]{Maal14a}
Maalej W, Tiarks R, Roehm T, Koschke R (2014) On the comprehension of program
  comprehension. ACM TOSEM 23(4):31:1--31:37, \doi{10.1145/2622669},
  \urlprefix\url{http://mobis.informatik.uni-hamburg.de/wp-content/uploads/2014/06/TOSEM-Maalej-Comprehension-PrePrint2.pdf}

\bibitem[{Marin(2005)}]{Mari05c}
Marin DP (2005) What motivates programmers to comment? Technical Report No
  UCB/EECS-2005018, University of California at Berkeley

\bibitem[{Moose(2020)}]{Moose}
Moose (2020) Moose. \urlprefix\url{https://moosetechnology.org/}, verified on
  10 Jan 2020

\bibitem[{Moreno et~al.(2013)Moreno, Aponte, Sridhara, Marcus, Pollock, and
  Vijay{-}Shanker}]{More13c}
Moreno L, Aponte J, Sridhara G, Marcus A, Pollock LL, Vijay{-}Shanker K (2013)
  Automatic generation of natural language summaries for {Java} classes. In:
  {IEEE} 21st International Conference on Program Comprehension, {ICPC} 2013,
  San Francisco, CA, USA, 20-21 May, 2013, pp 23--32

\bibitem[{Nurvitadhi et~al.(2003)Nurvitadhi, Leung, and Cook}]{Nurv03a}
Nurvitadhi E, Leung WW, Cook C (2003) Do class comments aid {Java} program
  understanding? In: 33rd Annual Frontiers in Education, 2003. FIE 2003., IEEE,
  vol~1, pp T3C--T3C

\bibitem[{Oracle Documentation guidelines(verified on 10 Sep 2020)}]{Orac20a}
Oracle Documentation guidelines (verified on 10 Sep 2020) Oracle documentation
  guidelines.
  \urlprefix\url{https://www.oracle.com/technical-resources/articles/java/javadoc-tool.html},
  https://www.oracle.com/technical-resources/articles/java/javadoc-tool.html

\bibitem[{Padioleau et~al.(2009)Padioleau, Tan, and Zhou}]{Padi09a}
Padioleau Y, Tan L, Zhou Y (2009) Listening to programmers --- taxonomies and
  characteristics of comments in operating system code. In: Proceedings of the
  31st International Conference on Software Engineering, IEEE Computer Society,
  pp 331--341

\bibitem[{Pascarella and Bacchelli(2017)}]{Pasc17a}
Pascarella L, Bacchelli A (2017) Classifying code comments in {Java}
  open-source software systems. In: Proceedings of the 14th International
  Conference on Mining Software Repositories, IEEE Press, MSR '17, pp 227--237,
  \doi{10.1109/MSR.2017.63},
  \urlprefix\url{https://doi.org/10.1109/MSR.2017.63}

\bibitem[{Petrosyan et~al.(2015)Petrosyan, Robillard, and De~Mori}]{Petr15a}
Petrosyan G, Robillard MP, De~Mori R (2015) Discovering information explaining
  {API} types using text classification. In: Proceedings of the 37th
  International Conference on Software Engineering - Volume 1, IEEE Press,
  Piscataway, NJ, USA, ICSE '15, pp 869--879

\bibitem[{Pharo(2020)}]{Pharo}
Pharo (2020) Pharo consortium. \urlprefix\url{http://consortium.pharo.org},
  verified on 10 Jan 2020

\bibitem[{Ratol and Robillard(2017)}]{Rato17a}
Ratol IK, Robillard MP (2017) Detecting fragile comments. In: Proceedings of
  the 32Nd IEEE/ACM International Conference on Automated Software Engineering,
  IEEE Press, pp 112--122

\bibitem[{Robbes et~al.(2010)Robbes, Pollet, and Lanza}]{Robb10a}
Robbes R, Pollet D, Lanza M (2010) Replaying {IDE} interactions to evaluate and
  improve change prediction approaches. In: Proceedings of the 7th IEEE Working
  Conference on Mining Software Repositories, IEEE, MSR '10, pp 161--170,
  \doi{10.1109/MSR.2010.5463278}

\bibitem[{RPackage(2019)}]{RPackage}
RPackage (2019) Replication package.
  \urlprefix\url{https://figshare.com/s/6d039cebc6c2609de11a}, verified on 20
  Nov 2019

\bibitem[{Scalabrino et~al.(2016)Scalabrino, Linares-Vasquez, Poshyvanyk, and
  Oliveto}]{Scal16a}
Scalabrino S, Linares-Vasquez M, Poshyvanyk D, Oliveto R (2016) Improving code
  readability models with textual features. In: 2016 IEEE 24th International
  Conference on Program Comprehension (ICPC), IEEE, pp 1--10

\bibitem[{Schreck et~al.(2007)Schreck, Dallmeier, and Zimmermann}]{Schr07a}
Schreck D, Dallmeier V, Zimmermann T (2007) How documentation evolves over
  time. In: IWPSE '07: Ninth international workshop on Principles of software
  evolution, ACM, New York, NY, USA, pp 4--10, \doi{10.1145/1294948.1294952}

\bibitem[{Shinyama et~al.(2018)Shinyama, Arahori, and Gondow}]{Shin18a}
Shinyama Y, Arahori Y, Gondow K (2018) Analyzing code comments to boost program
  comprehension. In: 2018 25th Asia-Pacific Software Engineering Conference
  (APSEC), IEEE, pp 325--334

\bibitem[{Siegmund and Schumann(2015)}]{Sieg15a}
Siegmund J, Schumann J (2015) Confounding parameters on program comprehension:
  a literature survey. Empirical Software Engineering 20(4):1159--1192

\bibitem[{Soetens et~al.(2017)Soetens, Robbes, and Demeyer}]{Soet17a}
Soetens QD, Robbes R, Demeyer S (2017) Changes as first-class citizens: A
  research perspective on modern software tooling. ACM Comput Surv
  50(2):18:1--18:38, \doi{10.1145/3038926},
  \urlprefix\url{http://doi.acm.org/10.1145/3038926}

\bibitem[{de~Souza et~al.(2005)de~Souza, Anquetil, and de~Oliveira}]{Souz05a}
de~Souza SCB, Anquetil N, de~Oliveira KM (2005) A study of the documentation
  essential to software maintenance. In: Proceedings of the 23rd annual
  international conference on Design of communication: documenting \& designing
  for pervasive information, ACM, New York, NY, USA, SIGDOC '05, pp 68--75,
  \doi{10.1145/1085313.1085331}

\bibitem[{de~Souza et~al.(2006)de~Souza, Anquetil, and de~Oliveira}]{Souz06a}
de~Souza SCB, Anquetil N, de~Oliveira KM (2006) Which documentation for
  software maintenance? Journal of the Brazilian Computer Society 12(3):31--44

\bibitem[{Steidl et~al.(2013)Steidl, Hummel, and Juergens}]{Stei13b}
Steidl D, Hummel B, Juergens E (2013) Quality analysis of source code comments.
  In: Program Comprehension (ICPC), 2013 IEEE 21st International Conference on,
  IEEE, pp 83--92

\bibitem[{Stylos et~al.(2009)Stylos, Myers, and Yang}]{Styl09a}
Stylos J, Myers BA, Yang Z (2009) Jadeite: Improving {API} documentation using
  usage information. In: CHI '09 Extended Abstracts on Human Factors in
  Computing Systems, ACM, New York, NY, USA, CHI EA '09, pp 4429--4434,
  \doi{10.1145/1520340.1520678}

\bibitem[{Tan et~al.(2007)Tan, Yuan, Krishna, and Zhou}]{Tan07c}
Tan L, Yuan D, Krishna G, Zhou Y (2007) /* {iComment}: Bugs or bad comments?*/.
  In: Proceedings of twenty-first ACM SIGOPS symposium on Operating systems
  principles, pp 145--158

\bibitem[{Tenny(1985)}]{Tenn85a}
Tenny T (1985) Procedures and comments vs. the banker's algorithm. ACM SIGCSE
  Bulletin 17(3):44--53

\bibitem[{Tenny(1988)}]{Tenn88a}
Tenny T (1988) Program readability: Procedures versus comments. IEEE
  Transactions on Software Engineering 14(9):1271--1279

\bibitem[{Tomassetti and Torchiano(2014)}]{Toma14a}
Tomassetti F, Torchiano M (2014) An empirical assessment of polyglot-ism in
  {GitHub}. In: Proceedings of the 18th International Conference on Evaluation
  and Assessment in Software Engineering, pp 1--4

\bibitem[{Triola(2006)}]{Trio06a}
Triola M (2006) Elementary Statistics. Addison-Wesley

\bibitem[{Vargha and Delaney(2000)}]{Varg00a}
Vargha A, Delaney HD (2000) A critique and improvement of the {CL} common
  language effect size statistics of {McGraw} and {Wong}. Journal of
  Educational and Behavioral Statistics 25(2):101--132,
  \doi{10.3102/10769986025002101},
  \urlprefix\url{http://dx.doi.org/10.3102/10769986025002101},
  \eprint{http://dx.doi.org/10.3102/10769986025002101}

\bibitem[{Wen et~al.(2019)Wen, Nagy, Bavota, and Lanza}]{Wen19a}
Wen F, Nagy C, Bavota G, Lanza M (2019) A large-scale empirical study on
  code-comment inconsistencies. In: Proceedings of the 27th International
  Conference on Program Comprehension, IEEE Press, pp 53--64

\bibitem[{Woodfield et~al.(1981)Woodfield, Dunsmore, and Shen}]{Wood81a}
Woodfield SN, Dunsmore HE, Shen VY (1981) The effect of modularization and
  comments on program comprehension. In: Proceedings of the 5th international
  conference on Software engineering, IEEE Press, pp 215--223

\bibitem[{Ying et~al.(2005)Ying, Wright, and Abrams}]{Ying05a}
Ying ATT, Wright JL, Abrams S (2005) Source code that talks: An exploration of
  {Eclipse} task comments and {Their Implication} to repository mining. SIGSOFT
  Softw Eng Notes 30(4):1--5, \doi{10.1145/1082983.1083152},
  \urlprefix\url{http://doi.acm.org/10.1145/1082983.1083152}

\bibitem[{Zaidman et~al.(2008)Zaidman, Van~Rompaey, Demeyer, and van
  Deursen}]{Zaid08c}
Zaidman A, Van~Rompaey B, Demeyer S, van Deursen A (2008) Mining software
  repositories to study co-evolution of production and test code. In: Software
  Testing, Verification, and Validation, 2008 1st International Conference on,
  pp 220 --229, \doi{10.1109/ICST.2008.47}

\bibitem[{Zhang et~al.(2018)Zhang, Xu, and Li}]{Zhan18a}
Zhang J, Xu L, Li Y (2018) Classifying python code comments based on supervised
  learning. In: Meng X, Li R, Wang K, Niu B, Wang X, Zhao G (eds) Web
  Information Systems and Applications - 15th International Conference, {WISA}
  2018, Taiyuan, China, September 14-15, 2018, Proceedings, Springer, Lecture
  Notes in Computer Science, vol 11242, pp 39--47,
  \doi{10.1007/978-3-030-02934-0\_4},
  \urlprefix\url{https://doi.org/10.1007/978-3-030-02934-0\_4}

\bibitem[{Zhou et~al.(2017)Zhou, Gu, Chen, Huang, Panichella, and
  Gall}]{Zhou17a}
Zhou Y, Gu R, Chen T, Huang Z, Panichella S, Gall H (2017) Analyzing {APIs}
  documentation and code to detect directive defects. In: Proceedings of the
  39th International Conference on Software Engineering, IEEE Press, pp 27--37

\end{thebibliography}
\newpage
\appendix
\section{Template Models}
\label{appendix:template-models}

\begin{figure}[ht]
    \centering
    \includegraphics[width=\linewidth]{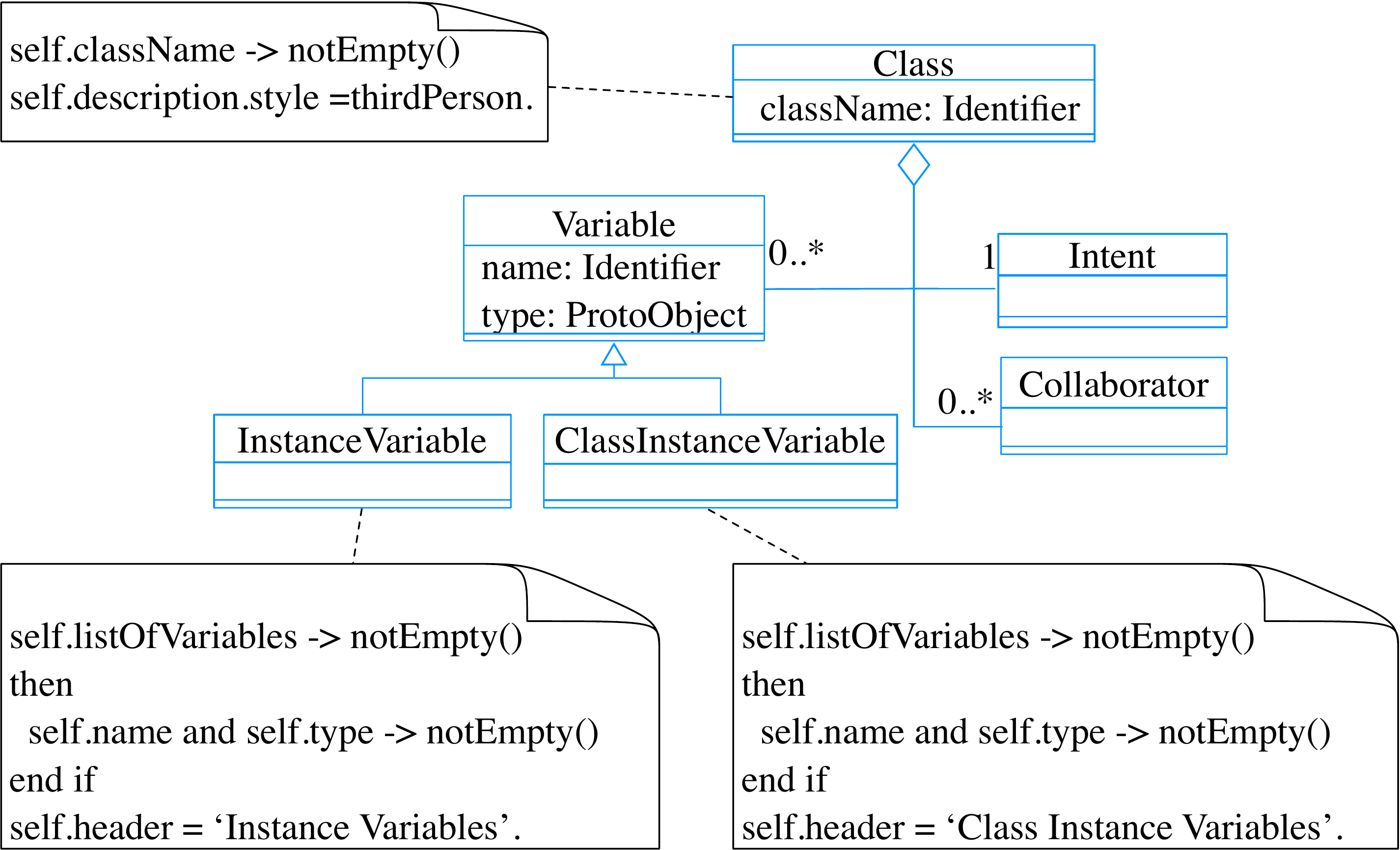}
    \caption{Writing style constraints formulated for Pharo 1 template}
    \figlabel{Pharo-1-writing-style-constraints}
\end{figure}

\begin{figure}[ht]
    \centering
    \includegraphics[width=\linewidth]{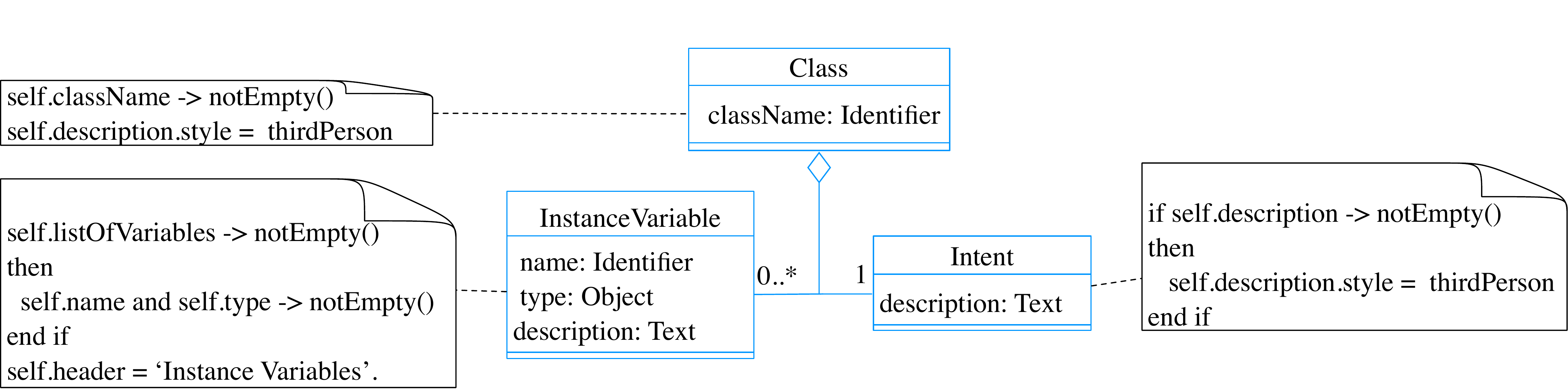}
    \caption{Writing style constraints formulated for Pharo 2 and Pharo 3 template}
    \figlabel{Pharo-2-3-writing-style-constraints}
\end{figure}

\begin{figure}[ht]
    \centering
    \includegraphics[width=\linewidth]{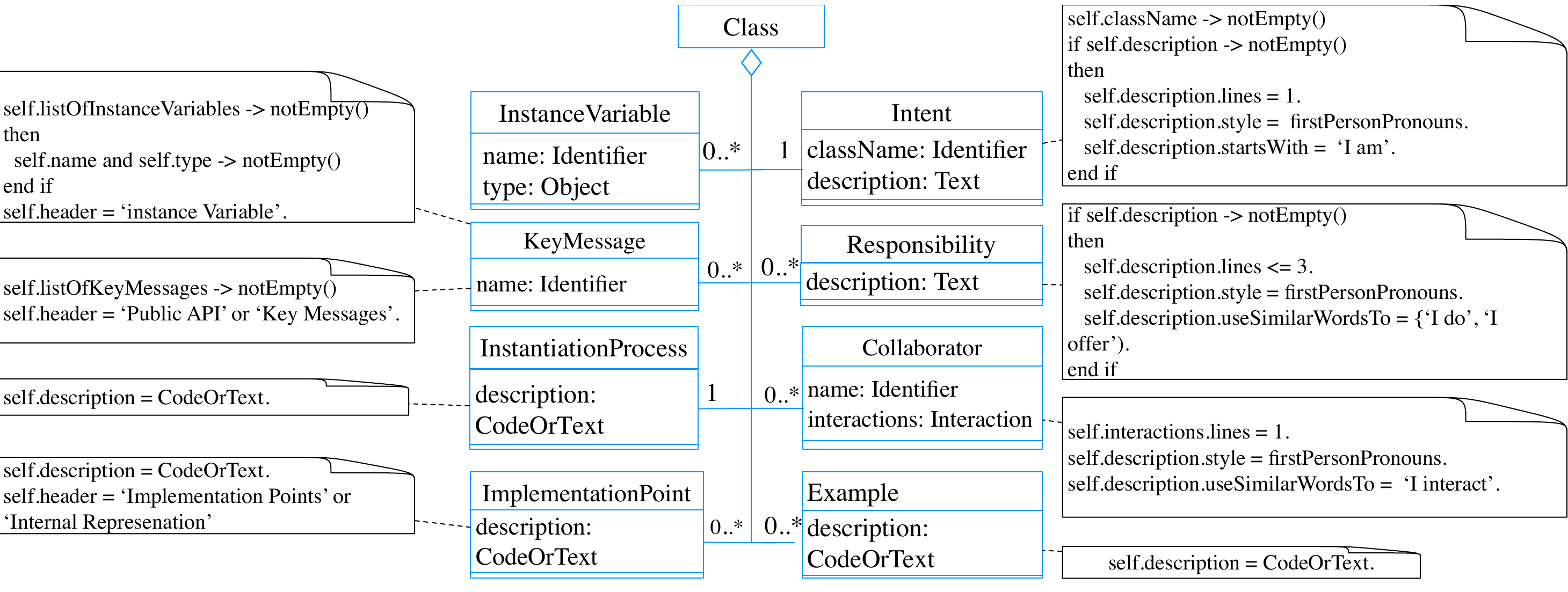}
    \caption{Writing style constraints formulated for Pharo 4 template}
    \figlabel{Pharo-4-writing-style-constraints}
\end{figure}

\begin{figure}[ht]
    \centering
    \includegraphics[width=\linewidth]{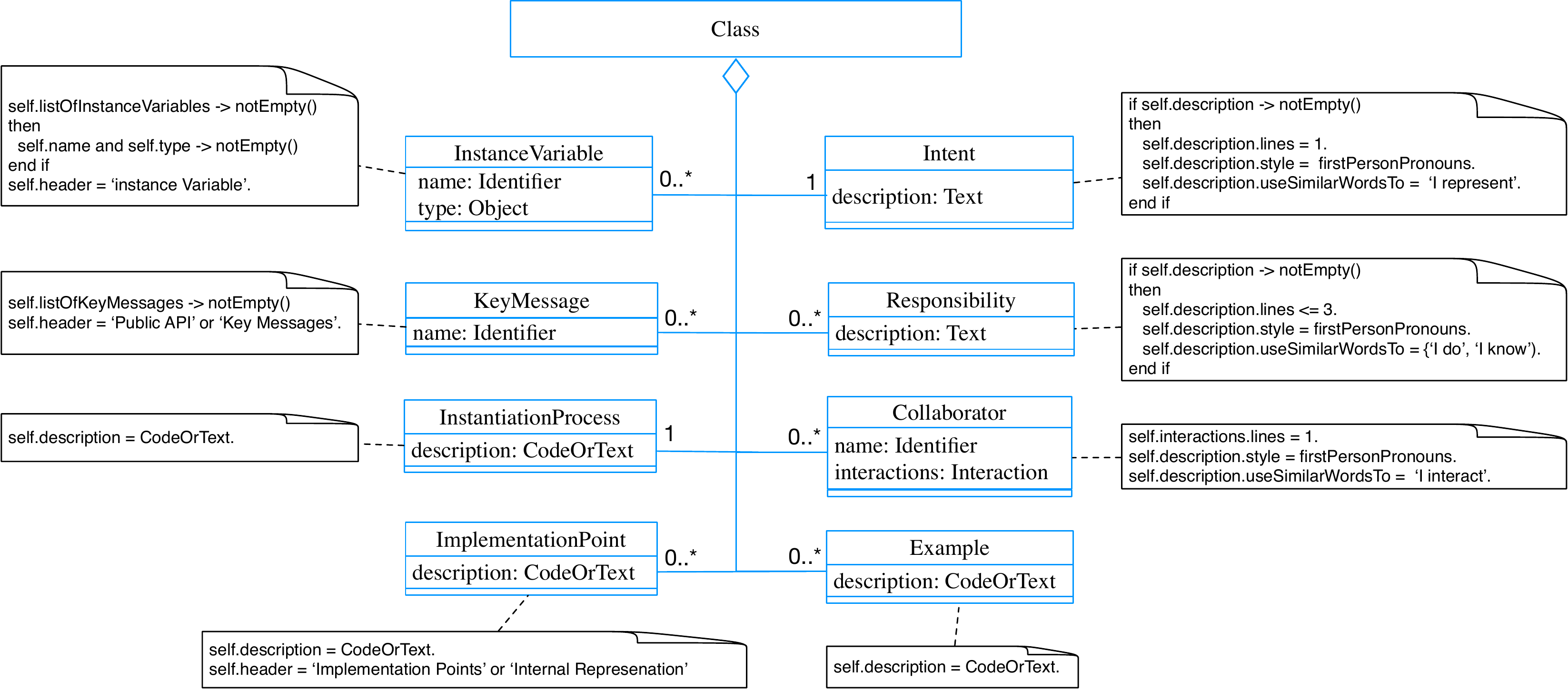}
    \caption{Writing style constraints formulated for Pharo 5,6,7 template}
    \figlabel{Pharo-5-6-7-writing-style-constraints}
\end{figure}

\end{document}